\PassOptionsToPackage{unicode}{hyperref}
\PassOptionsToPackage{hyphens}{url}
\documentclass[
  12pt,
]{article}
\usepackage{amsmath,amssymb}
\usepackage{iftex}
\ifPDFTeX
  \usepackage[T1]{fontenc}
  \usepackage[utf8]{inputenc}
  \usepackage{textcomp} 
\else 
  \usepackage{unicode-math} 
  \defaultfontfeatures{Scale=MatchLowercase}
  \defaultfontfeatures[\rmfamily]{Ligatures=TeX,Scale=1}
\fi
\usepackage{lmodern}
\ifPDFTeX\else
\fi
\IfFileExists{upquote.sty}{\usepackage{upquote}}{}
\IfFileExists{microtype.sty}{
  \usepackage[]{microtype}
  \UseMicrotypeSet[protrusion]{basicmath} 
}{}
\makeatletter
\@ifundefined{KOMAClassName}{
  \IfFileExists{parskip.sty}{%
    \usepackage{parskip}
  }{
    \setlength{\parindent}{0pt}
    \setlength{\parskip}{6pt plus 2pt minus 1pt}}
}{
  \KOMAoptions{parskip=half}}
\makeatother
\usepackage{xcolor}
\usepackage[margin=1in]{geometry}
\usepackage{longtable,booktabs,array}
\usepackage{calc} 
\usepackage{etoolbox}
\makeatletter
\patchcmd\longtable{\par}{\if@noskipsec\mbox{}\fi\par}{}{}
\makeatother
\IfFileExists{footnotehyper.sty}{\usepackage{footnotehyper}}{\usepackage{footnote}}
\makesavenoteenv{longtable}
\usepackage{graphicx}
\makeatletter
\def\maxwidth{\ifdim\Gin@nat@width>\linewidth\linewidth\else\Gin@nat@width\fi}
\def\maxheight{\ifdim\Gin@nat@height>\textheight\textheight\else\Gin@nat@height\fi}
\makeatother
\setkeys{Gin}{width=\maxwidth,height=\maxheight,keepaspectratio}
\makeatletter
\def\fps@figure{htbp}
\makeatother
\providecommand{\tightlist}{%
  \setlength{\itemsep}{0pt}\setlength{\parskip}{0pt}}
\usepackage{enumerate}
\usepackage{natbib}
\usepackage{hyperref}
\usepackage{mathtools}
\usepackage{amsthm}
\usepackage{amsmath}
\usepackage{amssymb}
\usepackage{physics}
\usepackage{booktabs}
\usepackage{longtable}
\usepackage{lscape}
\usepackage{subcaption}
\usepackage{setspace}
\usepackage{placeins}

\setcitestyle{aysep={}}

\newcommand\numberthis{\addtocounter{equation}{1}\tag{\theequation}}
\newcommand{\R}{\ensuremath{\mathbb{R}}}

\newcommand{\eps}{\varepsilon}
\newcommand{\dfeq}{\coloneqq}

\newcommand{\ind}{\mathbf{1}}

\newcommand{\Bernoulli}{\mathrm{Bernoulli}\qty}
\newcommand{\Norm}{\mathcal{N}\qty}

\usepackage{booktabs}
\usepackage{longtable}
\usepackage{array}
\usepackage{multirow}
\usepackage{wrapfig}
\usepackage{float}
\usepackage{colortbl}
\usepackage{pdflscape}
\usepackage{tabu}
\usepackage{threeparttable}
\usepackage{threeparttablex}
\usepackage[normalem]{ulem}
\usepackage{makecell}
\usepackage{xcolor}
\ifLuaTeX
  \usepackage{selnolig}  
\fi
\IfFileExists{bookmark.sty}{\usepackage{bookmark}}{\usepackage{hyperref}}
\IfFileExists{xurl.sty}{\usepackage{xurl}}{} 
\hypersetup{
  pdftitle={Measuring and Modeling Neighborhoods},
  pdfauthor={Cory McCartan; Jacob R. Brown; Kosuke Imai},
  pdfkeywords={hierarchical model, measurement, partisanship, race, residential segregation, surveys},
  hidelinks,
  pdfcreator={LaTeX via pandoc}}

\title{Measuring and Modeling Neighborhoods\footnote{The authors thank Ryan Enos for sharing TargetSmart voter file data, members of the Imai Research Group for input on the survey design, helpful feedback at the Political Methodology Society 2022 Summer Meeting and the Harvard University's Applied Statistics Workshop, through IQSS's Magaro Peer Pre-review Program, and Justin de Benedictis-Kessner, Michael Hankinson, Aaron Kaufman, Melissa Sands, and Kim Sønderskov for detailed comments. The open-source survey instrument for measuring subjective neighborhoods is available at \url{https://github.com/CoryMcCartan/neighborhood-survey} while the software package for analyzing the collected data is available at \url{https://github.com/CoryMcCartan/nbhdmodel}. The experimental analysis was pre-registered at Open Science Foundation, pre-registration materials can be found at \url{https://osf.io/xdumv/}.}}
\author{Cory McCartan\footnote{Faculty Fellow, Center for Data Science, New York University, Email \texttt{corymccartan@nyu.edu}; ORCID 0000-0002-6251-669X} \and Jacob R. Brown\footnote{Assistant Professor, Department of Political Science, Boston University, Email \texttt{jbrown13@bu.edu}; ORCID 0000-0001-8037-5360} \and Kosuke Imai\footnote{Corresponding author. Professor, Department of Government and Department of Statistics, Harvard University; URL \url{https://imai.fas.harvard.edu/}; Email \texttt{imai@harvard.edu}; ORCID 0000-0002-2748-1022}}
\date{December 7, 2023}

\begin{document}
\maketitle
\begin{abstract}
Granular geographic data present new opportunities to understand how neighborhoods are formed, and how they influence politics. At the same time, the inherent subjectivity of neighborhoods creates methodological challenges in measuring and modeling them. We develop an open-source survey instrument that allows respondents to draw their neighborhoods on a map. We also propose a statistical model to analyze how the characteristics of respondents and local areas determine subjective neighborhoods. We conduct two surveys: collecting subjective neighborhoods from voters in Miami, New York City, and Phoenix, and asking New York City residents to draw a community of interest for inclusion in their city council district. Our analysis shows that, holding other factors constant, White respondents include census blocks with more White residents in their neighborhoods. Similarly, Democrats and Republicans are more likely to include co-partisan areas. Furthermore, our model provides more accurate out-of-sample predictions than standard neighborhood measures.
\end{abstract}

\setstretch{1.1}

\newpage

\hypertarget{introduction}{%
\section{Introduction}\label{introduction}}

The availability of granular geographical data, together with increasing computing power, provide researchers with new opportunities to gain insights on how local geography influences politics.
Recent research uses such data to study the effects of neighborhoods on political behavior \citep{larsen_hjorth_dinesen_sonderskov_2019}, racial politics \citep{Enos:2017space, nuamah_ogorzalek_2021}, partisan sorting \citep{martin:webster, brown_measurement_2021}, public goods provision \citep{Wong:2010, Trounstine:2015segregation}, housing \citep{hankinson_2018}, and political representation \citep{rodden2019cities}.
These works have brought novel empirical evidence and new substantive arguments to a long-standing literature on the political and socio-economic consequences of local geography \citep{HuckfeldtSprague:1987, putnam2000bowling}.

At the same time, these complex data pose methodological challenges of \emph{measuring} and \emph{modeling} one's neighborhood.
Foundational work conceptualizes neighborhoods as sub-units of larger geographies (such as cities or towns) that arise from population grouping, infrastructure, land use, and economic forces \citep{ParkBurgess:city, Suttles:1972}.
However, neighborhoods are inherently subjective because they are shaped by personal experiences and views \citep{Paddison:1983, chaskin1997, keller2003community}.\footnote{In this paper, we focus on neighborhoods defined by people and do not study neighborhoods officially defined by cities or other administrative units. The proposed statistical model, however, can be applied to official neighborhoods as well.}
Thus, two individuals who live at the same address may identify different local communities as their neighborhoods.
This intrinsic subjectivity of neighborhoods leads to a substantial variation not just across places, but also across people \citep{coulton_mapping_2001}.

Unfortunately, most studies do not account for this subjectivity when measuring neighborhoods.
Many researchers approximate neighborhoods by administrative units such as census tracts and ZIP Codes \citep{Gay:2006, Hopkins:2010, BaxterKingEtAl:2022, hamel_wa:2022}.
These approaches implicitly assume that all individuals who live in the same unit would define their neighborhood in the same way that exactly matches with an administrative boundary \citep{Openshaw:1983, white1983measurement, coulton_how_2013}.
More recent work improves upon this shortcoming by using metrics based on distance and population density that are specific to an individual residence location \citep{DinesenSonderskov:2015, brown_measurement_2021}.
While such measures vary across people and places, they cannot directly account for factors that influence subjective neighborhoods.
Such factors include the demographic and other characteristics of individuals, their behaviors and opinions, administrative boundaries, and physical objects in local areas such as buildings, parks, and roads.

These measurement challenges thus create persistent problems for a wide range of contemporary research. For research on the effect of neighborhoods on political behavior, researchers must rely on definitions of local context that do not fully capture the influential areas around each voter \citep{nathan_sands:2023}.
For research on racial politics, residential proximity is frequently used as a proxy for intergroup contact but these studies cannot discern whether people perceive themselves as sharing geographic space with other racial groups \citep{Enos:2017space}.
The measurement of segregation is similarly limited since the standard measures of local exposure may understate the extent to which people encounter other racial or ethnic groups \citep{athey_et_al:2021, hamel_wa:2022}.
To understand how geography may shape public goods provision, researchers must investigate how local governments view the areas they govern when making decision about where to allocate resources \citep{Trounstine:2015segregation}.
Research on housing and NIMBYism requires accurate measurement of the areas where residents might be opposed to new development \citep{hankinson_2018}.
Lastly, one common requirement of legislative redistricting is to keep ``communities of interest'' intact when drawing district boundaries, but these communities are subjectively defined \citep{chambers_et_al}.

Therefore, novel methods are required to better understand the variation in perceptions of local geography across places and people. The main goal of this paper is to provide new methodological tools that address these limitations of existing approaches and further facilitate empirical studies of neighborhoods. In a pioneering study, \citet{WongBowersWilliamsDrake:2012} address this measurement problem by asking survey respondents to draw their own neighborhoods on a map \citep[see also][]{wong_bowers_rubenson_fredrickson_rundlett_2020}. We follow their innovative measurement strategy, developing software to collect such data and formalizing a model for analyzing how neighborhoods are defined.

\hypertarget{methodological-contributions}{%
\subsection{Methodological contributions}\label{methodological-contributions}}

The methodological contribution of this paper is twofold. First, we develop an easy-to-use online survey instrument to measure subjective neighborhoods. Our instrument is customizable and easily incorporated into standard online survey platforms such as Qualtrics, facilitating its use by other researchers. With this tool, researchers can collect maps drawn by survey respondents. As we illustrate in our empirical applications, the researcher may choose to ask respondents to draw their neighborhood or community of interest on a map. Our survey instrument can also be used for other purposes, for example, asking respondents to highlight their route to work, or, for scholars of civil war, collecting citizen perceptions of which areas different militia groups control. The tool is completely open-source and can incorporate different researcher design decisions (\url{https://github.com/CoryMcCartan/neighborhood-survey}).

As our empirical applications demonstrate, the most direct use of this survey instrument is to measure how people define their neighborhoods, communities of interest, or other subjective geographic definitions. Our statistical model then allows researchers to model these maps as the outcome and quantify the predictive influence of aggregate and individual characteristics on the characteristics of one's subjective neighborhood.

The survey instrument can be used for broader purposes as well. Many survey studies utilize contextual variables in their analyses. These include summaries of racial demographics \citep{BoboHutchings:1996,  Gay:2006, Hopkins:2010, Newman:2012, anoll2018}, neighborhood economic conditions \citep{Michener:2013, larsen_hjorth_dinesen_sonderskov_2019}, local political context or partisan composition \citep{mason_wronski_kane_2021, BaxterKingEtAl:2022}, and many other variables that are measured at some geographic unit. For each of these studies, researchers must choose the relevant geography at which to calculate aggregate summaries. Instead of relying on such proxy variables, our survey instrument enables researchers to directly measure respondent-defined local context and compute its aggregate characteristics of interest within the neighborhood drawn by each individual.

For example, a survey on how neighborhood racial composition drives political attitudes could collect individual-defined neighborhoods, calculate percent Black, White, and Hispanic in these drawn neighborhoods, and predict attitudes as a function of these demographics.\footnote{In the Additional Supplementary Information Table 3, we demonstrate such analyses, modeling the relationship between percent same race, percent same party, and percent college educated in a drawn neighborhood on trust in one's neighbors.} This use of subjective neighborhoods as an improved measure of contextual variables has been advocated for in previous work \citep[see][]{wong_bowers_rubenson_fredrickson_rundlett_2020}, but our open-source survey instrument allows for any researcher to adopt this approach.

The second methodological contribution is the development of a new statistical model that takes full advantage of this new measurement tool. Existing studies, including those that measure subjective neighborhoods, do not directly model how respondent and geographic characteristics, and their interactions, relate to one's neighborhood.
Instead, they almost exclusively rely upon descriptive statistics of observed neighborhoods such as racial and economic demographics, neighborhood size, and agreement with administrative boundaries to describe subjective neighborhood definitions \citep{wong_bowers_rubenson_fredrickson_rundlett_2020}.
The absence of a formal statistical model makes it difficult to systematically analyze the characteristics of respondents and places that together determine subjective neighborhoods.

We propose a Bayesian hierarchical model based on the probability that survey respondents include each small local area (e.g., census block) at the margin of their neighborhood.
The proposed model quantifies the degree to which characteristics of respondents, those of local areas, and their interactions shape subjective neighborhoods. Respondent characteristics can include demographic attributes and any attitudinal or behavioral measures, whereas the area characteristics may include census statistics, administrative boundaries, and the location of landmarks such as churches, parks, schools, and highways.
Like the survey instrument, this model is implemented as part of an open-source software package (\url{https://github.com/CoryMcCartan/nbhdmodel}).

This new statistical model offers several improvements over simpler methods such as regressing neighborhood summary statistics on a set of predictors.

\begin{itemize}
\item
  \emph{The model uses more information.} We model the probability that each constituent Census block is included in a neighborhood, leveraging granular block-level characteristics. A summary-statistic-based regression typically average these characteristics, losing valuable information contained in the respondent's decision to include or exclude each Census block, especially those near the boundaries of subjective neighborhoods.
\item
  \emph{Estimation uncertainty is quantified.} Our proposed Bayesian approach naturally accounts for estimation uncertainty in the model parameters, which is propagated through posterior predictions and other post-analysis summaries.
\item
  \emph{All characteristics of the neighborhood can be modeled simultaneously.} The model can incorporate any characteristics of local areas, those of respondents, and their interactions. Because the model is at the level of the actual Census blocks, any higher-level neighborhood summary statistic can be calculated for the model predictions and fitted values, allowing formal statistical quantification of differences in these statistics.
\item
  \emph{The model can be used to make individual-level predictions for neighborhoods or portions thereof.} We can sample new neighborhoods from the posterior of the model, including those for out-of-sample respondents or for counterfactual covariate values. For example, the model allows one to sample possible neighborhoods that would be drawn by a survey respondent with a certain set of characteristics if they lived at a different address.
\end{itemize}

In our empirical applications, we use the model to understand how individuals define their neighborhoods (broadly defined), and how they perceive their communities of interest as it relates to city council redistricting. While our contributions are methodological rather than theoretical, we believe that the proposed methodology can facilitate conceptual development by enabling researchers to empirically study a host of questions about neighborhoods. For example, how do infrastructure and buildings such as churches, community centers, high-ways, and libraries shape the way in which people view their local areas \citep{putnam2000bowling, HuckefeldtEtAl:1993}? How do new zoning rules and administrative boundaries affect neighborhoods of different people \citep{shlay1981}? Individual characteristics of neighbors may also be influential. People may define their local geography differently based on the race, religion, class, or even partisanship of the people they live around \citep{HuckfeldtSprague:1987,  Enos:2017space}. The proposed model can quantify the extent to which these individual and contextual characteristics together influence their subjective neighborhoods.

For those who are more interested in institutions than individuals, our modeling strategy can be used to study any geographic unit related to governance, where it can illuminate how resources and political power are allocated.
For example, our methodology could be applied to analyze political districts \citep{laraja:2009}, school districts \citep{fischel2009making, monarrez2022effect}, annexation and city incorporation \citep{AUSTIN1999501, LeonMoreta:2015}, the allocation of public goods across geography---i.e., which areas are the focus of urban renewal programs or grant investment \citep{Trounstine:2015segregation}---which areas receive more or less policing \citep{SossWeaver:2017}, or historical redlining \citep{AaronsonEtAl:2021}.
Each of these applications can be implemented if researchers have map data on the relevant geographic units and accompanying demographic data or other covariates.

\hypertarget{empirical-applications}{%
\subsection{Empirical applications}\label{empirical-applications}}

We apply the proposed methodology to two original surveys.
First, we examine whether people will define their neighborhoods in exclusionary terms, giving preference to in-group members and excluding out-group members---focusing on race and party as the salient group categories.
The racial composition of one's neighborhood is a powerful determinant of how individuals perceive the space around them \citep{Wong:2010}, influencing residential sorting, neighborhood trust, exclusionary attitudes, and group conflict \citep{MasseyDenton:1993, Enos:2017space}.
Likewise, Democrats and Republicans are increasingly likely to live separate from one another \citep{rodden2019cities, brown_measurement_2021}, and this partisan homogeneity influences political attitudes and behaviors \citep{perez_conformity, handan-nader_ho_morantz_rutter_2021}.
As such, when people consider their neighborhood or local community, they may define it along racial dimensions.
Existing research also demonstrates growing affective partisan polarization, where voters of each political party increasingly express dislike for out-partisans \citep{IyengarWestwood:2014}. Thus, partisan composition, similar to racial composition, may be an important social dimension upon which people will define their neighborhood.

To test these hypotheses, we analyze the responses from 2,508 registered voters across three major metropolitan areas: Miami, New York City, and Phoenix.
We demonstrate the model's ability to quantify the degree to which racial and partisan compositions influence how individuals draw their neighborhoods.
In addition to respondent characteristics (e.g., race, party, age, gender, education, and home-ownership), the model also incorporates various contextual variables that are known to affect the way in which local communities are formed \citep{HopkinsWilliamson:2010}.
They include local institutions (e.g., schools, parks, and places of worship), physical characteristics (e.g., land area, population, major roadways), administrative boundaries, and racial, partisan, and economic demographics of local areas.

Our analysis shows that race and partisanship are significant predictors of subjective neighborhoods.
Net of other factors, White respondents are 6.0 to 13.6 percentage points more likely to include in their neighborhoods a marginal census block composed entirely of White residents compared to one with no White residents.
Democratic and Republican respondents are 9.4 to 23.4 percentage points more likely to include an entirely co-partisan marginal census block compared to one consisting entirely of out-partisans.
These predictive effects are found even after accounting for other socio-economic demographics, local infrastructure, and administrative boundaries and survey respondent characteristics.

In our second application, we examine how residents define their communities of interest (COI) by conducting an additional survey of 627 New York City residents.
When drawing political districts for Congress, state legislatures, and city councils, many states and cities require inclusion of COI in the same district \citep{laraja:2009}.
The exact definition of COI varies, but they refer to groups of people who live in geographic proximity to one another and share political, economic, and other interests.
The \cite{nyc_charter}, for example, stipulates that city council districts shall ``keep intact neighborhoods and communities with established ties of common interest and association, whether historical, racial, economic, ethnic, religious or other.''
Preservation of these communities within single districts may reduce inequities or imbalances in the redistricting process \citep{barabas_jerit_2004}.
Some states and cities have collected citizen-drawn maps of communities of interest, and recent work has collected these data and introduced methods for classifying them \citep{chambers_et_al}.
Our methodology can be applied to understand communities of interest, quantifying which factors influence residents to define their political communities and demonstrating whether existing districts reflect these communities.

We ask these respondents to draw on a map the areas around where they live that reflect their community of interest and thus should be included in their city council district.
Thus, unlike the first survey, we give respondents a specific definition of neighborhood to be elicited.
To the best of our knowledge, we are the first to analyze citizens' preferences in defining their communities of interest, which represent a key factor in many legislative redistricting cases.

We use the same model specification as the first survey, inferring how individual and contextual characteristics influence definitions of communities of interest.
Our analysis shows that race plays even a stronger role in predicting one's community of interest than for the subjective neighborhood survey.
Both White respondent and minority respondents demonstrate strong tendencies to include census blocks with more co-ethnic residents in their city council district.
We also find similar co-partisan preferences for Democrats and stronger co-partisan preferences for Republicans.

After analyzing both surveys, we examine the out-of-sample prediction performance of the proposed model. We find that our model generally out-performs conventional neighborhood definitions based on distance and administrative units such as Census tracts or ZIP Codes. The proposed model has a higher out-of-sample prediction performance for communities of interest than for subjective neighborhoods, suggesting that a more specific definition of subjective neighborhood may yield model prediction with greater precision. Additionally, we illustrate how these predictions can be used to incorporate communities of interest into the redistricting process.

In both surveys, we find substantial individual heterogeneity in the size of drawn neighborhoods and communities.
These findings, while making out-of-sample prediction more difficult, also underscore the limitations of one-size-fits-all approaches to empirically studying neighborhoods, such as using administrative units.

In the Supplementary Information (SI), we provide three additional applications of the proposed methodology to illustrate its wide applicability.
First, we conduct a survey experiment by randomly assigning respondents to draw their neighborhoods on maps with or without information about the racial and partisan makeup of surrounding areas.\footnote{Due to space constraints, the experimental results can be found in the Additional Supplementary Information that will be made available in the replication repository upon publication.}
We find that while the aforementioned patterns of racial and partisan homophily are present across experimental conditions, such information does not fundamentally change how voters draw their neighborhoods.
Second, we collect respondent attitudes on the construction of new housing in their neighborhoods and test whether opposition to new housing intensifies exclusionary preferences.
Lastly, we measure attitudes on trust in one's neighbors, and test how these attitudes shape the influence of different factors on neighborhood definitions.
These additional analyses further illustrate the wide applicability of the proposed methodology.

\hypertarget{sec:measuring}{%
\section{Measuring neighborhoods}\label{sec:measuring}}

In this section, we use one of our two original surveys to explain how our mapping tool measures subjective neighborhoods.

\hypertarget{survey-setup}{%
\subsection{Survey setup}\label{survey-setup}}

Data for this study comes from an original survey of 2,508 respondents across three U.S. cities: Miami (\(n=473\)), New York City (\(n=450\)), and Phoenix (\(n=1,585\)).
These cities were chosen to provide a variety of political and regional contexts, with the aim to collect large enough samples for each city to conduct within-city analysis.

Survey respondents were recruited via e-mail using a list of email addresses attached to registered voter records.\footnote{This particular sampling frame is a result of our reliance on voter file data that has been merged with emails of voters. While the availability of email addresses made it possible for us to recruit survey respondents, the proposed methodology can be used with other sampling frames and recruitment methods. For example, one could work with a survey firm that already has secured a panel of online survey participants.  It is also possible to use an in-person survey with computer-assisted interviews where respondents use a tablet to draw neighborhoods.  These alternative sampling frames and collection methods will be more expensive but are likely to increase the response rate.} The list was provided to the researchers by the vendor Target Smart.
Those who did not respond to the initial invitation were sent up to 3 weekly reminder e-mails.
No compensation was offered or provided to respondents and no deception was employed in this study.
Section S1 in the SI contains more information on the sampling process.

Among voters who received a survey invitation, response rate to the survey invitations was 0.8\% (0.5\% among total voters sampled). This is only slightly lower than previous survey recruitments using another email list a similar vendor \citep[see e.g.,][]{brown_measurement_2021}. In total, we collected 7,691 responses, split evenly between 5 experimental conditions (see the Additional Supplementary Information for the details of the experimental conditions), but limit analysis to the 2,508 that drew their neighborhood on our mapping tool. We use the data from the control group to introduce and illustrate the statistical model while we discuss the details of the experiment and present experimental results in Additional Supplementary Information.

Respondents who accepted the invitation to take the survey were presented with a consent form informing them they were taking part in a research study. The consent form was followed by demographic questions including partisanship, race, age, income, employment, homeowner status, whether they had children, marital status, and how long they had lived at their current residence.

Section S1 of the SI presents the summary statistics of the survey sample (control group).
Across cities, the sample is approximately evenly divided between self-identified Democrats and Republicans, and a majority of respondents reported voting for President Biden in the 2020 general election.
Tables S1 and S3 of the SI compare the summary statistics of the sample with those of the overall adult population of the three metropolitan regions under study.
We find that our sample is more predominantly white, wealthier, educated, and more likely to be a homeowner than the population of each of the cities in our sample.

\hypertarget{mapping-tool}{%
\subsection{Mapping tool}\label{mapping-tool}}

Next, respondents were presented with an embedded mapping application where they enter their residential address, at which point the maps zooms to a centered view of their address.
Then, the underlying census block grid was shown on the map over the road base map, and respondents used the brush tool to select the census blocks that they considered a part of their ``local community.''\footnote{Respondents were able to alter the brush size if they desired. We do not have information from data collected on the brush sizes that people chose. Future versions of this survey instrument could build in this functionality.}
Our first survey used this terminology, mirroring previous surveys that asked respondents to draw their own neighborhoods \citep{wong_bowers_rubenson_fredrickson_rundlett_2020}.
These authors have shown that the phrase ``local community'' is a tangible concept in people's minds, and further demonstrate the consistency of drawn neighborhoods when re-contacting survey respondents.

\begin{figure}[t]

{\centering \includegraphics[width=4in]{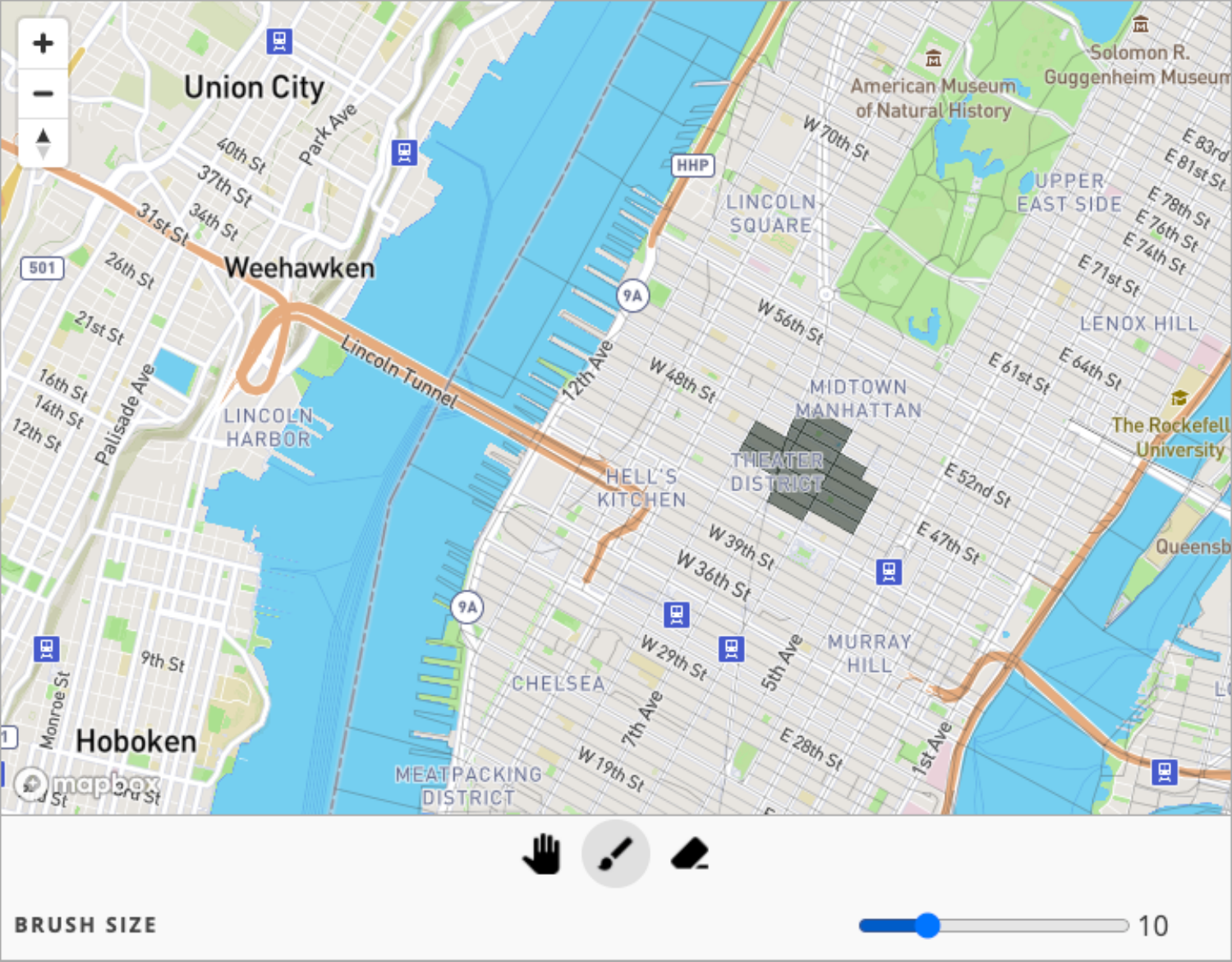} 

}

\caption[Map-drawing interface to collect respondent neighborhoods]{Map with brush tool used to draw neighborhoods.}\label{fig:map-interface}
\end{figure}

Our mapping application is comparable to these previous surveys in functionality.
One difference is that in this case, rather than having respondents use a drawing tool to draw a circle around their residence that constitutes their neighborhood, the application offers a brush tool to shade in the census blocks around the residence that are included in the neighborhood.
Respondents could zoom in or out on the map, and were able to make edits to their neighborhood after the initial shading.
The only constraint was that neighborhoods had to be contiguous.
Figure \ref{fig:map-interface} shows a screen shot of the map drawing tool.
We make our map drawing tool publicly available so that other researchers can use it for their own surveys (\url{https://github.com/CoryMcCartan/neighborhood-survey}).
In particular, the tool can be embedded into a popular survey platform such as Qualtrics as done in our survey.

\hypertarget{descriptive-statistics-of-drawn-neighborhoods}{%
\subsection{Descriptive statistics of drawn neighborhoods}\label{descriptive-statistics-of-drawn-neighborhoods}}

\begin{figure}[htb]

{\centering \includegraphics[width=1\linewidth]{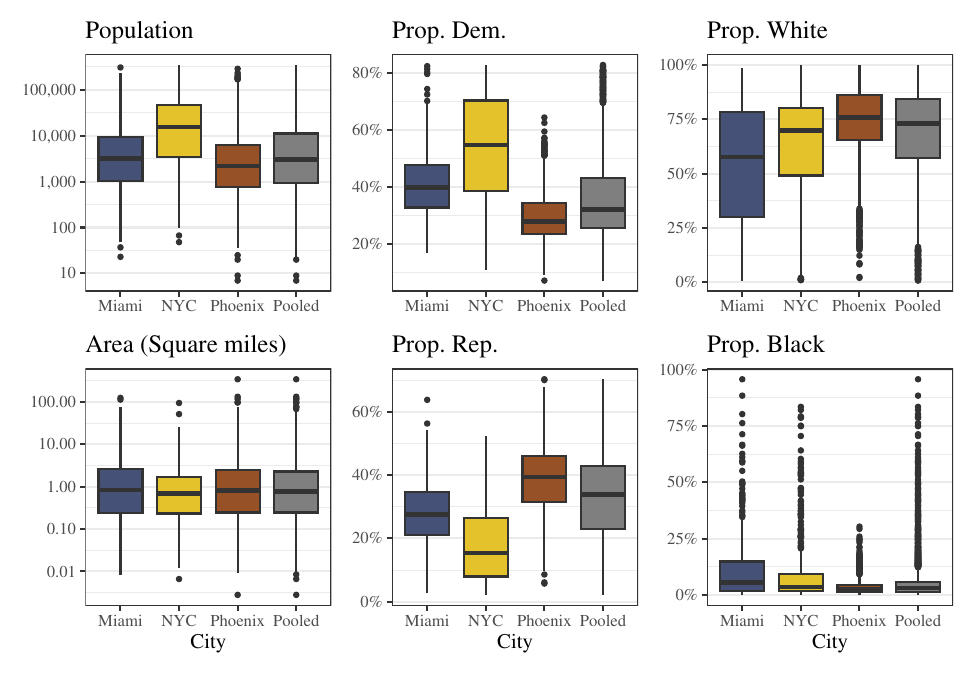} 

}

\caption[Descriptive statistics for respondent neighborhoods]{Descriptive statistics for respondent neighborhoods.}\label{fig:nbhd-stats-sum-full}
\end{figure}

Figure \ref{fig:nbhd-stats-sum-full} shows the central tendencies and range of the population, area, proportion Democrat, proportion Republican, proportion White, and proportion Black of all drawn neighborhoods, broken out by city as well as pooled.
Partisanship is measured using TargetSmart voter records of every registered voter in the three cities, using information on the residential location of each voter to create aggregate counts by census block.
The population, area, and racial demographics are measured using the 2010 census.

We find a wide range of neighborhood sizes, both in terms of population and land area.
The median number of residents contained in a drawn neighborhood is 3,020 residents, while its range extends from single digits to over 340,000 residents.
Similarly, the median area is 0.78 square miles, but the entire distribution ranges from 0.01 to 344.74 square miles.
This variation in size of neighborhoods is indicative of the variation in how different individuals experience their local geography and define their neighborhood.
Accounting for this heterogeneity is critical in our statistical model we introduce in the next section.

We also find substantial variation in the demographics of the drawn neighborhoods (see Section S2 of the SI).
Figure S1 of the SI shows the distribution of proportion White and proportion Black in drawn neighborhoods separately for White and Non-White survey respondents.
Drawn neighborhoods from White respondents are on average 21.1 percentage points whiter than those from Non-White respondents.
Figure S2 of the SI contains the distribution of proportion Democratic and Republican separately by respondent party registration, with Democrats drawing consistently more Democratic neighborhoods and Republicans drawing more Republican neighborhoods.

These differences likely reflect objective differences in racial and partisan exposure across race and party, but may also be influenced by conscious or subconscious motivations for respondents to construct their subjective neighborhoods to include more members of their own racial or partisan in-group.
Our statistical model can quantify the extent to which, net of other variables that may determine whether someone includes an area in their neighborhood, the added predictive effect of racial or partisan demographics on neighborhood inclusion.
We now turn to our proposed statistical model of subjective neighborhoods.

\hypertarget{sec:modeling}{%
\section{Modeling neighborhoods}\label{sec:modeling}}

To analyze the data collected through our survey tool, we propose a Bayesian model for neighborhood drawing that incorporates respondent characteristics, geographic factors, and their interaction.
The model predicts the likelihood of including a given census block in a voter's neighborhood.
In addition, the coefficients of the model represent the direction and magnitude of predictive effects different variables have on this inclusion probability.
Using this model, one can measure the degree to which the characteristics of respondents and geographic factors together predict subjective neighborhoods of different types.

Though the model is developed from explicitly spatial principles, as we show in the section motivating the methodology, ultimately it reduces to a generalized linear mixed-effects model (GLMM) with a particular link function, where every observation is a census block.
Spatial information enters the model explicitly through the use of distance as a covariate, and implicitly in deciding which census blocks are included in model fitting and which are excluded.
The simplification to a GLMM means that even without specialized software, practitioners can implement the model using existing statistical packages.
Nevertheless, we provide an open-source implementation that is computationally efficient and tailored to the particular use case here (\url{https://github.com/CoryMcCartan/nbhdmodel}).

\hypertarget{notation-and-setup}{%
\subsection{Notation and setup}\label{notation-and-setup}}

\begin{figure}[htb]

{\centering \includegraphics[width=6.5in]{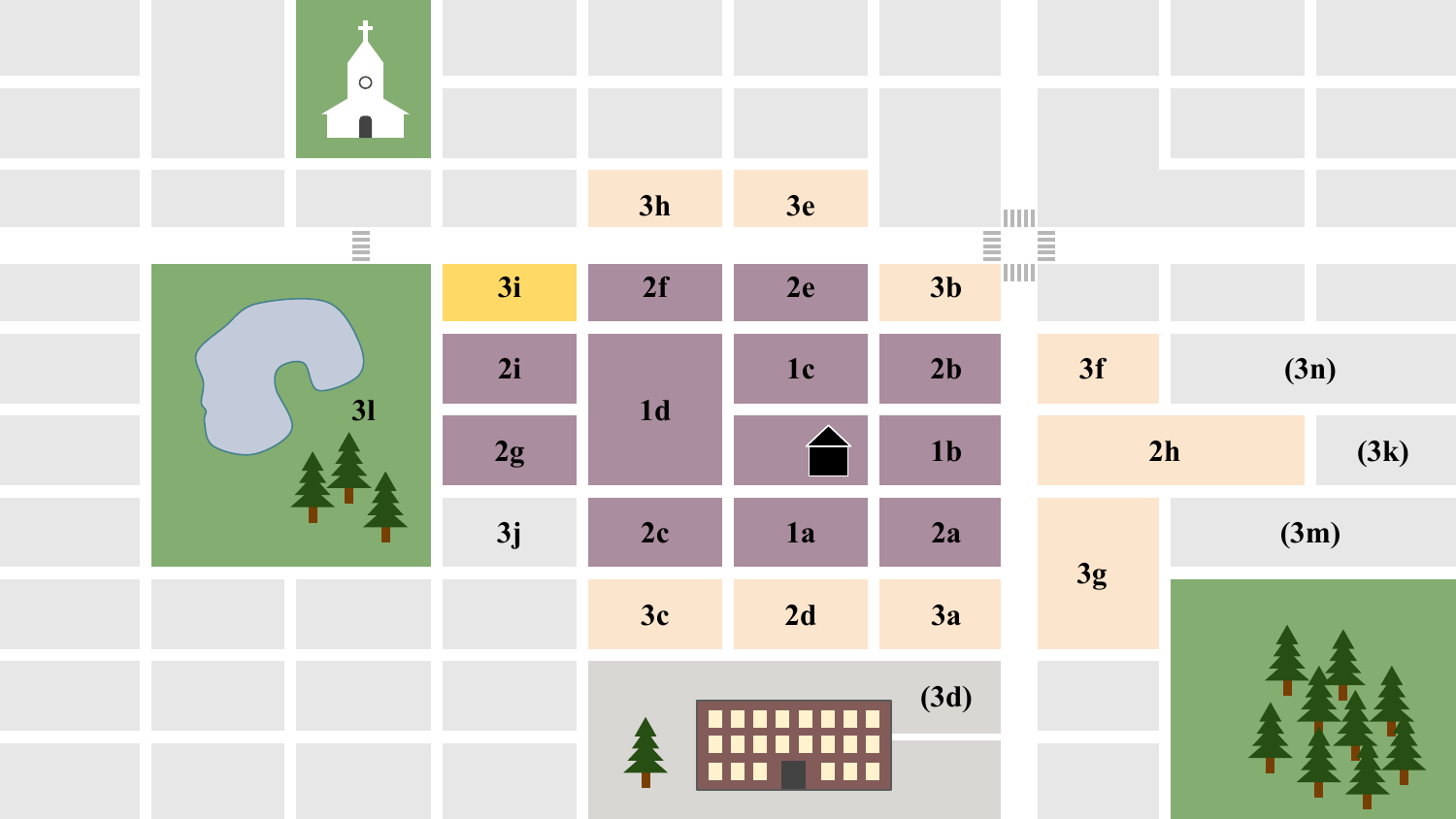} 

}

\caption[Neighborhood model schematic]{Model schematic.
The respondent's location is indicated by the black house in the center.
Blocks are labeled in the order in which they are considered for inclusion
in the neighborhood, with the number indicating the graph-theoretical
distance and the letter the spatial distance tiebreaker.
Blocks shaded purple have been included in the neighborhood,
while blocks shaded light orange have been excluded.
Parentheses around a block label indicate that it will not be considered
for inclusion in the neighborhood because none of its neighboring blocks
which are closer to the respondent belong to the neighborhood.
}\label{fig:schematic}
\end{figure}

For ease of notation, we begin by describing the model for a single neighborhood drawn by one respondent.
We start with an undirected graph \(G=(B,E)\) representing the layout of the city or town, with each vertex \(B_i\in B\) corresponding to a census block and the edges \(E\) corresponding to block adjacency.
We write \(B_i\sim B_j\) if \(B_i\) and \(B_j\) are adjacent, i.e., the edge \((i,j)\in E\).
We use \(K\dfeq|B|\) to denote the total number of blocks.
In Figure \ref{fig:schematic}, for example, the block with respondent's residence is adjacent to four blocks labeled as ``1a'', ``1b'', ``1c'', and ``1d''.
We do not consider two blocks that are touching one another with a single point as adjacent blocks (i.e., no point contiguity).
Thus, the block with respondent's residence is not adjacent to blocks ``2a'', ``2b'', and ``2c''.

Without loss of generality, we number the blocks in order of their distance from the block where the survey respondent resides, according to some distance function \(d:B\to\R\), so that \(B_0\) is this block, \(B_1\) its closest neighbor, and so on.
Thus, if \(i<j\) then \(d(B_i)<d(B_j)\).
In this application, we take \(d\) to be the graph-theoretical distance (i.e., the minimum number of edges between two nodes), with ties broken by spatial distance (i.e., the distance between centroids of census blocks).
Figure \ref{fig:schematic} illustrates this ordering scheme.
For example, block ``1d'' in the figure is numbered ``1'' because it is one step away from the respondent's block, and ``d'' because among all the one-step-away blocks, it is the fourth closest to the respondent spatially.
We note that the use of graph-theoretical distance introduces some dependency between the distance measure and population density: areas with higher population density generally contain smaller blocks.
Graph-theoretical distance will increase faster in these areas over a given spatial distance compared to areas wtih lower population density.

Let \(Y_i\) be an indicator variable for the inclusion of \(B_i\) in the respondent's neighborhood, so that the neighborhood itself may be defined as the set of blocks with \(Y_i=1\), i.e., \(Y=\{B_j:Y_j=1\}\).
Since we require any neighborhood to contain the block where respondent's residence is located, we always have \(Y_0=1\).
We define a following connectivity indicator \(C_i\) to be 1 if block \(B_i\) is connected to the respondent's neighborhood by way of any closer blocks.
Formally, \[
    C_i = \ind\{\text{there exists a } j < i : B_i\sim B_j \text{ and } Y_j = 1\}.
\]
This indicator checks whether, among the blocks which are closer to \(B_0\) than \(B_i\) is, if any are in the respondent's neighborhood.
In Figure \ref{fig:schematic}, blocks with parenthetic labels have \(C_i=0\).
For instance, block ``3d'' would be expected to be considered before ``3e'', ``3f'', etc. based on its location, but since none of ``2d'', ``3a'', and ``3c'' are in the neighborhood, ``3d'' is never considered.

\hypertarget{subsec:model}{%
\subsection{The model}\label{subsec:model}}

Under the proposed model, a neighborhood is generated sequentially, starting with \(B_0\) and adding blocks in order of increasing distance from \(B_0\) according to a probability.
This probability is heavily influenced by the graph-theoretic distance between the block under consideration \(B_i\) and \(B_0\), and its connectivity.
In particular, we assume that the neighborhood is connected and our survey tool does not allow respondents to draw disconnected neighborhoods.

The core of the model is \[
    Y_i\mid Y_0,\dots Y_{i-1} \sim \Bernoulli(\pi_i \cdot C_i),
\] where \(\pi_i\) is the inclusion probability of block \(B_i\) into one's neighborhood provided that it is connected.
As long as \(\pi_i \to 0\) as \(d(B_i)\to\infty\), the generated neighborhoods will be bounded around \(B_0\) almost surely.
Figure \ref{fig:schematic} illustrates the state of the neighborhood partway through the generation process, when block ``3i'' (shaded gold) is under consideration.
The process concludes once the neighborhood is surrounded by light orange blocks, since then there are no blocks left which could be added while keeping the neighborhood contiguous.

The specification of \(\pi_i\) determines the type of neighborhoods that are generated.
Let \(\vb X\) be a \(m \times K\) matrix of predictors, not including an intercept, with \(\vb x_i\) the column vector of \(m\) predictors for block \(i=1,2,\ldots,K\).
These may include the characteristics of the respondent, those of the graph or map (e.g., the demographics of blocks, locations of landmarks and roads), and their interactions.
The inclusion probability can also depend on the inclusion of blocks whose distance to \(B_0\) is less than that of the block under consideration \(B_i\).
This means that the predictors can include the information about the partially drawn neighborhood.
The factorization formulation above, however, precludes the possibility that \(\vb x_i\) depends on \(\{Y_j: j \ge i\}\), i.e., the inclusion of farther-out blocks.

\begin{figure}[t]

{\centering \includegraphics[width=1\linewidth]{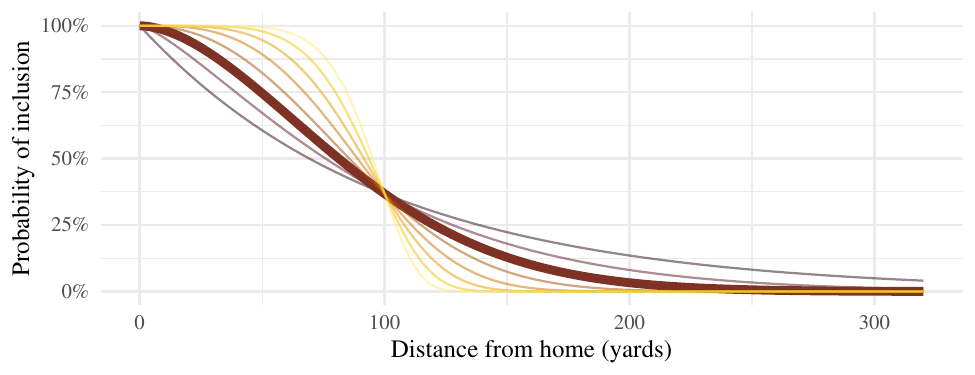} 

}

\caption[Illustration of kernel function across a range of values of the $\alpha$ parameter]{Illustration of kernel function across a range of values of the
$\alpha$ parameter, indicated by different colors.
The length scale shown here is arbitrary; in the model,
it is estimated as the $L$ parameter.
}\label{fig:kernel}
\end{figure}

We model the inclusion probability using a kernel function that smoothly decays as the distance between \(B_i\) and \(B_0\) grows: \begin{equation*}
    \pi_i \ = \ \pi(\vb x_i,\vb*\beta,\alpha,L,\sigma)\ = \ \exp(-\qty|\frac{d_{\mathrm{sp}}(B_i)}{L}
        \exp(\vb x_i^\top \vb*\beta+\eps)|^\alpha) \qand \eps \ \sim \ \Norm(0, \sigma^2),
\end{equation*} where \(\eps\) is the respondent random effect, \(d_{\mathrm{sp}}(B_i)\) represents spatial distance between \(B_0\) and \(B_i\), \(L\) controls the scale of decay, and \(\alpha\) controls its rate.
In particular, \(\alpha\) represents the sharpness of the neighborhood boundary.
Along with \(L\), the random effect \(\eps\) plays an important role in addressing the heterogeneity of neighborhood size across individual respondents.
Figure \ref{fig:kernel} visualizes the kernel function we use where we choose an arbitrary length scale (horizontal axis) for illustration.
As the value of \(\alpha\) increases, the inclusion probability decays faster as a function of distance.

Conveniently, the model reduces to a Bernoulli generalized linear mixed-effects model (GLMM) with complementary log-log (cloglog) link function for the exclusion probability, \begin{equation*}
 1-\pi_i \ = \ 1-\exp\left\{-\exp(\alpha\log d_{\mathrm{sp}}(B_i)-\alpha\log L
        +\alpha \vb x_i^\top\vb*\beta+\alpha\eps)\right\}. \numberthis\label{p-glm}
\end{equation*}
That is, we can fit the model by regressing the non-inclusion indicators on the log distance, any covariates, and an individual random effect, using the cloglog link function.
As discussed above, we do not include every census block in the model---just those that are included in the neighborhood, and those which are not but border the neighborhood (i.e., those for which \(C_i=1\)).
This reflects the sequential generation process that underlies the model.

Since the link function is nonlinear, the marginal effect of each covariate varies by its underlying value, and is not simply equal to the value of the coefficient.
We choose to focus our interpretation on the effect of each covariate on the ``margin'' of the neighborhood, where the probability of a block's inclusion is 50\%.
We interpret the coefficient estimates by calculating,\[
\exp(-\exp(\mu_{0.5} + \beta_j)) - \exp(-\exp(\mu_{0.5}))
\] for each coefficient \(\beta_j\), where \(\mu_{0.5}\dfeq\log\log 2\) is the value of the linear predictor which corresponds to an inclusion probability of \(0.5\).
This quantity reflects the percentage point change in the probability of inclusion for a one-unit change in the covariate \(\vb X_j\), at the margin.

The model is completed with the following prior distributions, \begin{align*}
    \alpha\log L &\sim \mathrm{t}_3(0, 2.5) \qand
    \sigma \sim \mathrm{t}_3(0, 2.5).
\end{align*} Prior distributions for the coefficients are formed indirectly by taking a QR decomposition of the centered covariate matrix (including the log distance variable) \citep{goodall199313}.
The coefficients on the QR-decomposed (centered) covariates are given a \(\mathrm{t}_2(0, 2.5)\) prior.
This setup implies priors for the actual coefficients of interest which are weakly informative and adapted to the scale and correlation of the covariates.
We have used importance sampling to fit the model under alternative prior specifications and found no measurable changes in the posterior distribution of the parameters.

Because of the sequential generation and the indicator function, the posterior distribution simplifies to, \begin{align*}
    p(\theta\mid G, Y, \vb X) &\propto p(\theta)\prod_{i=1}^K
        p(Y_i\mid Y_0,\dots,Y_{i-1},\vb x_i, \theta) \\
    &= p(\theta)\prod_{i:C_i=1}
        p(Y_i\mid Y_0,\dots,Y_{i-1}, \vb x_i, \theta) \\
    &= p(\theta)\prod_{i:C_i=1}
        \pi(\vb x_i,\theta)^{Y_i} \{1-\pi(\vb x_i,\theta)\}^{1-Y_i},
\end{align*} where \(\theta=(\vb*\beta,\alpha,L,\sigma)\) is shorthand for the parameters, and \(p(\theta)\) is its prior distribution.
This formulation only requires the computation of the likelihood for the blocks in the drawn neighborhood and all of their adjacent blocks.

We assume individual responses are exchangeable, allowing us to simply multiply their likelihoods to create a joint model for all responses.
The \(\beta\), \(L\), \(\sigma\) and \(\alpha\) are shared across responses, but each respondent has its own error term \(\eps\) common to all of its blocks.
As appropriate, \(\beta\) may also contain hierarchical terms that vary by demographic categories, or metropolitan area or subdivision.
Computational details for fitting models are described in Section S3 of the SI.

\hypertarget{sec:spec}{%
\subsection{Model specification}\label{sec:spec}}

To illustrate the proposed model, we apply it to our survey data using the control group alone.
We limit our analysis to the 468 respondents in the control group who drew a map that consisted of more than one census block.
Because the census block that contained the residential address is highlighted by default, we cannot distinguish between respondents who selected single block neighborhoods and those who entered a residential address but decided not to draw a neighborhood.
In Table S4 of the SI, we show that neither respondent characteristics nor the experimental conditions are powerful predictors of who draw usable maps.

We fit a full model, which includes demographic information, as well as a baseline model, which includes only geographic information. We fit this model to a random sample of 400 respondents from the 468 in the control group. The unsampled 68 will be the test set for our later prediction analysis.
Comparing the predictions of these two models will allow us to quantify the extent to which demographics contribute to the prediction of subjective neighborhoods.
In the full model, we include as predictors individual characteristics consisting of voter race, political party, homeowner status, educational attainment, income, age, retirement status, and length of residence in current home.
Individuals that differ along these characteristics may view their local area differently, and we quantify the predictive power of these factors about drawn neighborhoods.

We also include geographic characteristics of census blocks including race, party, and education demographics, whether the block contains a school, park, or church, and the distance to the closest of each of these features, whether the block is in the same block group as the voter's residence, the same census tract, whether the block is bounded by the same major roads and railroads as the respondent's residence, block population, and block land area.
These aggregate characteristics account for features of place that may influence whether respondents include census blocks in their neighborhood.
In particular, indicators for \emph{same block group}, \emph{same census tract}, and \emph{same road/rail regions} should help disentangle demographic effects from the effects of physical boundaries, which can often align with sharp transitions in demographic composition.
Block groups and tracts generally group blocks together following natural boundaries like existing neighborhood designations, highways, or bodies of water.
The custom indicator for road/rail regions is designed to have the same effect.

In our analysis, we limited these infrastructure variables to those which could be computed from national data such as Census TIGER shapefiles, but researcher could also incorporate more specific geographic data, often available from municipalities, such as by using road speed or width data, or the locations of community centers and city facilities.

The model specification also includes interactions between respondent race and racial demographics, respondent party and party demographics, and respondent educational attainment and education demographics.
Table S5 of the SI contains detailed model specifications, including transformations and interactions of our covariates.

Our main coefficients of interest correspond to the three variables that measure the fraction of people in each block who belong to the same racial, partisan, and educational category as the respondent, respectively.
We allow these coefficients to vary by the categories of each variable as well, to understand differences between groups.
For example, the coefficient for the \emph{same race category} variable can differ between white and minority respondents.

We fit a separate model to each city's data.
This decision is in part based on the fact that there are a sufficient number of respondents for each city, leading to relatively precise parameter estimates.
Linking all three cities through a single hierarchical model is also possible, but fitting such a model to the entire data would substantially increase computational cost.

\hypertarget{subsec:firstsurvey}{%
\subsection{Empirical findings}\label{subsec:firstsurvey}}

\begin{figure}[htb]

{\centering \includegraphics[width=1\linewidth]{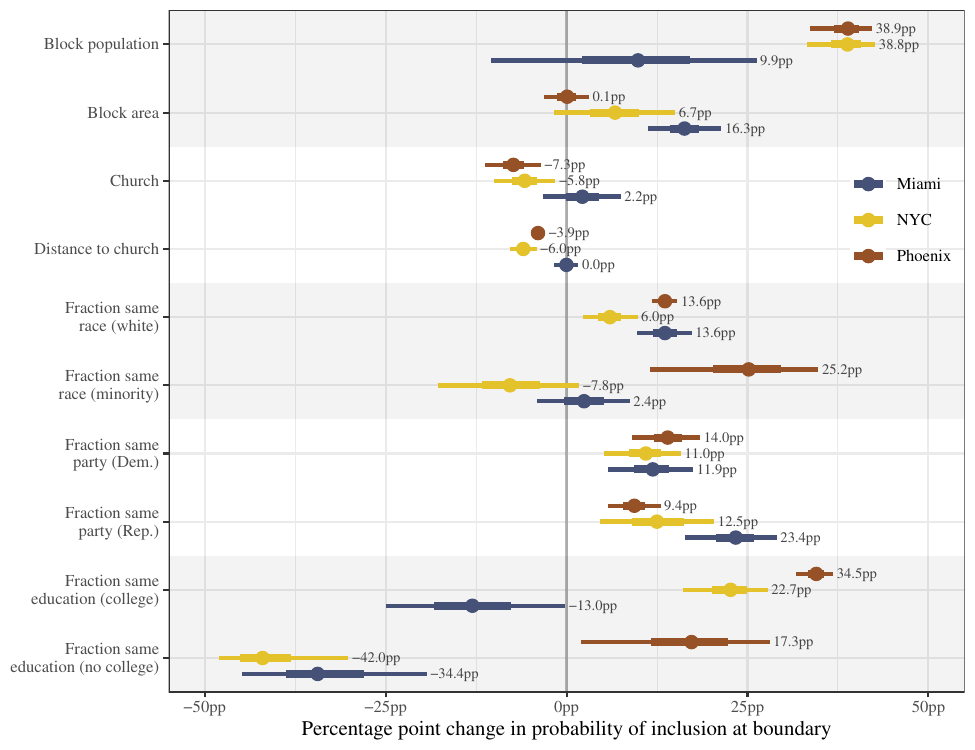} 

}

\caption[Selected full model posterior estimates for the control sample]{Selected full model coefficient posteriors, scaled to show the percentage
point change in probability of a block's inclusion for a baseline
probability of 50\%. Plotted are 90\% and 50\% credible intervals,
with posterior medians displayed to the right of each interval.
Section S5 of the SI contains the full results table
for the other variables specified in the model specification.
}\label{fig:fit-coef-full}
\end{figure}

To interpret the fitted models, we use the marginal effect calculation described above in the methodology section.
We compute the posterior estimate of the percentage point change in the predicted probability that a respondent will include a census block in their neighborhood when increasing the value of the corresponding covariate by one unit, over a baseline probability of 50\%, while holding other variables in the model constant.
Figure \ref{fig:fit-coef-full} presents these posterior means and credible intervals (90\% and 50\%) for selected coefficients from the full model (see Section S5 of the SI for the posterior summaries of all coefficients from the full and baseline models).
Posterior summaries are plotted separately by city.

Holding other variables in the model constant, a White respondent is 6.1 to 16.9 percentage points more likely to include a census block composed entirely of White residents compared to one with no White residents.
We cannot be confident that this preference for racial homogeneity occurs for neighborhoods drawn by minority (non-White) respondents, as the credible intervals for each city sample overlap with zero.
Partisan similarity exerts analogous predictive power.
Democrats are more likely to include Democratic neighbors in their neighborhoods and Republicans are more likely to include Republican neighbors, holding other variables in the model constant.\footnote{In the Additional Supplementary Information, we present results from an experiment that randomizes different information embedded in the map. The main findings about racial and partisan homophily hold even if we add the information about racial and partisan compositions on the map. }

We do not observe consistent results for educational similarity.
College educated and non-college educated respondents include areas with more college educated residents, although the credible intervals for some of the city samples overlap with zero and the medians vary considerably across cities.
Different results across cities could be due to contextual differences between cities but could also be due to sampling noise.
Additionally, it is difficult to determine with the data what specific characateristics of cities might produce differential results.

Size and population of local areas also influence inclusion probability.
In the New York and Phoenix, respondents are more likely to include populous census blocks in their neighborhoods.
This interval is less precise and smaller in magnitude in Miami.
In all three city samples, larger census blocks are more likely to be included in neighborhoods.

The presence of a church in a census block is negatively associated with inclusion of the census block in a neighborhood in the New York and Phoenix samples, with no predictive effect in the Miami sample.
Distance to a church is also negatively associated with inclusion in each of the samples, meaning that census blocks that are closer to churches are more likely to be included.
These results may speak to respondents opting to include residential census blocks over ones with churches in them, but their neighborhoods still being shaped by proximity to churches.

In Table S6 of the SI, we report all the estimates from the full model described in our model specification (see Table S7 for the estimates from the baseline model).
These include administrative variables such as roadways, census block groups, and census tracts. We find that these administrative definitions and physical characteristics, net of other factors in the model, influence whether people include areas in their neighborhoods. For example, respondents are more likely to include areas in their neighborhood that fall on the same side of major roadways as their residence. Similarly, they are more likely to include areas that fall in the same census tract. These estimates demonstrate how objective features of neighborhoods influence subjective definitions.

\hypertarget{sec:predicting}{%
\section{Predicting neighborhoods}\label{sec:predicting}}

The fitted model can also be used for posterior prediction of neighborhoods, for both in-sample and out-of-sample respondents.
We first examine the ability of the model to predict respondent's neighborhoods out-of-sample.
While we find a large amount of individual heterogeneity makes highly accurate model predictions difficult, the model's predictions still improve on naive methods such as using Census tracts as stand-ins for respondent neighborhoods.

We then demonstrate possible uses of neighborhood predictions in-sample to visualize and understand the effect of various factors on a single respondent's drawn neighborhood.
Section S6 of the SI takes this predictive framework one step further and connects aggregate-level model predictions to the substantive findings on co-racial and co-partisan preferences described above.

\hypertarget{out-of-sample-predictive-ability}{%
\subsection{Out-of-sample predictive ability}\label{out-of-sample-predictive-ability}}

First, we examine the quality of model fit as measured by its out-of-sample predictive ability.
There is significant heterogeneity in respondents' neighborhoods, as reflected in the wide range of neighborhood areas and demographics shown in Figure \ref{fig:nbhd-stats-sum-full}.
Neighborhoods range in size from less than 0.01 to over 100 square miles, and much of this variation in size is not captured by demographic variables.
Any model will consequently struggle to make accurate predictions, especially for respondents not included in the data to which the model was fitted.

Despite these challenges, both the full and baseline models are more effective in predicting respondents' neighborhoods than a naive approach based on circles centered around their residence locations.
We measure predictive accuracy by first generating 100 posterior predictions for each respondent's neighborhood.
This is accomplished by taking a random sample of parameter values from the posterior, and then sequentially sampling census block inclusions according to the model's data generating process.

For each neighborhood prediction, we compute the precision and recall for the constituent census blocks, and then take their median values over predictions.
Precision measures the fraction of the predicted neighborhood that is in the original neighborhood, while recall measures the fraction of the original neighborhood that is in the prediction.
The baseline and full models have in-sample median precision of 0.32 and 0.34, respectively, and recall of 0.75 and 0.71.
Out of sample, precision increases moderately to 0.38 and 0.48 for the baseline and full models, respectively.
The out-of-sample recall falls to 0.69 and 0.64.

However, due to the contiguity requirement and the sequential nature of the neighborhood model, precision and recall in this context are driven largely by the size of the predicted neighborhood.
As the neighborhood grows larger, the recall will increase at the cost of precision.
In addition, by shifting the intercept of our model, we can grow or shrink the predicted neighborhood while maintaining the same discrimination with regards to predictive covariates.

Thus, to better contextualize this performance, we compare each posterior prediction (before averaging) to a circular neighborhood of the same radius as the prediction.
We find this radius by taking the smallest circle centered on the respondent's home which covers their drawn neighborhood.
Using the same fixed-radius circle for comparison with all model predictions would not be appropriate given the wide variation in neighborhood sizes, but allowing the circular neighborhood to exactly match the modeled radius gives this naive approach a significant leg up---it can leverage all the information learned in the model about the neighborhood's radius.
We might therefore expect the baseline and full models to only minimally improve upon the circular neighborhoods, especially for out-of-sample predictions, where individual random effects have not been fit.

\begin{figure}[!t]

{\centering \includegraphics[width=1\linewidth]{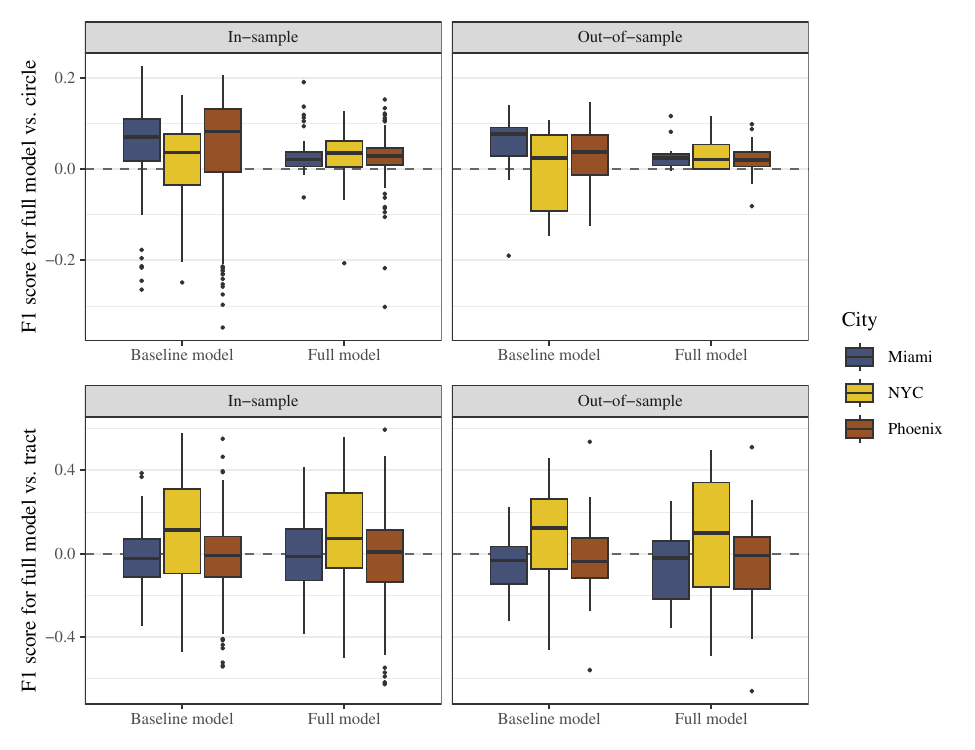} 

}

\caption[Posterior median of the difference in F1 scores between a neighborhood
predicted by the model and a circular or tract-based neighborhoods]{Posterior median of the difference in F1 scores between a neighborhood
predicted by the model and a circular neighborhood of the same radius
of the same radius (top) or a census tract (bottom).
The boxplot shows the variation in this median difference accros the
respondents included in the model fitting (left plot) and excluded from
the model fitting (right plot).
Positive values indicate the model outperforming the ciruclar baseline,
on average, for a particular respondent.
The baseline model includes geographic information only while the
full model also includes demographic information.
Section S5 of the SI contains the full results tables for the full and baseline models.
}\label{fig:gof-paired}
\end{figure}

For the 100 predictions for each respondent, we take the median of the difference in the F1 score, which is the harmonic mean of precision and recall, between the prediction and the matching circle. We also calculate this difference between the prediction and a census tract, the most common unit at which researchers measure local context (see Figure S8 of the SI for comparison to ZIP Code Tabulation Areas).
Figure \ref{fig:gof-paired} shows the results of this comparison, broken out by city.
Both models outperform the circular-neighborhood approach by around 0.04 in-sample, and 0.03 out-of-sample on average.
Importantly, both in and out of sample, only for a few respondents do the naive approaches meet or outperform the full model, as indicated by the bulk of each boxplot lying above the x-axis.
And even in these cases, as shown above, the model is able to provide uncertainty quantification and coefficient estimates, which a naive approach cannot. Compared to census tracts, the predictive performance is similar, and only in New York City do we see consistent out-performance of the predictions compared to tracts.

The substantial heterogeneity in neighborhood sizes makes accurate neighborhood predictions difficult in general.
However, the model's use of local and individual covariate information allows it to improve on purely distance-based measures, even when these are well calibrated by matching the radius of a circular neighborhood to that of the model-based prediction.

\hypertarget{in-sample-respondent-level-prediction}{%
\subsection{In-sample respondent-level prediction}\label{in-sample-respondent-level-prediction}}

\begin{figure}[htb]

{\centering \includegraphics[width=1\linewidth]{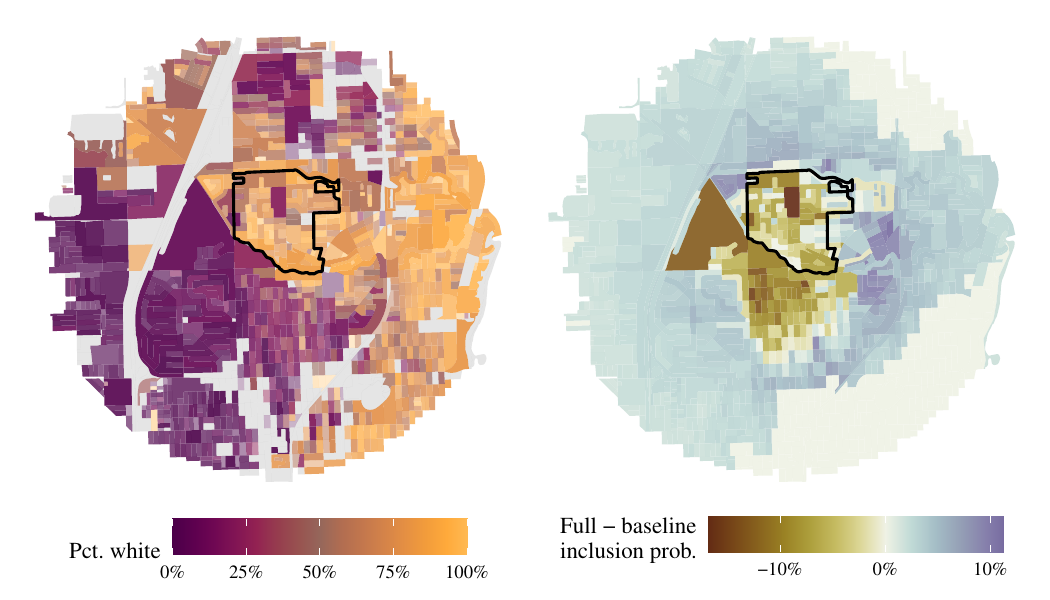} 

}

\caption[Example drawn and model-predicted neighborhood for a respondent outside Miami]{The left plot shows the racial demographics of area surrounding the
example respondent.
The subjective neighborhood drawn by this resopndent is indicated by the
solid black line and each census block is shaded based on the percent
White of its population.
The right plot shows the difference in the posterior probability
of a block being included in the respondent's neighborhood between
the full and baseline models. The baseline model includes geographic
information only while the full model also includes demographic information.
Blue areas are relatively more likely to be included under the full model,
while orange areas are relatively less likely to be included.
}\label{fig:indiv-nbhd}
\end{figure}

Model predictions can be useful in-sample as well.
Here, we demonstrate the predictive influence of race on census block inclusion probability using a single respondent in Miami.
This voter is white, female, and is not registered to a major political party.
Figure \ref{fig:indiv-nbhd} maps the racial demographics surrounding the respondent's residential address in the left panel (each census block shaded based on the percent White of its population), and the change in the posterior probability of inclusion for each census block comparing the full model to the baseline model in the right plot.
This respondent lives in a mixed but majority White area (indicated by light orange color in the left plot) that is just to the north of areas comprised largely of minority residents (dark purple color).
Her drawn neighborhood (represented by a black solid line) adheres sharply to this stark southern boundary.
The posterior probability map shows how these majority non-White areas are less likely to be included when demographics are accounted for in the model (indicated by dark brown color in the right plot).

Depending on one's substantive questions of interest, other quantities may be of interest and can be directly computed from the fitted model though it may require additional causal and other assumptions.
Examples include the probability of including one block, given that another block is or is not included; the change in an individual respondent's posterior predictive neighborhood if their demographics were different; or how a change in the demographics of one block (say, by a new housing development) could influence the shape and size of a respondent's neighborhood.

\hypertarget{sec:COI}{%
\section{Measuring and modeling communities of interest}\label{sec:COI}}

This section uses an an additional survey we conducted in New York City to demonstrate how the proposed methodology can be used to study political representation and communities of interest as they relate to redistricting.
The survey asked New York City residents to consider City's official guideline for considering ``communities of interest'' when redrawing city council districts.
This guideline directs the City to ``Keep intact neighborhoods and communities with established ties of common interest and association, whether historical, racial, economic, ethnic, religious or other'' \citep{nyc_charter}.
Respondents are then directed to shade in on an interactive map ``your community that should be kept together in your city council district.''
This map drawing exercise is followed by similar demographic questions to the previous survey.
Finally, we explore the ability to construct ``consensus'' neighborhoods from model predictions.

\hypertarget{city-council-survey-of-new-york-city-residents}{%
\subsection{City council survey of New York City residents}\label{city-council-survey-of-new-york-city-residents}}

The survey was administered in two ways. First, similar to the previous survey, we contacted New York City residents via email, randomly drawing residents off of registered voter lists.
Those who did not respond were sent a reminder email each week for three weeks.
Of the 277,641 registered voters who were successfully contacted, we received 1,102 responses, for a response rate of 0.40\%.
Section S1 in the SI contains more information on the sampling process.

The second method of survey administration was through targeted advertisements on Meta from December 6, 2022 to February 21, 2023, we ran advertisements targeting New York City residents inviting them to draw their neighborhood on a map.
Facebook users who clicked on the advertisement were led to the Qualtrics survey instrument.
Based on statistics from Meta, 25,767 Facebook users clicked on our advertisements during this time period, of which 1,086 chose to take the survey.\footnote{Unlike the email survey, the Meta survey does not condition on respondents being already registered to vote. In Table S3 of the SI, we compare results across the two samples.}

\begin{figure}[t]

{\centering \includegraphics[width=1\linewidth]{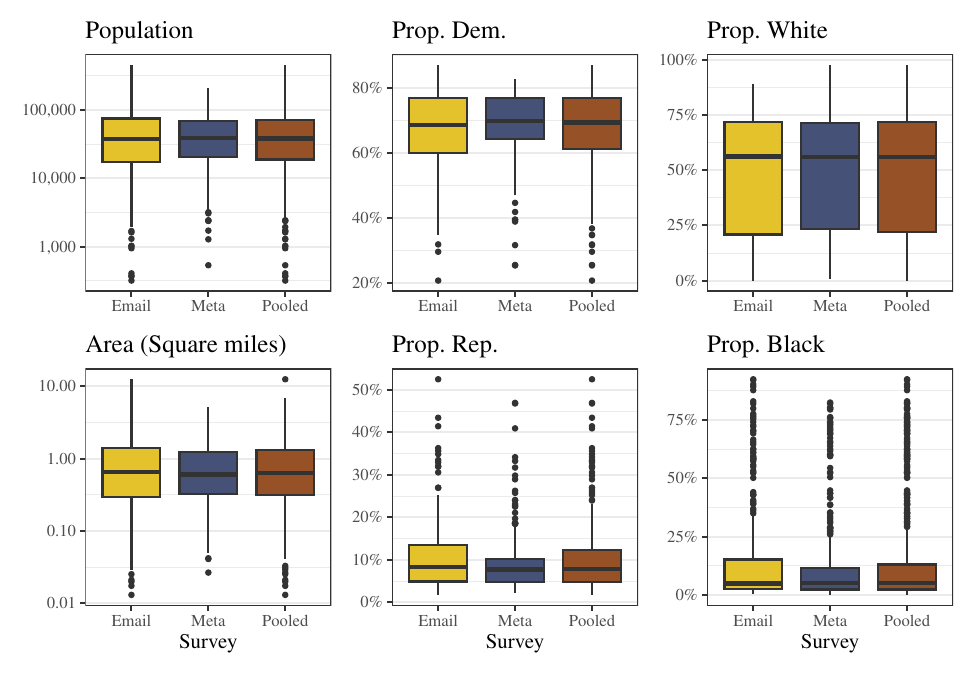} 

}

\caption[Descriptive statistics for New York respondent communities of interest]{Descriptive statistics for respondent communities of interest.}\label{fig:nbhd-stats-sum-full-nycc}
\end{figure}

In our analysis, we focus on the 627 respondents who drew maps consisting of more than one census block.
Figure \ref{fig:nbhd-stats-sum-full-nycc} shows the central tendencies of the population, area, proportion White and Black, and proportion Democratic and Republican for the drawn communities of interest.
These statistics are shown broken out by the email and Meta surveys, as well as the pooled sample.
The median population of these drawn community of interests is 38,070 people, with the full distribution ranging from 319 to 455,384.
The median area is 0.63 square miles (range: 0.01-12.41 square miles).
The median percentage White is 56\%, median percentage Black is 5\%, and the median values for percent Democratic and Republican are 69\% and 8\% (out of registered voters).
Figures S3 and S4 of the SI show the breakdowns of map demographics by respondent race and partisan lean.
The results show clear descriptive differences between Whites and non-Whites and between Democrats and Republicans in the racial and partisan demographics of their drawn communities of interest.

\hypertarget{determinants-of-communities-of-interest}{%
\subsection{Determinants of communities of interest}\label{determinants-of-communities-of-interest}}

Next, we fit the proposed model with the same specification as the one used for the analysis of the first survey.
We examine the way in which respondent traits, aggregate characteristics, and their interactions influence the inclusion of different areas into their communities of interest.
Similar to the previous analysis, we first take a random sample of 500 (approximately 80\% of the sample) responses as our training data and fit the model to this training set.
All coefficients reported below are on this sample, while predictive comparisons in the next section conduct out-of-sample predictions on the 127 neighborhoods in the test set.

\begin{figure}[!t]

{\centering \includegraphics[width=1\linewidth]{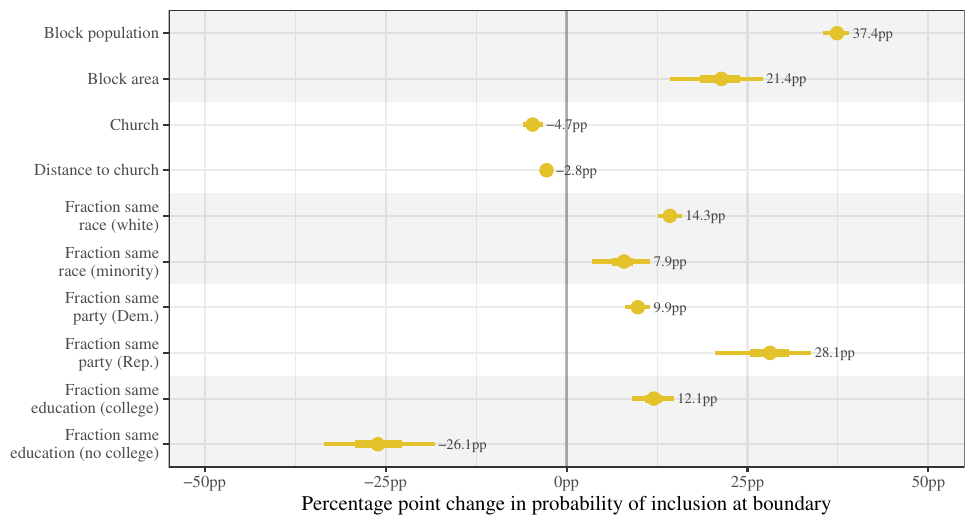} 

}

\caption[Selected full model coefficient estimates for the New York city council survey]{Selected full model coefficient posteriors, scaled to show the percentage
point change in probability of a block's inclusion for a baseline probability
of 50\%. Plotted are 90\% and 50\% credible intervals, with posterior
medians displayed to the right of each interval. Section S5
of the SI contains the full results table.
}\label{fig:fit-coef-full-nycc}
\end{figure}

Figure \ref{fig:fit-coef-full-nycc} presents the coefficients of interest from the model fit to the city council data (see Table S8 of the SI for the full results).
Holding other factors constant, White respondents are 14.3 percentage points more likely to include a census block comprised entirely of white residents in their community of interest.
Minority respondents are 7.9 percentage points more likely to include census blocks comprised entirely of minority residents.
The analogous estimates from the New York City sample in the subjective neighborhoods survey were 6.1 percentage points for Whites and -7.5 percentage points for minority respondents (but not statistically significant).
Therefore, the preference for racial homophily is stronger when respondents define the areas that should be included in their city council district than when drawing subjective neighborhoods without explicit direction as to the political implications of these definitions.

We also find that Democratic respondents are 9.9 percentage points more likely to include Democratic census blocks, (slightly lower than that 11.1 percentage point estimate in the first survey), while Republicans are 28.1 percentage points more likely to include Republican census blocks (much higher than the 12.7 percentage point estimate in the subjective neighborhoods survey).
The estimates for education are consistent in sign but smaller in magnitude as in the previous survey.
Respondents, regardless of whether they graduated college or not, tend to include census blocks containing more residents who graduated college.

Looking at the other coefficients in the model, we find high levels of consistency between effect sizes and direction between the city council survey and the subjective neighborhood survey.
For example, block population and block area are again positively associated with inclusion, while presence of a church and distance to the nearest church are both negatively associated with inclusion.

In sum, these results demonstrate that citizen conceptions of how they should be represented are shaped by local racial and partisan demographics, as well as by infrastructural and institutional characteristics of the places in which they live.
Specifically, respondents are influenced by racial and partisan compositions of local areas when drawing subjective neighborhoods regardless of whether they are given specific definitions of neighborhoods.
The magnitude of racial influence is particularly greater when drawing communities of interest as they relate to legislative redistricting.

\hypertarget{subsec:modelfit}{%
\subsection{Quality of model fit}\label{subsec:modelfit}}

\begin{figure}[!t]

{\centering \includegraphics[width=1\linewidth]{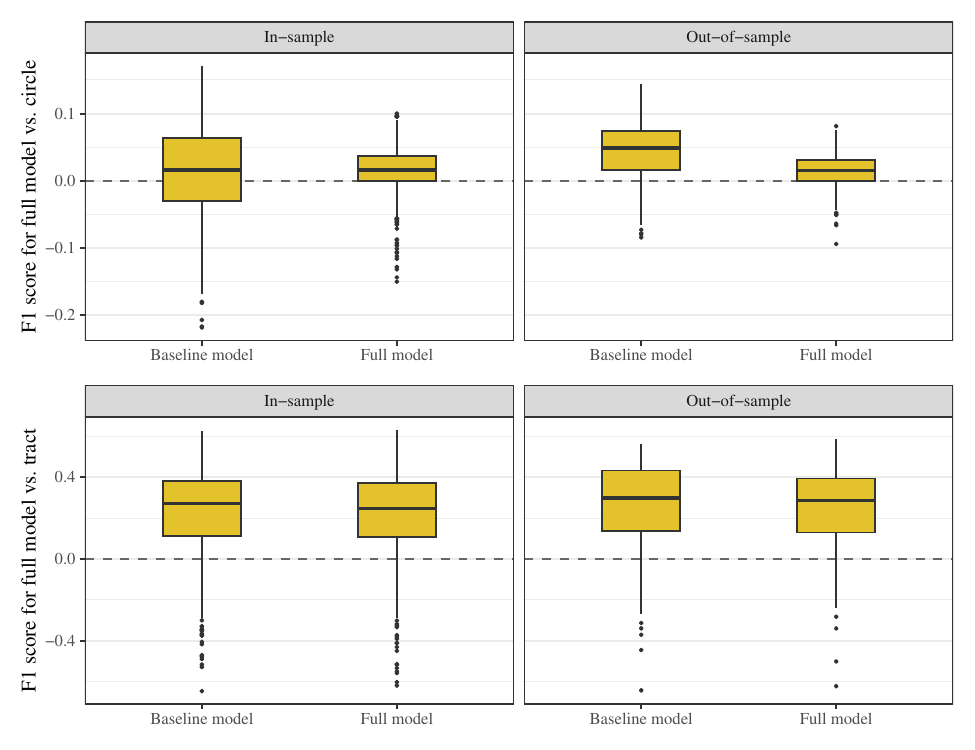} 

}

\caption[Posterior median of the difference in F1 scores between a neighborhood
predicted by the New York city council model and a circular o
r tract-based neighborhoods]{Posterior median of the difference in F1 scores between a community of
interest predicted by the model prediction and a circular neighborhood
of the same radius (top) and a census tract (bottom).
The boxplot shows the variation in this median difference accros the
respondents included in the model fitting (left plot) and excluded from
the model fitting (right plot).
Positive values indicate the model outperforming the circle (tract),
on average, for a particular respondent.
The baseline model includes geographic information only while the
full model also includes demographic information.
}\label{fig:gof-paired-nycc}
\end{figure}

As before, we examine the quality of model fit using the city council survey.
The top row of Figure \ref{fig:gof-paired-nycc} shows the distribution of the median difference in the F1 score between predicted communities of interest and the matching circle, using the baseline and full model.
The baseline model outperforms the circular-neighborhood approach by approximately 0.016 in-sample, and 0.049 out-of-sample on average.
The full model outperforms the circular neighborhood by approximately 0.014 in-sample and 0.015 out-of-sample.

The bottom row of the figure shows that compared to tracts, the model shows a much more notable improvement.
The median difference in F1 scores between the baseline model and tracts is 0.27 in-sample and 0.30 out-of-sample.
For the full model, the median difference is 0.25 in-sample and 0.29 out-of-sample.
The performance advantage is much higher than that observed in the tract comparison from the first survey.
This improvement in predictive performance suggests that drawn maps are easier to predict when respondents are provided with a more concrete prompt related to redistricting.

\hypertarget{subsec:consensus}{%
\subsection{Building consensus neighborhoods}\label{subsec:consensus}}

A major challenge for map drawers in redistricting city council boundaries is to
incorporate many communities of interest at the same time.
As we have consistently found, different people living in the same location may define their local community or neighborhood differently.
When it comes time to select a ``community of interest'' for redistricting purposes, these varying individual communities must be somehow aggregated.
We can use individual predictions to explore options for this aggregation process, and to understand how our substantive findings on same-race preference affect the difficulty of building an aggregate or consensus neighborhood.

We begin by sampling a synthetic residential population for a particular Census block.
We generate a random race, party, homeownership status, and educational level for 100 individuals according to the Census-reported demographics for the block.
Then for each synthetic resident, we estimate the posterior predictive distribution over Census blocks by simulating 20 neighborhoods from the posterior predictive distribution of the city council model.\footnote{We remove other individual-level covariates (e.g., retirement status) from the city council model for this exercise, since the Census Bureau does not provide such variables.}
Since we are simulating neighborhoods for synthetic residents who did not take the survey, we draw new random effects for each resident.
While all the synthetic residents live in the same Census block, they differ in their covariates, and so their posterior predictive neighborhoods are different as well.

\begin{figure}[!t]

{\centering \includegraphics{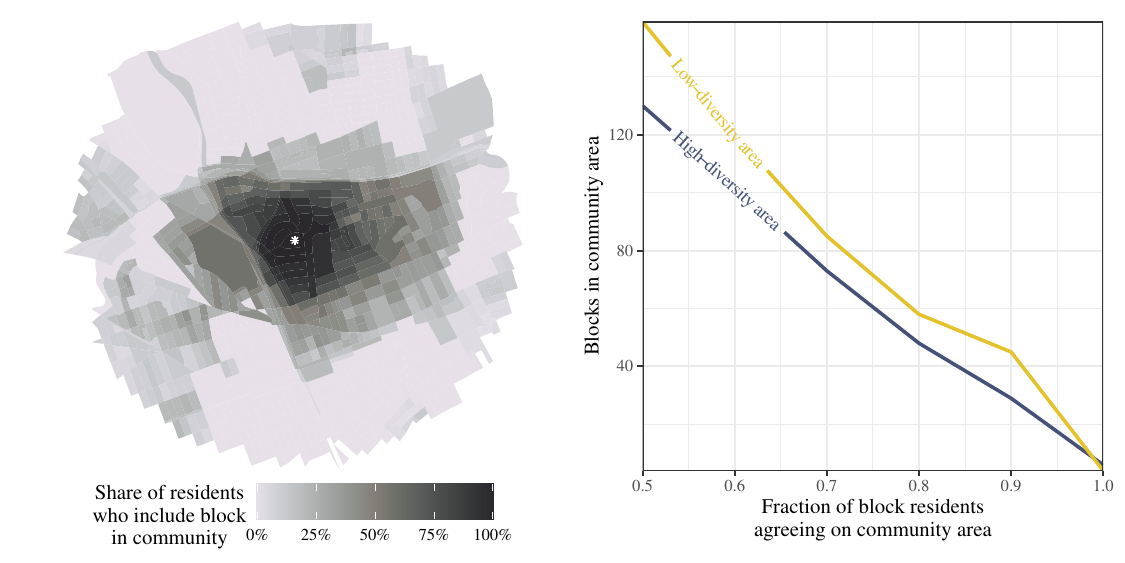} 

}

\caption[Consensus community of interest map and comparison between high and low diversity areas.]{On the left, a map visualizing the consensus community of a synthetic
residential population of a single Census block, which is marked with a white asterisk.
Darker blocks are those which are included in a higher proportion of
synthetic resident's predicted neighborhoods.
On the right, the tradeoff between the size of the community of interest
and the degree of consensus is plotted for communities in two areas:
one with high and one with low racial diversity.
}\label{fig:coi-pred}
\end{figure}

We can now aggregate the 100 residents' posterior predictive distributions by calculating, for each block, the fraction of the synthetic respondents who assign at least 50\% posterior probability to that block.
Blocks which belong, with high probability, to almost everyone's predicted neighborhood will have high values, while blocks that generally belong to only one or two residents' neighborhood will have low values.
These values are plotted on a map in the left of Figure \ref{fig:coi-pred}, for a block in a highly racially diverse area of the borough of Queens.

Unsurprisingly, there is strong agreement for blocks close to the block where all the residents live, with the share of residents including a block falling with distance.
This illustrates a fundamental trade-off in building consensus neighborhoods: all else being equal, a smaller neighborhood will have a higher level of consensus.
We can visualize this tradeoff directly by changing the threshold used to decide whether a block belongs to the consensus neighborhood in the left map of Figure \ref{fig:coi-pred}.
In other words, we consider modifying a minimum share of residents who include the block in their neighborhood.
As we vary this threshold value (horizontal axis), the size of the consensus neighborhood changes as well.

This is visualized in the right plot of Figure \ref{fig:coi-pred} where the blue line represents the result for this highly racially diverse block in Queens.
When we repeat this prediction exercise in an area of Brooklyn with low racial diversity, we obtain a different community consensus-size curve, which is represented by a yellow line in the plot.
Specifically, in the the low-diversity area, it is easier to build a consensus neighborhood: for any given neighborhood size, a higher fraction of block residents can agree on a neighborhood of that size.
Conversely, for a given share of agreement, the neighborhoods in the the low-diversity area are larger, on average.

These patterns reflect the individual-level findings from our fitted models: residents prefer racially homogeneous neighborhoods and communities.
As a result, racially diverse areas will find it harder to agree on a common definition of a neighborhood.

\FloatBarrier

\hypertarget{concluding-remarks}{%
\section{Concluding Remarks}\label{concluding-remarks}}

The study of political and social geography is often impeded by persistent measurement challenges.
Progress on substantive questions necessitates methodological advancements in the construction and analysis of geographic data.
We provide an open-source survey tool that allows researchers to measure subjective social and political geography. This survey tool can be used to collect any type of drawn map from survey respondents, and these collected data can be used as an outcome -- to quantify why respondents drew the map they did -- or can serve as more appropriate measures (compared to common practices of using administrative units such as Census tracts or ZIP Codes) of local context from which to measure geographic variables. Our survey module can be easily modified to directly measure various geographies of interest using different prompts, designs, and instructions. For example, researchers could use different sub-geographic units (such as housing parcels, rather than census blocks) as the building blocks for subjective geographies.
Another possibility is to add or remove certain information about geographical units, buildings, and landmarks.

We also propose a statistical model that can be used to analyze the data obtained from our survey module. The model helps us better understand how people perceive their local geography, and this perception in turn informs the investigation of how perceived geography may influence social, political, and economic behaviors. Researchers can also use our model to analyze any type of geographic unit -- creating opportunities for enchanced understanding of administrative boundaries, local governance, and the interaction of political institutions and geography. For example, one could take a dataset of census tracts and take the centroid of each tract as the point from which to operationalize distance, and then quantify how much aggregate characteristics of census blocks or any smaller geography predicts census how tracts are drawn. In terms of explanatory variables of the model, any information that is spatially measured can be incorporated into the analysis, and the model can produce estimates of its influence on neighborhood inclusion.

Our substantive applications illustrate the potential uses of this methodology, and demonstrates a striking relationship between racial and partisan demographics and subjective neighborhoods. Even after accounting for individual characteristics, aggregate socio-economic variables, and infrastructural characteristics, voters are more likely to include census blocks that consist of greater numbers of same-race or same-party residents.
Variation in racial or partisan homophily produces sizable changes in inclusion probability, which in turn produce substantive differences in the kinds of subjective neighborhoods for respondents of different parties and races. These patterns spur further questions about the role of inter-ethnic and inter-party relations in shaping social geography.

Lastly, we demonstrate that our methodology can be used to make better out-of-sample predictions of subjective neighborhoods than distance-based measures, Census tracts, or ZIP Codes. This result suggests that researchers could, under certain circumstances, use our methodology to generate likely neighborhoods for individuals where drawn maps are not collected. To do so, researchers would need to collect drawn neighborhoods from a representative sample of their target population of interest. Even with a representative sample, researchers should still be mindful of the possibility that the measurement which is necessarily present in predicted neighborhoods could be correlated with the outcome of interest, leading to biased inference \citep{fong_tyler_2021, KnoxLucasTam:2022, mccartan2023estimating, egami2023using}. This concern is present in any use of predicted data, and researchers should use caution when applying our methodology in this manner. But, once these conditions are met, researchers could use our model to improve the measure of local context in larger datasets.

\hypertarget{human-subjects}{%
\section{Human Subjects}\label{human-subjects}}

The authors declare the human subjects research in this article was reviewed and approved by Harvard University's Institutional Review Board and certificate numbers are provided in the appendix. The authors affirm that this article adheres to the APSA's Principles and Guidance on Human Subject Research.

\hypertarget{ethics-and-conflicts-of-interest}{%
\section{Ethics and Conflicts of Interest}\label{ethics-and-conflicts-of-interest}}

The authors declare no ethical issues or conflicts of interest in this research. This research was in part funded by Meta, who provided advertisement credits to collect some of the survey data. The survey advertisement content went through the Meta policy review. Meta had no other role in the design or conduct of the research and no role in the interpretation of the data or preparation of the manuscript.

\hypertarget{data-transparency}{%
\section{Data Transparency}\label{data-transparency}}

Research documentation and code that support the findings of this study are openly available in the APSR Dataverse at \url{https://doi.org/10.7910/DVN/SDSUQG} \citep{DVN/SDSUQG_2023}. Although our survey tool is publicly available, we are not able to publicly release our survey data to protect the privacy of respondents.

\newpage
\bibliography{bibliography}

\newpage

\clearpage
\appendix
\renewcommand\thesection{S\arabic{section}}
\renewcommand\thefigure{S\arabic{figure}}
\renewcommand\thetable{S\arabic{table}}
\setcounter{figure}{0} \setcounter{table}{0} \setcounter{page}{1}
\fontsize{10}{10.5}\selectfont
\captionsetup[figure]{font=footnotesize}
\captionsetup[table]{font=footnotesize}

\begin{center}
 {\Large \bf Supplementary Information for \\ Cory McCartan, Jacob Brown, and Kosuke Imai. ``Measuring and Modeling Neighborhoods.''}
\end{center}

\hypertarget{app:summary}{%
\section{Summary of the survey samples}\label{app:summary}}

Each of our surveys were approved by the Harvard Institutional Review Board (IRB-20-0938). Neighborhood survey respondents were recruited via e-mail using a list of email addresses attached to registered voter records.
We randomly sampled a total of 1,514,612 potential respondents from this list within each city (533,333 in Miami, 476,605 in New York City, and 504,674 in Phoenix).
We sent an email invitation to the sampled registered voters on non-holiday weekdays between December 21, 2020 and February 19, 2021.
The invitation informed the potential respondent of the purpose of the survey, that they would be asked to draw their neighborhood on a map, and provided information on the researcher's affiliations and contact information.
Of these e-mails, 38.5\% failed to be delivered to the potential respondent's inbox, due to the email address either being invalid or the receiving email server rejecting the email.
In total, 930,839 voters received survey invitations (329,624 in Miami, 275,449 in New York, and 325,766 in Phoenix).
The Phoenix sample exhibited a higher response rate than the Miami and New York City sample, with a 0.5\% response rate in Miami and New York and a 1.3\% response rate in Phoenix.
Although our survey tool is publicly available, we are not able to publicly release our survey data to protect the privacy of respondents.

Here we present summary statistics describing the demographics of the first survey sample (Table \ref{tab:surv-sum}), the overall demographics of the cities in our sample (Table \ref{tab:city-sum}) and the demographics of the city council survey sample (Table \ref{tab:surv-sum-full-nycc}).

Table \ref{tab:surv-sum} shows the overall summary statistics and those broken out by city for our first survey. The demographic comparison across cities also may inform the differences in response rates across cities. The higher response rate in Phoenix is likely due to higher quality of email lists in this city than in New York or Miami and due to an older sampling population being more likely to respond to the surveys.

Comparison of Table \ref{tab:city-sum} to Tables \ref{tab:surv-sum} and \ref{tab:surv-sum-full-nycc} gives a sense of the representativeness of the sample relative to the adult population of the three metropolitan areas of the study. We find that our sample is more predominantly white, wealthier, educated, and more likely to be a homeowner than the population of each of the cities in our sample.

\begin{table}[h]
\centering
\footnotesize
\caption{Survey Sample Summary Statistics -- Full Sample}
\label{tab:surv-sum}
\begin{tabular}[t]{lrrrrrr|rr}
\toprule
\multicolumn{1}{c}{ } & \multicolumn{2}{c}{Miami} & \multicolumn{2}{c}{NYC} & \multicolumn{2}{c}{Phoenix} & \multicolumn{2}{|c}{ Pooled} \\
\multicolumn{1}{c}{ } & \multicolumn{2}{c}{(n = 473)} & \multicolumn{2}{c}{(n = 450)} & \multicolumn{2}{c}{(n = 1,585)} & \multicolumn{2}{|c}{(n = 2,508) } \\
\cmidrule(l{3pt}r{3pt}){2-3} \cmidrule(l{3pt}r{3pt}){4-5} \cmidrule(l{3pt}r{3pt}){6-7}\cmidrule(l{3pt}r{3pt}){8-9}
  & Avg. & St. Dev. & Avg. & St. Dev. & Avg. & St. Dev.   & Avg. & St. Dev.   \\
\midrule
Democrat & 0.48 & 0.50 & 0.57 & 0.49 & 0.38 & 0.49 & 0.44 & 0.50 \\
Republican &0.40 & 0.49 & 0.33 & 0.47 & 0.52 & 0.50 & 0.46 & 0.50 \\
Vote Biden 2020 &  0.57 & 0.50 & 0.65 & 0.48 & 0.49 & 0.50 & 0.53 & 0.50 \\
Age & 61.78 & 14.16 & 60.13 & 13.37 & 64.11 & 12.98 & 62.89 & 13.41 \\
Female & 0.44 & 0.50 & 0.45 & 0.50 & 0.48 & 0.50 & 0.46 & 0.50 \\
White & 0.65 & 0.48 & 0.72 & 0.45 & 0.86 & 0.34 & 0.79 & 0.41\\
Income (1,000s) & 104.91 & 49.35 & 117.30 & 48.35 & 110.88 & 48.03 & 110.91 & 48.59 \\
College & 0.67 & 0.47 & 0.68 & 0.47 & 0.62 & 0.48 & 0.64 & 0.48 \\
Years Residence & 17.47 & 53.24 & 23.70 & 58.59 & 19.11 & 58.29 & 19.57 & 57.30 \\
Homeowner & 0.86 & 0.35 & 0.74 & 0.44 & 0.90 & 0.30 & 0.86 & 0.34 \\
Married & 0.60 & 0.49 & 0.60 & 0.49 & 0.67 & 0.47 & 0.64 & 0.48 \\
Children in Home & 0.33 & 0.47 & 0.39 & 0.49 & 0.27 & 0.45 & 0.31 & 0.46 \\
\bottomrule
\end{tabular}
\end{table}

\begin{table}[ht]
\centering
\footnotesize
\caption{Metropolitan Region Population Demographics.
Proportions are out of adult census population in each metropolitan region, except for Democrat and Republican, which are out of total registered voters.}
\label{tab:city-sum}
\begin{tabular}{lccc}
  \hline
City & Miami & NYC & Phoenix \\
  \hline
  College & 0.413 & 0.455 & 0.389 \\
  Homeowner & 0.608 & 0.449 & 0.641 \\
  Median Income & \$66,944 & \$85,267 & \$71,421 \\
  Registered & 0.650 & 0.614 & 0.656 \\
  White & 0.348 & 0.420 & 0.587 \\
  Black & 0.197 & 0.181 & 0.046 \\
  Hispanic & 0.416 & 0.263 & 0.295 \\
  Democrat & 0.450 & 0.587 & 0.313 \\
  Republican & 0.247 & 0.148 & 0.343 \\
   \hline
\end{tabular} \\
\end{table}

For the city council survey, a total of 490,000 registered voters were sampled from voter lists and sent an email invitation on non-holiday weekdays between December 22, 2022 and January 25, 2023.
Similar to the previous survey, the bounce rate for the email invitation was 43.3\%.

\begin{table}
\centering
\footnotesize
\caption{Survey Sample Summary Statistics - City Council Survey}
\label{tab:surv-sum-full-nycc}
\begin{tabular}[t]{lrrrrrr}
\toprule
\multicolumn{1}{c}{ } & \multicolumn{2}{c}{Email} & \multicolumn{2}{c}{Meta} & \multicolumn{2}{c}{Pooled } \\
\cmidrule(l{3pt}r{3pt}){2-3} \cmidrule(l{3pt}r{3pt}){4-5} \cmidrule(l{3pt}r{3pt}){6-7}
  & Avg. & St. Dev. & Avg. & St. Dev. &Avg. & St. Dev.\\
\midrule
Democrat & 0.71 & 0.45 & 0.78 & 0.41 & 0.75 & 0.44\\
Republican & 0.20 & 0.40 & 0.11 & 0.32 & 0.16 & 0.36\\
Eric Adams & 0.34 & 0.47 & 0.17 & 0.37 & 0.26 & 0.44\\
Age & 56.95 & 13.72 & 41.19 & 18.59 & 49.63 & 17.97\\
Female & 0.49 & 0.50 & 0.31 & 0.46 & 0.41 & 0.49\\
White & 0.64 & 0.48 & 0.77 & 0.42 & 0.70 & 0.46\\
Income & 114,008 & 49,553 & 103,972 & 52,130 & 109,001 & 51,076\\
College & 0.48 & 0.50 & 0.41 & 0.49 & 0.44 & 0.50\\
Years Residence & 16.10 & 8.71 & 9.00 & 9.42 & 12.82 & 9.71\\
Homeowner & 0.50 & 0.50 & 0.27 & 0.44 & 0.39 & 0.49\\
Married & 0.50 & 0.50 & 0.32 & 0.47 & 0.42 & 0.49\\
Children in Home & 0.37 & 0.48 & 0.19 & 0.39 & 0.28 & 0.45\\
\bottomrule
\end{tabular}
\end{table}

\FloatBarrier

\hypertarget{app:descriptive}{%
\section{Additional descriptive statistics of drawn neighborhoods}\label{app:descriptive}}

\hypertarget{first-survey}{%
\subsection{First survey}\label{first-survey}}

\begin{figure}[htb]

{\centering \includegraphics[width=0.75\linewidth]{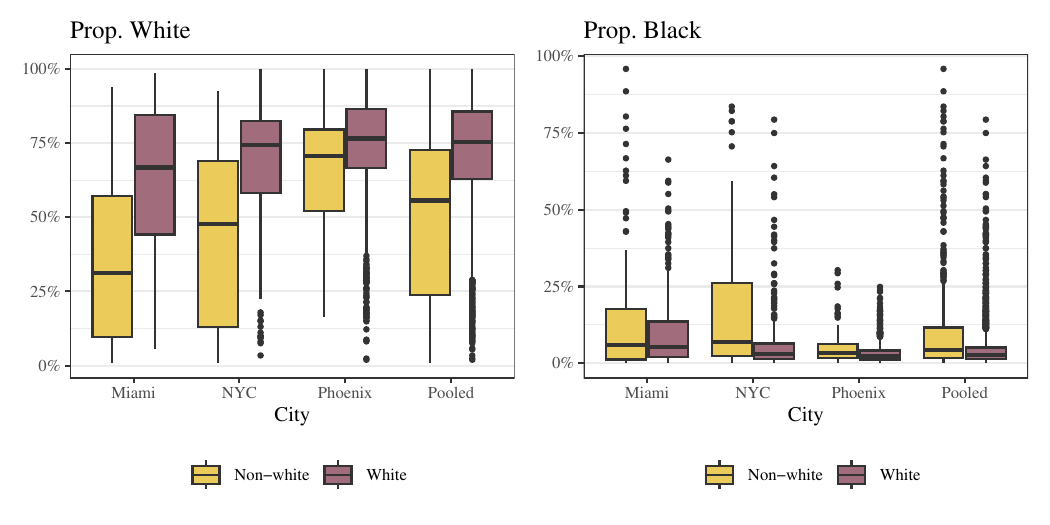} 

}

\caption[Racial demographics for respondent neighborhoods by race (first survey)]{Racial demographics for respondent neighborhoods by respondent race (first survey)}\label{fig:nbhd-stats-by-race-full}
\end{figure}

\begin{figure}[htb]

{\centering \includegraphics[width=0.75\linewidth]{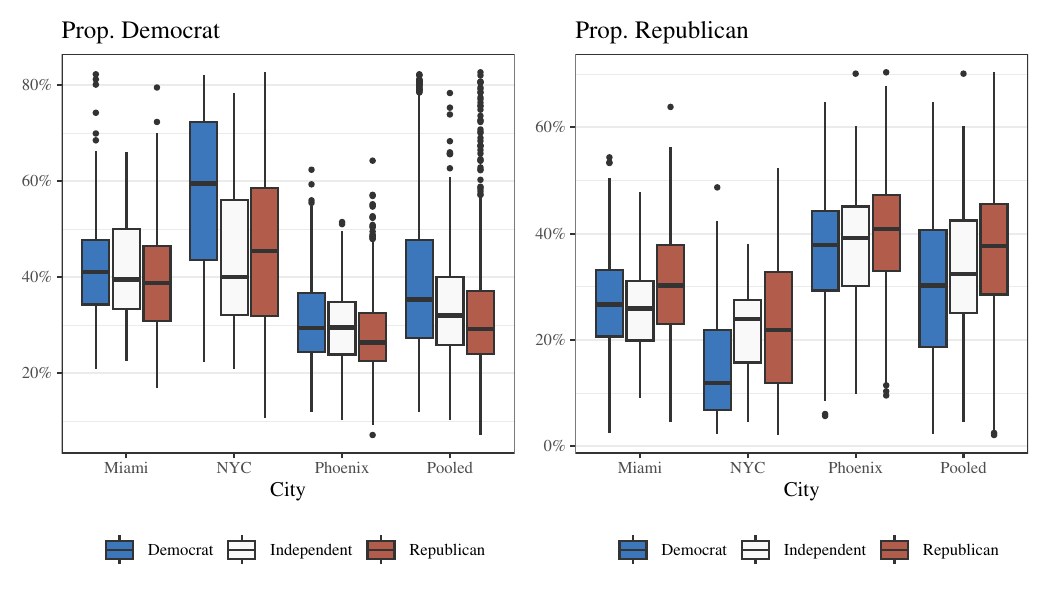} 

}

\caption[Party demographics for respondent neighborhoods by party (first survey)]{Party demographics for respondent neighborhoods by respondent party (first survey)}\label{fig:nbhd-stats-by-party-full}
\end{figure}

\begin{table}[!htbp] \centering
  \caption{Treatment Effect on Usable Neighborhoods}
  \label{tab:map-attrition}
  \footnotesize
\begin{tabular}{@{\extracolsep{0pt}}lccc}
\\\hline
\hline \\
 & \multicolumn{3}{c}{\textit{Dependent variable:}} \\
\cline{2-4}
\\& \multicolumn{3}{c}{Usable Neighborhood} \\
\\& Miami & New York City & Phoenix\\
\hline \\
 Party Condition & 0.034 & 0.081 & 0.035 \\
  & (0.052) & (0.057) & (0.036) \\

 Party Placebo Condition & 0.066 & $-$0.037 & $-$0.004 \\
  & (0.055) & (0.058) & (0.037) \\

 Race Condition & $-$0.004 & 0.093 & 0.050 \\
  & (0.056) & (0.058) & (0.037) \\

 Race Placebo Condition & 0.081 & 0.026 & 0.025 \\
  & (0.054) & (0.058) & (0.036) \\

 Age & $-$0.005$^{***}$ & $-$0.001 & $-$0.004$^{***}$ \\
  & (0.002) & (0.002) & (0.001) \\

 College & 0.021 & 0.053 & 0.045$^{*}$ \\
  & (0.043) & (0.050) & (0.027) \\

 Democrat & 0.119$^{*}$ & 0.006 & 0.053 \\
  & (0.064) & (0.077) & (0.047) \\

 Female & 0.011 & 0.007 & 0.011 \\
  & (0.035) & (0.037) & (0.024) \\

 Homeowner & 0.017 & $-$0.080$^{*}$ & 0.030 \\
  & (0.051) & (0.048) & (0.038) \\

 Income & 0.0004 & 0.002$^{***}$ & 0.0004 \\
  & (0.0004) & (0.0004) & (0.0003) \\

 Married & 0.047 & $-$0.063 & 0.006 \\
  & (0.039) & (0.043) & (0.027) \\

 Republican & 0.142$^{**}$ & $-$0.010 & 0.021 \\

 Vote Biden & 0.110$^{*}$ & 0.078 & 0.098$^{**}$ \\
  & (0.062) & (0.063) & (0.039) \\

 Years Residence & 0.002 & 0.003 & 0.002 \\
  & (0.002) & (0.002) & (0.001) \\

 Constant & 0.260$^{**}$ & 0.157 & 0.401$^{***}$ \\
  & (0.114) & (0.134) & (0.083) \\
 \hline \\
Observations & 1,468 & 1,193 & 4,028 \\
R$^{2}$ & 0.007 & 0.005 & 0.0004 \\
Adjusted R$^{2}$ & 0.005 & 0.001 & $-$0.001 \\
Residual Std. Error & 0.467 (df = 1463) & 0.485 (df = 1188) & 0.490 (df = 4023) \\
F Statistic & 2.735$^{**}$ (df = 4; 1463) & 1.366 (df = 4; 1188) & 0.430 (df = 4; 4023) \\
\hline
\hline \\
\textit{Note:}  & \multicolumn{3}{r}{ $^{*}$p$<$0.05; $^{**}$p$<$0.01} \\
\end{tabular}
\end{table}

\FloatBarrier

\hypertarget{city-council-survey}{%
\subsection{City council survey}\label{city-council-survey}}

\begin{figure}[htb]

{\centering \includegraphics[width=0.75\linewidth]{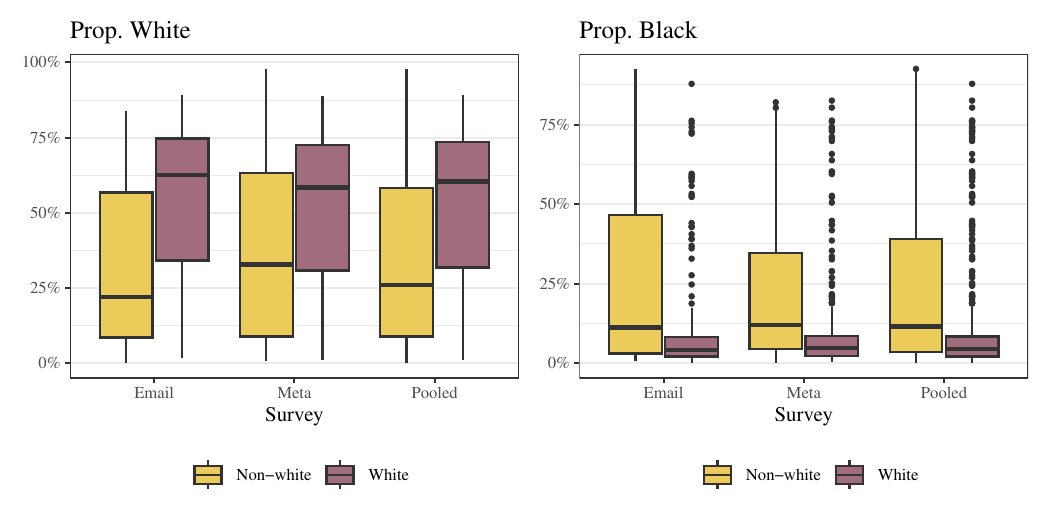} 

}

\caption[Racial demographics for New York respondent communities of interest by race (city council survey)]{Racial demographics for respondent communities of interest by respondent race (city council survey)}\label{fig:nbhd-stats-by-race-full-nycc}
\end{figure}

\begin{figure}[htb]

{\centering \includegraphics[width=0.75\linewidth]{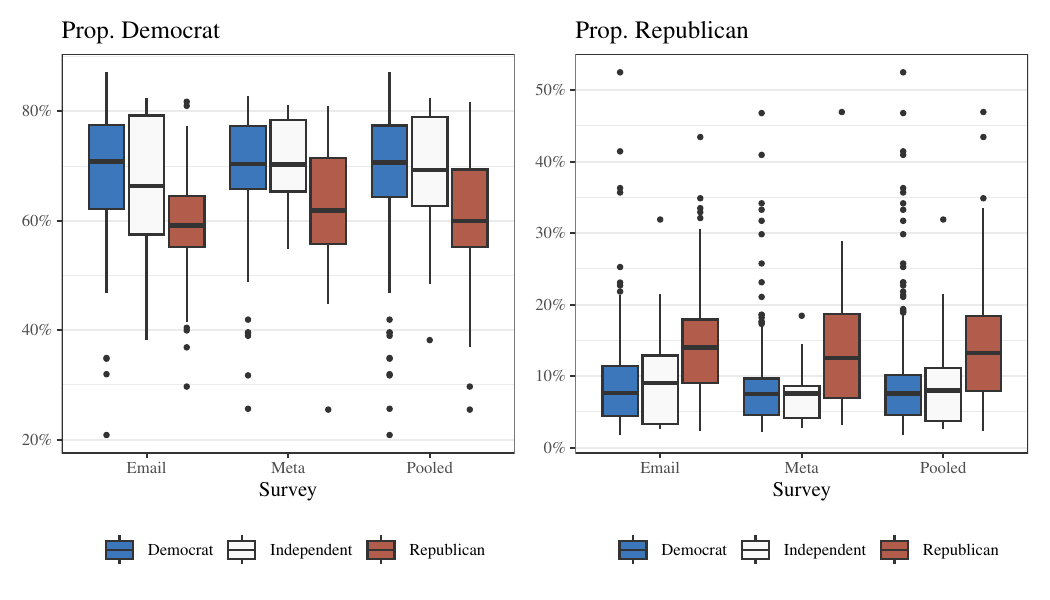} 

}

\caption[Party demographics for New York respondent communities of interest by party (city council survey)]{Party demographics for respondent communities of interest by respondent party (city council survey)}\label{fig:nbhd-stats-by-party-full-nycc}
\end{figure}

\FloatBarrier

\hypertarget{app:compute}{%
\section{Computational details}\label{app:compute}}

As shown by Equation \eqref{p-glm}, the model can be expressed as a Bernoulli GLMM with complementary log-log link function.
In this setup, every block that is included the neighborhood, as well as those blocks at the boundary of the neighborhood which could have been included (\(C_i=1\)) but are not, becomes a separate GLMM observation.
As a result, even moderate respondent sample sizes lead to many more block-level observations, although these block-level observations are of course dependent.
In our data, for example, 309 respondents in the control group from Phoenix translate to 57,242 block-level observations.

The large number of block-level observations means that estimates will generally be precise for coefficients which vary at the block level and are shared across all respondents.
At the same time, it can create computational efficiency problems for traditional Bayesian posterior sampling methods.
Consequently, for our analysis we use a Normal approximation corrected by importance resampling, as implemented by the Stan modeling package Rstan.
This approximation centers a Normal distribution at the posterior mode with covariance matrix the inverse of the curvature of the log posterior density at the mode.
It subsequently samples 1,000 draws from this Gaussian approximation and performs importance resampling so that the draws better approximate the posterior.

\fontsize{10}{10.5}\selectfont
\FloatBarrier

\hypertarget{app:model}{%
\section{Variable descriptions and model specifications}\label{app:model}}

\begin{longtable}[]{@{}
  >{\raggedright\arraybackslash}p{(\columnwidth - 6\tabcolsep) * \real{0.1605}}
  >{\raggedright\arraybackslash}p{(\columnwidth - 6\tabcolsep) * \real{0.5123}}
  >{\raggedright\arraybackslash}p{(\columnwidth - 6\tabcolsep) * \real{0.2099}}
  >{\centering\arraybackslash}p{(\columnwidth - 6\tabcolsep) * \real{0.1049}}@{}}
\caption{\label{tab:spec} Variable description and model specification.}\tabularnewline
\toprule\noalign{}
\begin{minipage}[b]{\linewidth}\raggedright
Variable
\end{minipage} & \begin{minipage}[b]{\linewidth}\raggedright
Description
\end{minipage} & \begin{minipage}[b]{\linewidth}\raggedright
Interaction(s)
\end{minipage} & \begin{minipage}[b]{\linewidth}\centering
Baseline only?
\end{minipage} \\
\midrule\noalign{}
\endfirsthead
\toprule\noalign{}
\begin{minipage}[b]{\linewidth}\raggedright
Variable
\end{minipage} & \begin{minipage}[b]{\linewidth}\raggedright
Description
\end{minipage} & \begin{minipage}[b]{\linewidth}\raggedright
Interaction(s)
\end{minipage} & \begin{minipage}[b]{\linewidth}\centering
Baseline only?
\end{minipage} \\
\midrule\noalign{}
\endhead
\bottomrule\noalign{}
\endlastfoot
Church & Block contains church & & \\
Distance to church & Logarithm of distance to nearest church from block, in meters & & \\
Park & Block contains park & & \\
School & Block contains school & Children & \\
Distance to school & Logarithm of distance to nearest school from block, in meters & Children & \\
Children & Respondent has children at home & School, distance to school & * \\
Same block group & Block in same block group as respondent & & \\
Same tract & Block in same tract as respondent & & \\
Same road region & Block in same region bounded by major roads and railroads as respondent & & \\
Population & Square root of the block population divided by 10,000 & & \\
Area & Square root of the area in square miles & & \\
Fraction same race & Fraction of block of the same race as respondent (White, Black, Hispanic, Other) & Minority & * \\
Minority & Whether respondent is non-white, or Hispanic of any race & Fraction same race & * \\
Fraction same party & Fraction of block of the same party as respondent & Party & * \\
Party & Self-reported political party: Democratic, Republican, or independent & Fraction same party & * \\
Fraction same ownership & Fraction of block with same home ownership or rental status as respondent & Homeowner & * \\
Homeowner & Whether respondent owns their home & Fraction same ownership & * \\
Fraction same education & Fraction of block with same education as respondent & Education & * \\
Education & Respondent education; either ``college'' or ``no college'' & Fraction same education, Income & * \\
Income & Logarithm of median income of block group & Education & * \\
Age & Respondent age group: 0--40, 41--55, 56--65, 66-75, or 76+ & & * \\
Retired & Whether respondent is retired & & * \\
Tenure & Square root of how long respondent has lived in current residence & & * \\
\end{longtable}

\fontsize{10}{10.5}\selectfont
\FloatBarrier

\hypertarget{app:coefs}{%
\section{Full and baseline model estimates}\label{app:coefs}}

\hypertarget{first-survey-1}{%
\subsection{First survey}\label{first-survey-1}}

Tables \ref{tab:model_coef_full} and \ref{tab:model_coef_base} contain posterior summaries for all model coefficients on the original model scale.
These models were fit using data from a random sample of 400 survey respondents from the control group, consisting of 78,771 individual block-level observations.

\begingroup\fontsize{10}{12}\selectfont

\begin{longtable}[t]{llrrrrr}
\caption{Full model estimates.}
\label{tab:model_coef_full}\\
\toprule
Coefficient & City & Mean & Std. Dev. & Q5 & Median & Q95\\
\midrule
\cellcolor{gray!6}{(Intercept)} & \cellcolor{gray!6}{Miami} & \cellcolor{gray!6}{-6.59} & \cellcolor{gray!6}{9.52} & \cellcolor{gray!6}{-22.30} & \cellcolor{gray!6}{-6.44} & \cellcolor{gray!6}{8.89}\\
\cellcolor{gray!6}{(Intercept)} & \cellcolor{gray!6}{NYC} & \cellcolor{gray!6}{-5.37} & \cellcolor{gray!6}{7.56} & \cellcolor{gray!6}{-17.36} & \cellcolor{gray!6}{-5.29} & \cellcolor{gray!6}{6.43}\\
\cellcolor{gray!6}{(Intercept)} & \cellcolor{gray!6}{Phoenix} & \cellcolor{gray!6}{-9.30} & \cellcolor{gray!6}{6.36} & \cellcolor{gray!6}{-19.82} & \cellcolor{gray!6}{-9.17} & \cellcolor{gray!6}{0.51}\\
Church & Miami & -0.04 & 0.07 & -0.16 & -0.04 & 0.07\\
Church & NYC & 0.11 & 0.05 & 0.03 & 0.11 & 0.19\\
Church & Phoenix & 0.14 & 0.04 & 0.07 & 0.14 & 0.21\\
\cellcolor{gray!6}{Distance} & \cellcolor{gray!6}{Miami} & \cellcolor{gray!6}{0.00} & \cellcolor{gray!6}{0.02} & \cellcolor{gray!6}{-0.03} & \cellcolor{gray!6}{0.00} & \cellcolor{gray!6}{0.03}\\
\cellcolor{gray!6}{Distance} & \cellcolor{gray!6}{Miami} & \cellcolor{gray!6}{0.13} & \cellcolor{gray!6}{0.03} & \cellcolor{gray!6}{0.08} & \cellcolor{gray!6}{0.13} & \cellcolor{gray!6}{0.18}\\
\cellcolor{gray!6}{Distance} & \cellcolor{gray!6}{NYC} & \cellcolor{gray!6}{0.12} & \cellcolor{gray!6}{0.02} & \cellcolor{gray!6}{0.08} & \cellcolor{gray!6}{0.12} & \cellcolor{gray!6}{0.15}\\
Distance & NYC & 0.01 & 0.03 & -0.03 & 0.01 & 0.06\\
Distance & Phoenix & 0.07 & 0.01 & 0.06 & 0.07 & 0.09\\
Distance & Phoenix & 0.16 & 0.02 & 0.14 & 0.16 & 0.19\\
\cellcolor{gray!6}{Park} & \cellcolor{gray!6}{Miami} & \cellcolor{gray!6}{-0.06} & \cellcolor{gray!6}{0.09} & \cellcolor{gray!6}{-0.21} & \cellcolor{gray!6}{-0.05} & \cellcolor{gray!6}{0.08}\\
\cellcolor{gray!6}{Park} & \cellcolor{gray!6}{NYC} & \cellcolor{gray!6}{0.15} & \cellcolor{gray!6}{0.05} & \cellcolor{gray!6}{0.08} & \cellcolor{gray!6}{0.15} & \cellcolor{gray!6}{0.23}\\
\cellcolor{gray!6}{Park} & \cellcolor{gray!6}{Phoenix} & \cellcolor{gray!6}{0.07} & \cellcolor{gray!6}{0.04} & \cellcolor{gray!6}{0.01} & \cellcolor{gray!6}{0.07} & \cellcolor{gray!6}{0.13}\\
School & Miami & 0.39 & 0.14 & 0.16 & 0.39 & 0.62\\
School & NYC & -0.15 & 0.12 & -0.34 & -0.15 & 0.06\\
School & Phoenix & 0.15 & 0.07 & 0.04 & 0.16 & 0.27\\
\cellcolor{gray!6}{Children} & \cellcolor{gray!6}{Miami} & \cellcolor{gray!6}{0.53} & \cellcolor{gray!6}{3.10} & \cellcolor{gray!6}{-4.48} & \cellcolor{gray!6}{0.57} & \cellcolor{gray!6}{5.39}\\
\cellcolor{gray!6}{Children} & \cellcolor{gray!6}{NYC} & \cellcolor{gray!6}{-0.77} & \cellcolor{gray!6}{2.79} & \cellcolor{gray!6}{-5.15} & \cellcolor{gray!6}{-0.78} & \cellcolor{gray!6}{3.73}\\
\cellcolor{gray!6}{Children} & \cellcolor{gray!6}{Phoenix} & \cellcolor{gray!6}{-1.09} & \cellcolor{gray!6}{2.05} & \cellcolor{gray!6}{-4.35} & \cellcolor{gray!6}{-1.14} & \cellcolor{gray!6}{2.39}\\
Same block group & Miami & 0.08 & 0.07 & -0.04 & 0.08 & 0.19\\
Same block group & NYC & 0.21 & 0.09 & 0.07 & 0.21 & 0.35\\
Same block group & Phoenix & 0.08 & 0.03 & 0.03 & 0.08 & 0.13\\
\cellcolor{gray!6}{Same tract} & \cellcolor{gray!6}{Miami} & \cellcolor{gray!6}{-0.17} & \cellcolor{gray!6}{0.09} & \cellcolor{gray!6}{-0.33} & \cellcolor{gray!6}{-0.17} & \cellcolor{gray!6}{-0.04}\\
\cellcolor{gray!6}{Same tract} & \cellcolor{gray!6}{NYC} & \cellcolor{gray!6}{0.06} & \cellcolor{gray!6}{0.12} & \cellcolor{gray!6}{-0.14} & \cellcolor{gray!6}{0.06} & \cellcolor{gray!6}{0.26}\\
\cellcolor{gray!6}{Same tract} & \cellcolor{gray!6}{Phoenix} & \cellcolor{gray!6}{-0.09} & \cellcolor{gray!6}{0.10} & \cellcolor{gray!6}{-0.27} & \cellcolor{gray!6}{-0.09} & \cellcolor{gray!6}{0.07}\\
Same road region & Miami & -0.49 & 0.05 & -0.57 & -0.49 & -0.41\\
Same road region & NYC & 0.13 & 0.04 & 0.07 & 0.13 & 0.20\\
Same road region & Phoenix & -0.09 & 0.03 & -0.14 & -0.09 & -0.04\\
\cellcolor{gray!6}{Population} & \cellcolor{gray!6}{Miami} & \cellcolor{gray!6}{-0.22} & \cellcolor{gray!6}{0.26} & \cellcolor{gray!6}{-0.66} & \cellcolor{gray!6}{-0.22} & \cellcolor{gray!6}{0.21}\\
\cellcolor{gray!6}{Population} & \cellcolor{gray!6}{NYC} & \cellcolor{gray!6}{-1.21} & \cellcolor{gray!6}{0.20} & \cellcolor{gray!6}{-1.53} & \cellcolor{gray!6}{-1.21} & \cellcolor{gray!6}{-0.90}\\
\cellcolor{gray!6}{Population} & \cellcolor{gray!6}{Phoenix} & \cellcolor{gray!6}{-1.18} & \cellcolor{gray!6}{0.16} & \cellcolor{gray!6}{-1.43} & \cellcolor{gray!6}{-1.19} & \cellcolor{gray!6}{-0.89}\\
Area & Miami & -0.37 & 0.08 & -0.51 & -0.37 & -0.24\\
Area & NYC & -0.14 & 0.11 & -0.32 & -0.14 & 0.03\\
Area & Phoenix & 0.00 & 0.04 & -0.06 & 0.00 & 0.06\\
\cellcolor{gray!6}{Fraction same race} & \cellcolor{gray!6}{Miami} & \cellcolor{gray!6}{-0.30} & \cellcolor{gray!6}{0.06} & \cellcolor{gray!6}{-0.39} & \cellcolor{gray!6}{-0.30} & \cellcolor{gray!6}{-0.21}\\
\cellcolor{gray!6}{Fraction same race} & \cellcolor{gray!6}{NYC} & \cellcolor{gray!6}{-0.12} & \cellcolor{gray!6}{0.05} & \cellcolor{gray!6}{-0.21} & \cellcolor{gray!6}{-0.12} & \cellcolor{gray!6}{-0.04}\\
\cellcolor{gray!6}{Fraction same race} & \cellcolor{gray!6}{Phoenix} & \cellcolor{gray!6}{-0.28} & \cellcolor{gray!6}{0.03} & \cellcolor{gray!6}{-0.33} & \cellcolor{gray!6}{-0.29} & \cellcolor{gray!6}{-0.24}\\
Fraction same party & Miami & -0.26 & 0.09 & -0.41 & -0.26 & -0.12\\
Fraction same party & NYC & -0.23 & 0.08 & -0.35 & -0.23 & -0.10\\
Fraction same party & Phoenix & -0.29 & 0.07 & -0.41 & -0.29 & -0.18\\
\cellcolor{gray!6}{Fraction same ownership} & \cellcolor{gray!6}{Miami} & \cellcolor{gray!6}{-0.24} & \cellcolor{gray!6}{0.15} & \cellcolor{gray!6}{-0.48} & \cellcolor{gray!6}{-0.24} & \cellcolor{gray!6}{0.02}\\
\cellcolor{gray!6}{Fraction same ownership} & \cellcolor{gray!6}{NYC} & \cellcolor{gray!6}{-0.27} & \cellcolor{gray!6}{0.13} & \cellcolor{gray!6}{-0.49} & \cellcolor{gray!6}{-0.27} & \cellcolor{gray!6}{-0.07}\\
\cellcolor{gray!6}{Fraction same ownership} & \cellcolor{gray!6}{Phoenix} & \cellcolor{gray!6}{-0.28} & \cellcolor{gray!6}{0.12} & \cellcolor{gray!6}{-0.47} & \cellcolor{gray!6}{-0.28} & \cellcolor{gray!6}{-0.10}\\
Fraction same education & Miami & 0.25 & 0.15 & 0.00 & 0.26 & 0.48\\
Fraction same education & NYC & -0.53 & 0.10 & -0.70 & -0.53 & -0.35\\
Fraction same education & Phoenix & -0.94 & 0.08 & -1.07 & -0.95 & -0.82\\
\cellcolor{gray!6}{Income} & \cellcolor{gray!6}{Miami} & \cellcolor{gray!6}{-0.12} & \cellcolor{gray!6}{0.05} & \cellcolor{gray!6}{-0.21} & \cellcolor{gray!6}{-0.12} & \cellcolor{gray!6}{-0.03}\\
\cellcolor{gray!6}{Income} & \cellcolor{gray!6}{NYC} & \cellcolor{gray!6}{-0.13} & \cellcolor{gray!6}{0.05} & \cellcolor{gray!6}{-0.21} & \cellcolor{gray!6}{-0.13} & \cellcolor{gray!6}{-0.06}\\
\cellcolor{gray!6}{Income} & \cellcolor{gray!6}{Phoenix} & \cellcolor{gray!6}{0.14} & \cellcolor{gray!6}{0.03} & \cellcolor{gray!6}{0.09} & \cellcolor{gray!6}{0.14} & \cellcolor{gray!6}{0.19}\\
Age & Miami & 0.01 & 0.16 & -0.25 & 0.00 & 0.28\\
Age & NYC & 0.01 & 0.13 & -0.20 & 0.01 & 0.23\\
Age & Phoenix & -0.01 & 0.11 & -0.17 & -0.02 & 0.17\\
\cellcolor{gray!6}{Education = no college} & \cellcolor{gray!6}{Miami} & \cellcolor{gray!6}{-2.20} & \cellcolor{gray!6}{3.20} & \cellcolor{gray!6}{-7.44} & \cellcolor{gray!6}{-2.23} & \cellcolor{gray!6}{3.13}\\
\cellcolor{gray!6}{Education = no college} & \cellcolor{gray!6}{NYC} & \cellcolor{gray!6}{0.06} & \cellcolor{gray!6}{2.83} & \cellcolor{gray!6}{-4.59} & \cellcolor{gray!6}{0.09} & \cellcolor{gray!6}{4.72}\\
\cellcolor{gray!6}{Education = no college} & \cellcolor{gray!6}{Phoenix} & \cellcolor{gray!6}{-0.33} & \cellcolor{gray!6}{2.26} & \cellcolor{gray!6}{-4.02} & \cellcolor{gray!6}{-0.28} & \cellcolor{gray!6}{3.31}\\
Retired & Miami & 0.50 & 3.10 & -4.57 & 0.51 & 5.29\\
Retired & NYC & 0.44 & 3.03 & -4.71 & 0.51 & 5.28\\
Retired & Phoenix & 0.11 & 2.44 & -4.01 & 0.20 & 4.11\\
\cellcolor{gray!6}{Tenure} & \cellcolor{gray!6}{Miami} & \cellcolor{gray!6}{-0.10} & \cellcolor{gray!6}{0.96} & \cellcolor{gray!6}{-1.71} & \cellcolor{gray!6}{-0.09} & \cellcolor{gray!6}{1.52}\\
\cellcolor{gray!6}{Tenure} & \cellcolor{gray!6}{NYC} & \cellcolor{gray!6}{-0.12} & \cellcolor{gray!6}{0.89} & \cellcolor{gray!6}{-1.55} & \cellcolor{gray!6}{-0.14} & \cellcolor{gray!6}{1.39}\\
\cellcolor{gray!6}{Tenure} & \cellcolor{gray!6}{Phoenix} & \cellcolor{gray!6}{0.24} & \cellcolor{gray!6}{0.77} & \cellcolor{gray!6}{-1.01} & \cellcolor{gray!6}{0.21} & \cellcolor{gray!6}{1.53}\\
Party = ind & Miami & -0.52 & 4.32 & -7.78 & -0.57 & 6.71\\
Party = ind & NYC & 0.41 & 4.73 & -7.33 & 0.30 & 8.66\\
Party = ind & Phoenix & 0.08 & 3.53 & -5.81 & 0.15 & 5.90\\
\cellcolor{gray!6}{Party = rep} & \cellcolor{gray!6}{Miami} & \cellcolor{gray!6}{-0.07} & \cellcolor{gray!6}{2.76} & \cellcolor{gray!6}{-4.57} & \cellcolor{gray!6}{-0.02} & \cellcolor{gray!6}{4.36}\\
\cellcolor{gray!6}{Party = rep} & \cellcolor{gray!6}{NYC} & \cellcolor{gray!6}{-0.27} & \cellcolor{gray!6}{2.60} & \cellcolor{gray!6}{-4.45} & \cellcolor{gray!6}{-0.30} & \cellcolor{gray!6}{4.09}\\
\cellcolor{gray!6}{Party = rep} & \cellcolor{gray!6}{Phoenix} & \cellcolor{gray!6}{-0.11} & \cellcolor{gray!6}{1.81} & \cellcolor{gray!6}{-3.08} & \cellcolor{gray!6}{-0.14} & \cellcolor{gray!6}{2.88}\\
Minority & Miami & -0.46 & 3.14 & -5.62 & -0.47 & 4.53\\
Minority & NYC & 0.03 & 2.69 & -4.47 & 0.04 & 4.59\\
Minority & Phoenix & -0.15 & 2.57 & -4.28 & -0.16 & 4.00\\
\cellcolor{gray!6}{Homeowner} & \cellcolor{gray!6}{Miami} & \cellcolor{gray!6}{-0.11} & \cellcolor{gray!6}{4.93} & \cellcolor{gray!6}{-8.35} & \cellcolor{gray!6}{0.02} & \cellcolor{gray!6}{7.57}\\
\cellcolor{gray!6}{Homeowner} & \cellcolor{gray!6}{NYC} & \cellcolor{gray!6}{-0.43} & \cellcolor{gray!6}{2.91} & \cellcolor{gray!6}{-5.14} & \cellcolor{gray!6}{-0.38} & \cellcolor{gray!6}{4.46}\\
\cellcolor{gray!6}{Homeowner} & \cellcolor{gray!6}{Phoenix} & \cellcolor{gray!6}{-0.48} & \cellcolor{gray!6}{3.64} & \cellcolor{gray!6}{-6.24} & \cellcolor{gray!6}{-0.42} & \cellcolor{gray!6}{5.40}\\
School * children & Miami & -0.35 & 0.22 & -0.70 & -0.34 & 0.01\\
School * children & NYC & 0.39 & 0.17 & 0.12 & 0.39 & 0.66\\
School * children & Phoenix & 0.24 & 0.11 & 0.07 & 0.24 & 0.42\\
\cellcolor{gray!6}{Children * distance} & \cellcolor{gray!6}{Miami} & \cellcolor{gray!6}{-0.03} & \cellcolor{gray!6}{0.05} & \cellcolor{gray!6}{-0.11} & \cellcolor{gray!6}{-0.03} & \cellcolor{gray!6}{0.05}\\
\cellcolor{gray!6}{Children * distance} & \cellcolor{gray!6}{NYC} & \cellcolor{gray!6}{0.07} & \cellcolor{gray!6}{0.04} & \cellcolor{gray!6}{0.01} & \cellcolor{gray!6}{0.07} & \cellcolor{gray!6}{0.15}\\
\cellcolor{gray!6}{Children * distance} & \cellcolor{gray!6}{Phoenix} & \cellcolor{gray!6}{0.12} & \cellcolor{gray!6}{0.03} & \cellcolor{gray!6}{0.08} & \cellcolor{gray!6}{0.12} & \cellcolor{gray!6}{0.16}\\
Same tract * same road region & Miami & 0.02 & 0.10 & -0.15 & 0.02 & 0.19\\
Same tract * same road region & NYC & -0.24 & 0.13 & -0.45 & -0.23 & -0.02\\
Same tract * same road region & Phoenix & -0.37 & 0.10 & -0.53 & -0.37 & -0.20\\
\cellcolor{gray!6}{Fraction same race * minority} & \cellcolor{gray!6}{Miami} & \cellcolor{gray!6}{0.25} & \cellcolor{gray!6}{0.09} & \cellcolor{gray!6}{0.10} & \cellcolor{gray!6}{0.25} & \cellcolor{gray!6}{0.40}\\
\cellcolor{gray!6}{Fraction same race * minority} & \cellcolor{gray!6}{NYC} & \cellcolor{gray!6}{0.27} & \cellcolor{gray!6}{0.12} & \cellcolor{gray!6}{0.07} & \cellcolor{gray!6}{0.27} & \cellcolor{gray!6}{0.46}\\
\cellcolor{gray!6}{Fraction same race * minority} & \cellcolor{gray!6}{Phoenix} & \cellcolor{gray!6}{-0.31} & \cellcolor{gray!6}{0.22} & \cellcolor{gray!6}{-0.66} & \cellcolor{gray!6}{-0.31} & \cellcolor{gray!6}{0.04}\\
Fraction same party * party = ind & Miami & -0.12 & 0.20 & -0.43 & -0.12 & 0.20\\
Fraction same party * party = ind & NYC & -0.04 & 0.32 & -0.57 & -0.05 & 0.49\\
Fraction same party * party = ind & Phoenix & 0.22 & 0.14 & -0.02 & 0.22 & 0.45\\
\cellcolor{gray!6}{Fraction same party * party = rep} & \cellcolor{gray!6}{Miami} & \cellcolor{gray!6}{-0.31} & \cellcolor{gray!6}{0.14} & \cellcolor{gray!6}{-0.54} & \cellcolor{gray!6}{-0.31} & \cellcolor{gray!6}{-0.07}\\
\cellcolor{gray!6}{Fraction same party * party = rep} & \cellcolor{gray!6}{NYC} & \cellcolor{gray!6}{-0.04} & \cellcolor{gray!6}{0.13} & \cellcolor{gray!6}{-0.25} & \cellcolor{gray!6}{-0.04} & \cellcolor{gray!6}{0.17}\\
\cellcolor{gray!6}{Fraction same party * party = rep} & \cellcolor{gray!6}{Phoenix} & \cellcolor{gray!6}{0.10} & \cellcolor{gray!6}{0.08} & \cellcolor{gray!6}{-0.02} & \cellcolor{gray!6}{0.10} & \cellcolor{gray!6}{0.23}\\
Fraction same ownership * homeowner & Miami & 0.52 & 0.18 & 0.22 & 0.53 & 0.83\\
Fraction same ownership * homeowner & NYC & 0.42 & 0.16 & 0.15 & 0.42 & 0.67\\
Fraction same ownership * homeowner & Phoenix & 0.26 & 0.13 & 0.05 & 0.26 & 0.47\\
\cellcolor{gray!6}{Fraction same education * educ = no college} & \cellcolor{gray!6}{Miami} & \cellcolor{gray!6}{0.44} & \cellcolor{gray!6}{0.25} & \cellcolor{gray!6}{0.05} & \cellcolor{gray!6}{0.44} & \cellcolor{gray!6}{0.85}\\
\cellcolor{gray!6}{Fraction same education * educ = no college} & \cellcolor{gray!6}{NYC} & \cellcolor{gray!6}{1.41} & \cellcolor{gray!6}{0.21} & \cellcolor{gray!6}{1.05} & \cellcolor{gray!6}{1.41} & \cellcolor{gray!6}{1.77}\\
\cellcolor{gray!6}{Fraction same education * educ = no college} & \cellcolor{gray!6}{Phoenix} & \cellcolor{gray!6}{0.57} & \cellcolor{gray!6}{0.21} & \cellcolor{gray!6}{0.24} & \cellcolor{gray!6}{0.57} & \cellcolor{gray!6}{0.91}\\
Income * education = no college & Miami & 0.18 & 0.09 & 0.04 & 0.18 & 0.33\\
Income * education = no college & NYC & -0.10 & 0.09 & -0.24 & -0.10 & 0.04\\
Income * education = no college & Phoenix & 0.05 & 0.08 & -0.07 & 0.05 & 0.18\\
\cellcolor{gray!6}{Alpha} & \cellcolor{gray!6}{Miami} & \cellcolor{gray!6}{1.42} & \cellcolor{gray!6}{0.05} & \cellcolor{gray!6}{1.34} & \cellcolor{gray!6}{1.42} & \cellcolor{gray!6}{1.50}\\
\cellcolor{gray!6}{Alpha} & \cellcolor{gray!6}{NYC} & \cellcolor{gray!6}{1.46} & \cellcolor{gray!6}{0.05} & \cellcolor{gray!6}{1.38} & \cellcolor{gray!6}{1.46} & \cellcolor{gray!6}{1.55}\\
\cellcolor{gray!6}{Alpha} & \cellcolor{gray!6}{Phoenix} & \cellcolor{gray!6}{1.50} & \cellcolor{gray!6}{0.03} & \cellcolor{gray!6}{1.46} & \cellcolor{gray!6}{1.50} & \cellcolor{gray!6}{1.54}\\
\bottomrule
\end{longtable}
\endgroup{}

\begingroup\fontsize{10}{12}\selectfont

\begin{longtable}[t]{llrrrrr}
\caption{Baseline model estimates.}
\label{tab:model_coef_base}\\
\toprule
Coefficient & City & Mean & Std. Dev. & Q5 & Median & Q95\\
\midrule
\cellcolor{gray!6}{(Intercept)} & \cellcolor{gray!6}{Miami} & \cellcolor{gray!6}{-7.72} & \cellcolor{gray!6}{1.12} & \cellcolor{gray!6}{-9.62} & \cellcolor{gray!6}{-7.71} & \cellcolor{gray!6}{-5.84}\\
\cellcolor{gray!6}{(Intercept)} & \cellcolor{gray!6}{NYC} & \cellcolor{gray!6}{-8.05} & \cellcolor{gray!6}{1.12} & \cellcolor{gray!6}{-9.91} & \cellcolor{gray!6}{-8.07} & \cellcolor{gray!6}{-6.27}\\
\cellcolor{gray!6}{(Intercept)} & \cellcolor{gray!6}{Phoenix} & \cellcolor{gray!6}{-8.82} & \cellcolor{gray!6}{0.89} & \cellcolor{gray!6}{-10.30} & \cellcolor{gray!6}{-8.83} & \cellcolor{gray!6}{-7.30}\\
Church & Miami & -0.04 & 0.07 & -0.15 & -0.04 & 0.07\\
Church & NYC & 0.13 & 0.05 & 0.05 & 0.13 & 0.21\\
Church & Phoenix & 0.18 & 0.04 & 0.11 & 0.18 & 0.24\\
\cellcolor{gray!6}{Distance} & \cellcolor{gray!6}{Miami} & \cellcolor{gray!6}{0.00} & \cellcolor{gray!6}{0.02} & \cellcolor{gray!6}{-0.04} & \cellcolor{gray!6}{0.00} & \cellcolor{gray!6}{0.03}\\
\cellcolor{gray!6}{Distance} & \cellcolor{gray!6}{Miami} & \cellcolor{gray!6}{0.11} & \cellcolor{gray!6}{0.02} & \cellcolor{gray!6}{0.07} & \cellcolor{gray!6}{0.11} & \cellcolor{gray!6}{0.15}\\
\cellcolor{gray!6}{Distance} & \cellcolor{gray!6}{NYC} & \cellcolor{gray!6}{0.12} & \cellcolor{gray!6}{0.02} & \cellcolor{gray!6}{0.09} & \cellcolor{gray!6}{0.12} & \cellcolor{gray!6}{0.16}\\
Distance & NYC & 0.05 & 0.02 & 0.01 & 0.05 & 0.08\\
Distance & Phoenix & 0.09 & 0.01 & 0.07 & 0.09 & 0.11\\
Distance & Phoenix & 0.20 & 0.01 & 0.18 & 0.20 & 0.23\\
\cellcolor{gray!6}{Park} & \cellcolor{gray!6}{Miami} & \cellcolor{gray!6}{-0.02} & \cellcolor{gray!6}{0.09} & \cellcolor{gray!6}{-0.17} & \cellcolor{gray!6}{-0.02} & \cellcolor{gray!6}{0.12}\\
\cellcolor{gray!6}{Park} & \cellcolor{gray!6}{NYC} & \cellcolor{gray!6}{0.18} & \cellcolor{gray!6}{0.05} & \cellcolor{gray!6}{0.10} & \cellcolor{gray!6}{0.18} & \cellcolor{gray!6}{0.26}\\
\cellcolor{gray!6}{Park} & \cellcolor{gray!6}{Phoenix} & \cellcolor{gray!6}{0.09} & \cellcolor{gray!6}{0.04} & \cellcolor{gray!6}{0.03} & \cellcolor{gray!6}{0.08} & \cellcolor{gray!6}{0.15}\\
School & Miami & 0.26 & 0.11 & 0.08 & 0.26 & 0.43\\
School & NYC & 0.05 & 0.09 & -0.11 & 0.05 & 0.20\\
School & Phoenix & 0.29 & 0.06 & 0.20 & 0.29 & 0.39\\
\cellcolor{gray!6}{Same block group} & \cellcolor{gray!6}{Miami} & \cellcolor{gray!6}{0.07} & \cellcolor{gray!6}{0.07} & \cellcolor{gray!6}{-0.04} & \cellcolor{gray!6}{0.07} & \cellcolor{gray!6}{0.19}\\
\cellcolor{gray!6}{Same block group} & \cellcolor{gray!6}{NYC} & \cellcolor{gray!6}{0.17} & \cellcolor{gray!6}{0.09} & \cellcolor{gray!6}{0.03} & \cellcolor{gray!6}{0.17} & \cellcolor{gray!6}{0.32}\\
\cellcolor{gray!6}{Same block group} & \cellcolor{gray!6}{Phoenix} & \cellcolor{gray!6}{0.09} & \cellcolor{gray!6}{0.03} & \cellcolor{gray!6}{0.04} & \cellcolor{gray!6}{0.09} & \cellcolor{gray!6}{0.14}\\
Same tract & Miami & -0.16 & 0.08 & -0.29 & -0.16 & -0.01\\
Same tract & NYC & 0.14 & 0.12 & -0.06 & 0.14 & 0.34\\
Same tract & Phoenix & -0.06 & 0.10 & -0.21 & -0.06 & 0.11\\
\cellcolor{gray!6}{Same road region} & \cellcolor{gray!6}{Miami} & \cellcolor{gray!6}{-0.51} & \cellcolor{gray!6}{0.05} & \cellcolor{gray!6}{-0.58} & \cellcolor{gray!6}{-0.51} & \cellcolor{gray!6}{-0.43}\\
\cellcolor{gray!6}{Same road region} & \cellcolor{gray!6}{NYC} & \cellcolor{gray!6}{0.16} & \cellcolor{gray!6}{0.04} & \cellcolor{gray!6}{0.10} & \cellcolor{gray!6}{0.16} & \cellcolor{gray!6}{0.22}\\
\cellcolor{gray!6}{Same road region} & \cellcolor{gray!6}{Phoenix} & \cellcolor{gray!6}{-0.13} & \cellcolor{gray!6}{0.03} & \cellcolor{gray!6}{-0.17} & \cellcolor{gray!6}{-0.13} & \cellcolor{gray!6}{-0.08}\\
Population & Miami & -1.23 & 0.23 & -1.60 & -1.23 & -0.84\\
Population & NYC & -1.74 & 0.18 & -2.03 & -1.74 & -1.46\\
Population & Phoenix & -2.44 & 0.15 & -2.68 & -2.45 & -2.17\\
\cellcolor{gray!6}{Area} & \cellcolor{gray!6}{Miami} & \cellcolor{gray!6}{-0.30} & \cellcolor{gray!6}{0.08} & \cellcolor{gray!6}{-0.42} & \cellcolor{gray!6}{-0.30} & \cellcolor{gray!6}{-0.17}\\
\cellcolor{gray!6}{Area} & \cellcolor{gray!6}{NYC} & \cellcolor{gray!6}{-0.08} & \cellcolor{gray!6}{0.11} & \cellcolor{gray!6}{-0.26} & \cellcolor{gray!6}{-0.08} & \cellcolor{gray!6}{0.10}\\
\cellcolor{gray!6}{Area} & \cellcolor{gray!6}{Phoenix} & \cellcolor{gray!6}{0.02} & \cellcolor{gray!6}{0.04} & \cellcolor{gray!6}{-0.04} & \cellcolor{gray!6}{0.02} & \cellcolor{gray!6}{0.08}\\
Same tract * same road region & Miami & 0.00 & 0.09 & -0.15 & 0.01 & 0.15\\
Same tract * same road region & NYC & -0.32 & 0.13 & -0.54 & -0.32 & -0.11\\
Same tract * same road region & Phoenix & -0.43 & 0.10 & -0.60 & -0.43 & -0.28\\
\cellcolor{gray!6}{Alpha} & \cellcolor{gray!6}{Miami} & \cellcolor{gray!6}{1.42} & \cellcolor{gray!6}{0.05} & \cellcolor{gray!6}{1.34} & \cellcolor{gray!6}{1.42} & \cellcolor{gray!6}{1.49}\\
\cellcolor{gray!6}{Alpha} & \cellcolor{gray!6}{NYC} & \cellcolor{gray!6}{1.38} & \cellcolor{gray!6}{0.05} & \cellcolor{gray!6}{1.31} & \cellcolor{gray!6}{1.38} & \cellcolor{gray!6}{1.47}\\
\cellcolor{gray!6}{Alpha} & \cellcolor{gray!6}{Phoenix} & \cellcolor{gray!6}{1.51} & \cellcolor{gray!6}{0.03} & \cellcolor{gray!6}{1.46} & \cellcolor{gray!6}{1.51} & \cellcolor{gray!6}{1.55}\\
\bottomrule
\end{longtable}
\endgroup{}

\FloatBarrier

\hypertarget{city-council-survey-1}{%
\subsection{City council survey}\label{city-council-survey-1}}

Tables \ref{tab:model_coef_full_nycc} and \ref{tab:model_coef_base_nycc} contain posterior summaries for all model coefficients on the original model scale.
These models were fit using a training sample of 500 survey respondents (out of 627), consisting of 94,349 individual block-level observations.

\begingroup\fontsize{10}{12}\selectfont

\begin{longtable}[t]{lrrrrr}
\caption{Full model estimates.}\label{tab:model_coef_full_nycc}\\
\toprule
Coefficient & Mean & Std. Dev. & Q5 & Median & Q95\\
\midrule
\cellcolor{gray!6}{(Intercept)} & \cellcolor{gray!6}{-6.03} & \cellcolor{gray!6}{3.24} & \cellcolor{gray!6}{-11.11} & \cellcolor{gray!6}{-5.97} & \cellcolor{gray!6}{-0.99}\\
\cellcolor{gray!6}{Church} & \cellcolor{gray!6}{0.10} & \cellcolor{gray!6}{0.02} & \cellcolor{gray!6}{0.07} & \cellcolor{gray!6}{0.10} & \cellcolor{gray!6}{0.12}\\
\cellcolor{gray!6}{Distance} & \cellcolor{gray!6}{0.06} & \cellcolor{gray!6}{0.01} & \cellcolor{gray!6}{0.04} & \cellcolor{gray!6}{0.06} & \cellcolor{gray!6}{0.07}\\
Distance & 0.05 & 0.01 & 0.04 & 0.05 & 0.07\\
Park & 0.19 & 0.02 & 0.17 & 0.19 & 0.22\\
School & 0.06 & 0.04 & 0.00 & 0.06 & 0.12\\
\cellcolor{gray!6}{Children} & \cellcolor{gray!6}{0.03} & \cellcolor{gray!6}{1.71} & \cellcolor{gray!6}{-2.89} & \cellcolor{gray!6}{0.02} & \cellcolor{gray!6}{2.78}\\
\cellcolor{gray!6}{Same block group} & \cellcolor{gray!6}{0.25} & \cellcolor{gray!6}{0.06} & \cellcolor{gray!6}{0.14} & \cellcolor{gray!6}{0.25} & \cellcolor{gray!6}{0.35}\\
\cellcolor{gray!6}{Same tract} & \cellcolor{gray!6}{0.19} & \cellcolor{gray!6}{0.09} & \cellcolor{gray!6}{0.04} & \cellcolor{gray!6}{0.19} & \cellcolor{gray!6}{0.33}\\
Same road region & -0.22 & 0.02 & -0.25 & -0.22 & -0.19\\
Population & -1.18 & 0.07 & -1.30 & -1.18 & -1.07\\
Area & -0.52 & 0.12 & -0.71 & -0.52 & -0.32\\
\cellcolor{gray!6}{Fraction same race} & \cellcolor{gray!6}{-0.32} & \cellcolor{gray!6}{0.03} & \cellcolor{gray!6}{-0.37} & \cellcolor{gray!6}{-0.32} & \cellcolor{gray!6}{-0.28}\\
\cellcolor{gray!6}{Fraction same party} & \cellcolor{gray!6}{-0.22} & \cellcolor{gray!6}{0.03} & \cellcolor{gray!6}{-0.26} & \cellcolor{gray!6}{-0.22} & \cellcolor{gray!6}{-0.18}\\
\cellcolor{gray!6}{Fraction same ownership} & \cellcolor{gray!6}{-0.11} & \cellcolor{gray!6}{0.04} & \cellcolor{gray!6}{-0.18} & \cellcolor{gray!6}{-0.11} & \cellcolor{gray!6}{-0.03}\\
Fraction same education & -0.27 & 0.04 & -0.34 & -0.27 & -0.20\\
Income & -0.11 & 0.02 & -0.14 & -0.11 & -0.08\\
Age & 0.01 & 0.06 & -0.09 & 0.00 & 0.10\\
\cellcolor{gray!6}{Education = no college} & \cellcolor{gray!6}{-3.58} & \cellcolor{gray!6}{1.96} & \cellcolor{gray!6}{-6.75} & \cellcolor{gray!6}{-3.60} & \cellcolor{gray!6}{-0.30}\\
\cellcolor{gray!6}{Retired} & \cellcolor{gray!6}{-0.14} & \cellcolor{gray!6}{2.25} & \cellcolor{gray!6}{-3.66} & \cellcolor{gray!6}{-0.20} & \cellcolor{gray!6}{3.63}\\
\cellcolor{gray!6}{Tenure} & \cellcolor{gray!6}{0.02} & \cellcolor{gray!6}{0.54} & \cellcolor{gray!6}{-0.85} & \cellcolor{gray!6}{0.03} & \cellcolor{gray!6}{0.87}\\
Party = ind & -0.23 & 2.75 & -4.73 & -0.21 & 4.26\\
Party = rep & 0.15 & 2.05 & -3.30 & 0.14 & 3.50\\
Minority & 0.10 & 1.66 & -2.61 & 0.08 & 2.79\\
\cellcolor{gray!6}{Homeownerother (please specify)} & \cellcolor{gray!6}{0.08} & \cellcolor{gray!6}{3.35} & \cellcolor{gray!6}{-5.40} & \cellcolor{gray!6}{0.13} & \cellcolor{gray!6}{5.49}\\
\cellcolor{gray!6}{Homeownerrenter} & \cellcolor{gray!6}{0.17} & \cellcolor{gray!6}{1.64} & \cellcolor{gray!6}{-2.56} & \cellcolor{gray!6}{0.14} & \cellcolor{gray!6}{2.76}\\
\cellcolor{gray!6}{School * children} & \cellcolor{gray!6}{-0.20} & \cellcolor{gray!6}{0.07} & \cellcolor{gray!6}{-0.32} & \cellcolor{gray!6}{-0.20} & \cellcolor{gray!6}{-0.08}\\
Children * distance & -0.02 & 0.02 & -0.06 & -0.02 & 0.01\\
Same tract * same road region & -0.40 & 0.09 & -0.55 & -0.40 & -0.25\\
Fraction same race * minority & 0.16 & 0.06 & 0.06 & 0.15 & 0.25\\
\cellcolor{gray!6}{Fraction same party * party = ind} & \cellcolor{gray!6}{0.12} & \cellcolor{gray!6}{0.13} & \cellcolor{gray!6}{-0.09} & \cellcolor{gray!6}{0.12} & \cellcolor{gray!6}{0.33}\\
\cellcolor{gray!6}{Fraction same party * party = rep} & \cellcolor{gray!6}{-0.53} & \cellcolor{gray!6}{0.15} & \cellcolor{gray!6}{-0.77} & \cellcolor{gray!6}{-0.53} & \cellcolor{gray!6}{-0.27}\\
\cellcolor{gray!6}{Fraction same ownership * homeownerother (please specify)} & \cellcolor{gray!6}{0.00} & \cellcolor{gray!6}{0.13} & \cellcolor{gray!6}{-0.22} & \cellcolor{gray!6}{0.01} & \cellcolor{gray!6}{0.22}\\
Fraction same ownership * homeownerrenter & 0.14 & 0.06 & 0.03 & 0.14 & 0.24\\
Fraction same education * educ = no college & 0.79 & 0.11 & 0.62 & 0.79 & 0.97\\
Income * education = no college & 0.27 & 0.04 & 0.21 & 0.27 & 0.34\\
\cellcolor{gray!6}{Alpha} & \cellcolor{gray!6}{1.39} & \cellcolor{gray!6}{0.02} & \cellcolor{gray!6}{1.35} & \cellcolor{gray!6}{1.39} & \cellcolor{gray!6}{1.42}\\
\bottomrule
\end{longtable}
\endgroup{}

\begingroup\fontsize{10}{12}\selectfont

\begin{longtable}[t]{lrrrrr}
\caption{Baseline model estimates.}\label{tab:model_coef_base_nycc}\\
\toprule
Coefficient & Mean & Std. Dev. & Q5 & Median & Q95\\
\midrule
\cellcolor{gray!6}{(Intercept)} & \cellcolor{gray!6}{-7.13} & \cellcolor{gray!6}{1.48} & \cellcolor{gray!6}{-9.57} & \cellcolor{gray!6}{-7.11} & \cellcolor{gray!6}{-4.64}\\
\cellcolor{gray!6}{Church} & \cellcolor{gray!6}{0.09} & \cellcolor{gray!6}{0.02} & \cellcolor{gray!6}{0.07} & \cellcolor{gray!6}{0.09} & \cellcolor{gray!6}{0.12}\\
\cellcolor{gray!6}{Distance} & \cellcolor{gray!6}{0.07} & \cellcolor{gray!6}{0.01} & \cellcolor{gray!6}{0.05} & \cellcolor{gray!6}{0.07} & \cellcolor{gray!6}{0.08}\\
Distance & 0.04 & 0.01 & 0.03 & 0.04 & 0.06\\
Park & 0.20 & 0.02 & 0.17 & 0.20 & 0.22\\
School & 0.00 & 0.03 & -0.05 & -0.01 & 0.05\\
\cellcolor{gray!6}{Same block group} & \cellcolor{gray!6}{0.22} & \cellcolor{gray!6}{0.06} & \cellcolor{gray!6}{0.12} & \cellcolor{gray!6}{0.22} & \cellcolor{gray!6}{0.32}\\
\cellcolor{gray!6}{Same tract} & \cellcolor{gray!6}{0.23} & \cellcolor{gray!6}{0.09} & \cellcolor{gray!6}{0.09} & \cellcolor{gray!6}{0.23} & \cellcolor{gray!6}{0.38}\\
\cellcolor{gray!6}{Same road region} & \cellcolor{gray!6}{-0.27} & \cellcolor{gray!6}{0.02} & \cellcolor{gray!6}{-0.29} & \cellcolor{gray!6}{-0.27} & \cellcolor{gray!6}{-0.23}\\
Population & -1.79 & 0.07 & -1.90 & -1.79 & -1.68\\
Area & -0.24 & 0.10 & -0.41 & -0.24 & -0.07\\
Same tract * same road region & -0.45 & 0.09 & -0.60 & -0.45 & -0.30\\
\cellcolor{gray!6}{Alpha} & \cellcolor{gray!6}{1.39} & \cellcolor{gray!6}{0.02} & \cellcolor{gray!6}{1.36} & \cellcolor{gray!6}{1.39} & \cellcolor{gray!6}{1.43}\\
\bottomrule
\end{longtable}
\endgroup{}

\hypertarget{app:pred-agg}{%
\section{Aggregate-level prediction}\label{app:pred-agg}}

\begin{figure}[htb]

{\centering \includegraphics[width=0.85\linewidth]{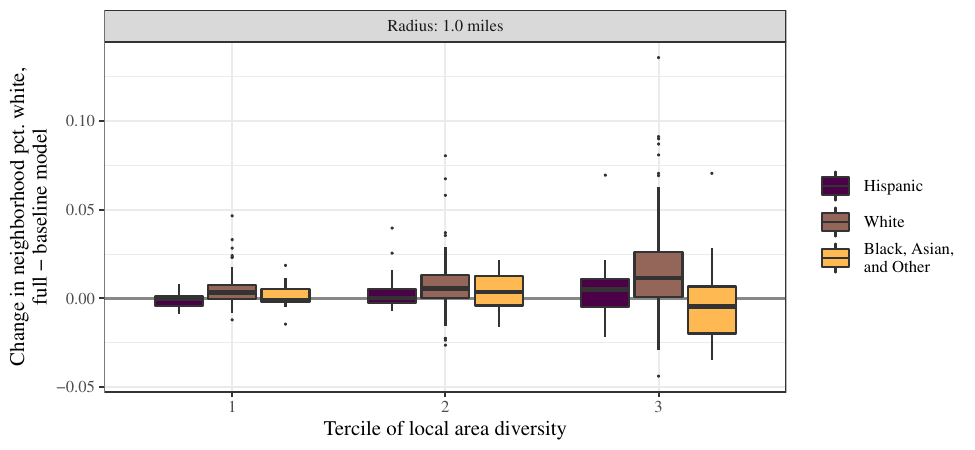} 

}

\caption[Change in the posterior predictive share of respondents'
neighborhood that is white between the full and baseline models]{For each respondent, we compare the posterior mean of the fraction white
in their predicted neighborhoods between the full and baseline models.
Positive values indicate that using demographic information (full model)
leads to a predicted neighborhood that is more white, on average.
These differences in neighborhood fraction white are broken out by
tercile of local area diversity and by the race of the respondent.
Section S5 of the SI contains the full results tables for the full and baseline models.
}\label{fig:agg-race}
\end{figure}

Here we demonstrate how, in aggregate, the modeled relationship between co-racial and co-partisan demographics and census block inclusion produces predicted neighborhoods with different racial and partisan makeups across voters of different races and parties.
Figure \ref{fig:agg-race} presents boxplots showing the median and interquartile ranges of the change in proportion White in predicted neighborhoods, comparing the baseline model to the full model.
The full model considers demographic information, while the baseline model does not, so the difference between the two is evidence of how much more homogeneous subjective neighborhoods become when demographics are considered.

\begin{figure}[p!]
\includegraphics[width=\textwidth]{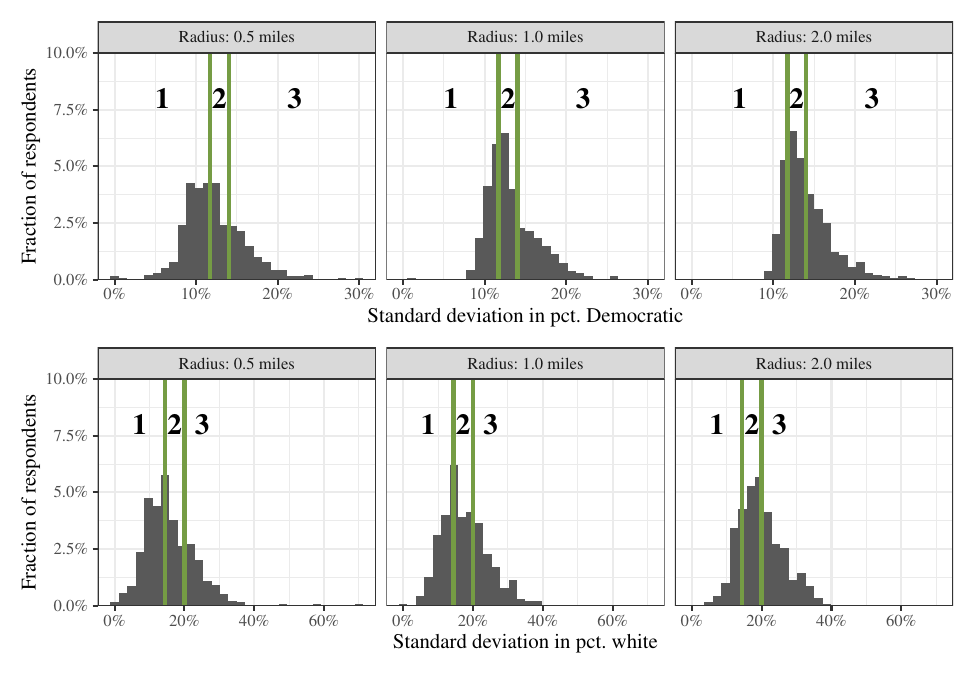}
\caption{ Local partisan and racial diversity for each respondent, as measured by the
block-level standard deviation of the respective variables for blocks within a 0.5, 1.0, and
2.0-mile radius. Terciles are indicated by vertical lines and bold labels.}
\label{fig:local-div}
\end{figure}

This comparison is plotted separately by tercile of local racial diversity, to illustrate that, when voters live in areas where it is plausible to include or exclude out-group neighbors, they tend to do so.
We measure local racial diversity as the standard deviation in the White percentage of each block within a fixed radius of a voter's residence.
Since respondents must include or exclude whole blocks, measuring diversity according to block-level statistics is appropriate.
Higher standard deviations indicate more block-level variation in racial composition and thus more opportunity to exclude out-group neighbors.
Figure \ref{fig:local-div} shows the variation in local diversity across respondents.

Three quarters of White respondents' predicted neighborhoods contain greater proportions of White residents under the full model compared to the baseline model, with the disparity increasing as neighborhood diversity increases.
We further see evidence, in the most mixed neighborhoods, of predicted neighborhoods for Black and Asian respondents with lower numbers of White residents, further evidence of the impact of demographics on subjective racial neighborhoods.
Figure \ref{fig:agg-race} shows the results for a one-mile radius, but
the results are robust to different specifications.

\begin{figure}[htb]

{\centering \includegraphics[width=0.85\linewidth]{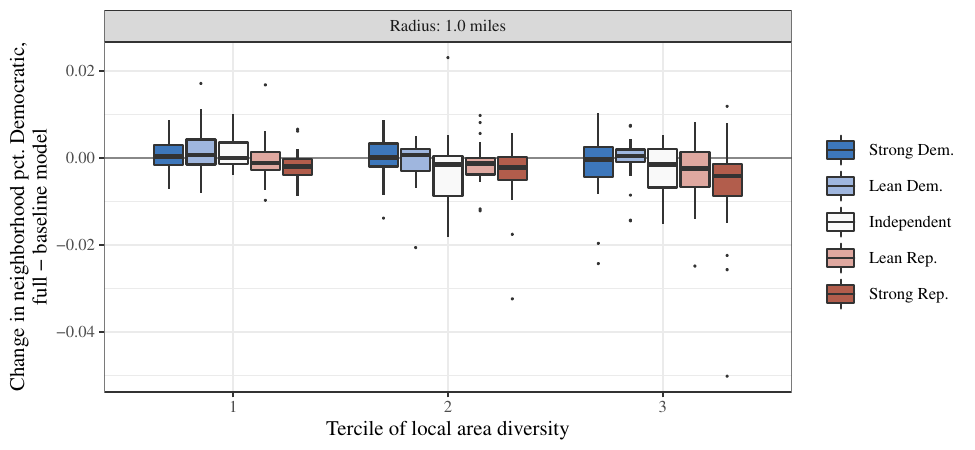} 

}

\caption[Change in the posterior predictive share of respondents'
neighborhood that is Democratic between the full and baseline models]{For each respondent, we compare the posterior mean of the fraction of
Democrats in their predicted neighborhoods between the full and baseline models.
Positive values indicate that using demographic information (full model)
leads to a predicted neighborhood that is more Democratic, on average.
These differences in neighborhood fraction Democratic are broken
out by tercile of local area diversity and by the party identification
of the respondent.
Section S5 of the SI contains the full results tables for the full and baseline models.
}\label{fig:agg-party}
\end{figure}

In Figure \ref{fig:agg-party}, we plot the same comparison for partisan demographics, showing the boxplots of the difference in proportion Democratic between the full and baseline models across terciles of partisan diversity.
Boxplots are shown separately by the self-reported partisan identification of the voter, ranging from strong Republican to strong Democrat.

On average, predicted neighborhoods for Democrats are slightly more Democratic in the full model compared to the baseline model, although the interquartile ranges overlap with zero across diversity terciles.
Predicted neighborhoods for Republicans, on the other hand, are noticeably less Democratic in the full model compared to the baseline model, and the largest disparity is seen for strong Republicans, where the gap reaches 0.37 percentage points in the most politically diverse areas.
Similar to the racial comparison, the degree of difference is increasing with partisan diversity, and thus the potential to draw neighborhoods more differentiated by partisanship.

For both race and partisanship, the changes in neighborhood composition as a result of factoring in demographics are small in magnitude.
This reflects, we believe, the overwhelming influence of residential segregation and sorting.
Voters' preferences for homogeneous neighborhoods are already reflected in their choice of residence; all that we measure here is the marginal predictive effect of this preference on their subjective definition of neighborhood, given that residence.

\fontsize{10}{10.5}\selectfont

\hypertarget{predictive-performance-compared-to-zctas}{%
\section{Predictive performance compared to ZCTAs}\label{predictive-performance-compared-to-zctas}}

\begin{figure}[htb]

{\centering \includegraphics[width=0.75\linewidth]{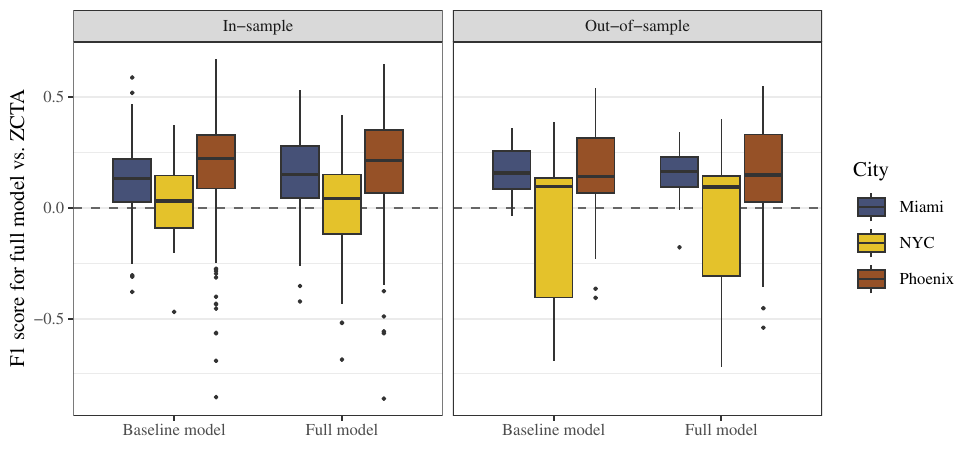} 

}

\caption{Posterior median of the difference in F1 scores between a neighborhood
 predicted by the model prediction and a census ZCTA.
 The boxplot shows the variation in this median difference accros the
 respondents included in the model fitting (left plot) and excluded from
 the model fitting (right plot).
Section S5 of the SI contains the full results tables for the full and baseline models.}\label{fig:gof-zcta}
\end{figure}

\FloatBarrier

\hypertarget{additional-model-fits}{%
\section{Additional model fits}\label{additional-model-fits}}

\hypertarget{model-fit-with-urban-suburban-indicator}{%
\subsection{Model fit with urban-suburban indicator}\label{model-fit-with-urban-suburban-indicator}}

We fit the full model specification again, but include an indicator for whether
a block belongs to the primary city in each metro area.
This indicator is also interacted with the same-tract indicator, and the
fraction-same-race variable.

\begin{figure}[htb]

{\centering \includegraphics[width=0.72\linewidth]{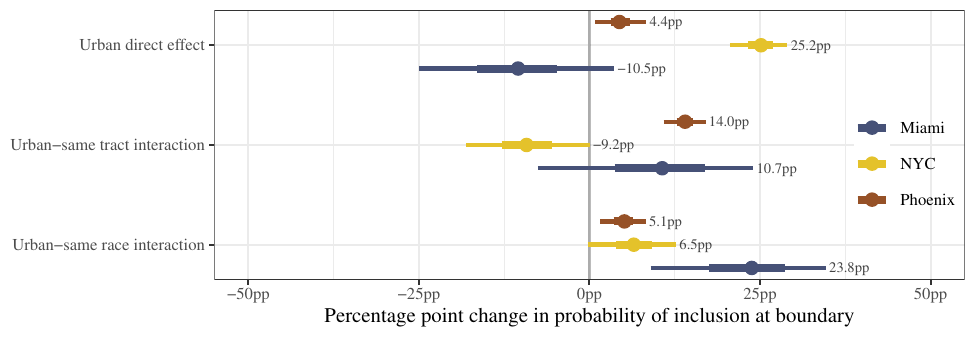} 

}

\caption[Full model estimates for urban indicator variable and interactions.]{Full model estimates for urban indicator variable and interactions.
The full coefficient estimates for this model are available in CSV format
as part of the APSR Dataverse files.
}\label{fig:full-coef-urban}
\end{figure}

These additional model coefficients are summarized in Figure \ref{fig:full-coef-urban}.
Results are mixed for the direct effect (which affects the size of the neighborhood)
and same-tract interaction.
In all 3 cities, the same-race preference is slightly stronger in the urban area
compared to suburban areas.
The main same-race effect (corresponding to the coefficient for suburban voters) remains positive for both white and minority voters.
Thus while there is some evidence of urban-suburban heterogeneity, directionally the results are consistent with the overall findings.

\hypertarget{model-fit-with-home-prices}{%
\subsection{Model fit with home prices}\label{model-fit-with-home-prices}}

We fit the full model specification again, but include the (log of the) median home value in each block group.
The home value covariate is also interacted with the same-race and same-party variables,
and with the respondent's educational group (the indicator for not having attended college).

\begin{figure}[htb]

{\centering \includegraphics[width=0.75\linewidth]{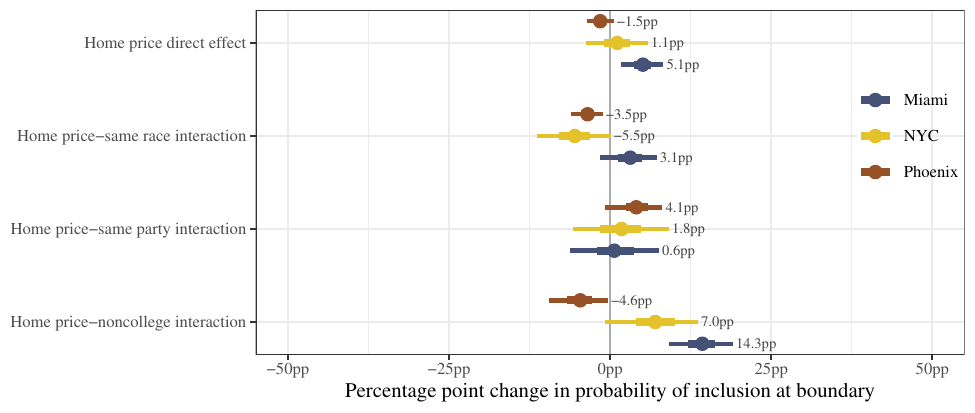} 

}

\caption[Full model estimates for urban indicator variable and interactions.]{Full model estimates with home prices control.
The full coefficient estimates for this model are available in CSV format
as part of the APSR Dataverse files.
}\label{fig:full-coef-homeprc}
\end{figure}

These additional model coefficients are summarized in Figure \ref{fig:full-coef-homeprc}.
Both the direct effect and interaction terms are relatively small.
In New York and Miami, there is some evidence that non-college respondents are more likely to include a region if it has higher home prices.
Overall the main conclusions about party and race in-group preference remain unchanged.

\FloatBarrier

\hypertarget{model-fit-with-community-centers-and-other-cultural-institutions}{%
\subsection{Model fit with community centers and other cultural institutions}\label{model-fit-with-community-centers-and-other-cultural-institutions}}

\begin{figure}[htb]

{\centering \includegraphics[width=0.75\linewidth]{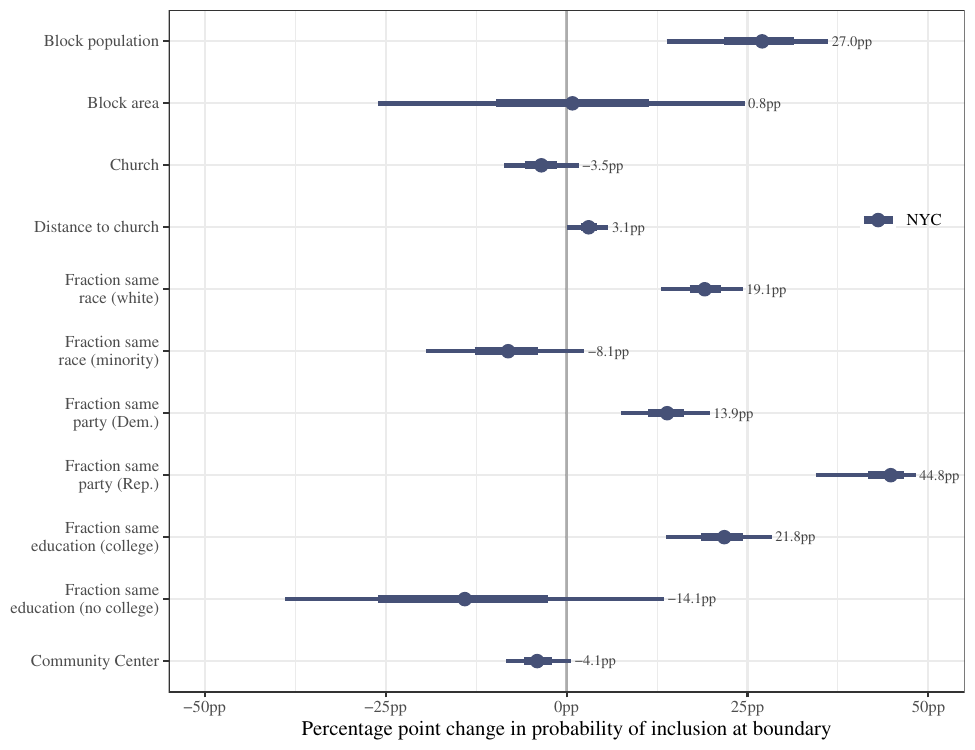} 

}

\caption[Full model estimates for community center indicator variable.]{Full model estimates for community center indicator variable.
The full coefficient estimates for this model are available in CSV format
as part of the APSR Dataverse files.
}\label{fig:full-coef-cc}
\end{figure}

We fit the full model specification again, this time only on the sample from New York City, where we acquired data on other types of cultural institutions besides churches and schools. These include: youth centers, senior centers, libraries, community centers, and other cultural institutions. We include an indicator for the presence of any of these buildings in a census block. We estimate the full model specification with this indicator as an additional covariate. Figure \ref{fig:full-coef-cc} reports the results, demonstrating the robustness of the original variables to the inclusion of this covariates. We further see that community centers and other cultural institutions exert a similar effect as churches on inclusion in subjective neighborhoods, with the present of any of these making it less likely a place is included.

\FloatBarrier

\hypertarget{model-fit-with-turnout-and-party-interactions}{%
\subsection{Model fit with turnout and party interactions}\label{model-fit-with-turnout-and-party-interactions}}

\begin{figure}[htb]

{\centering \includegraphics[width=0.75\linewidth]{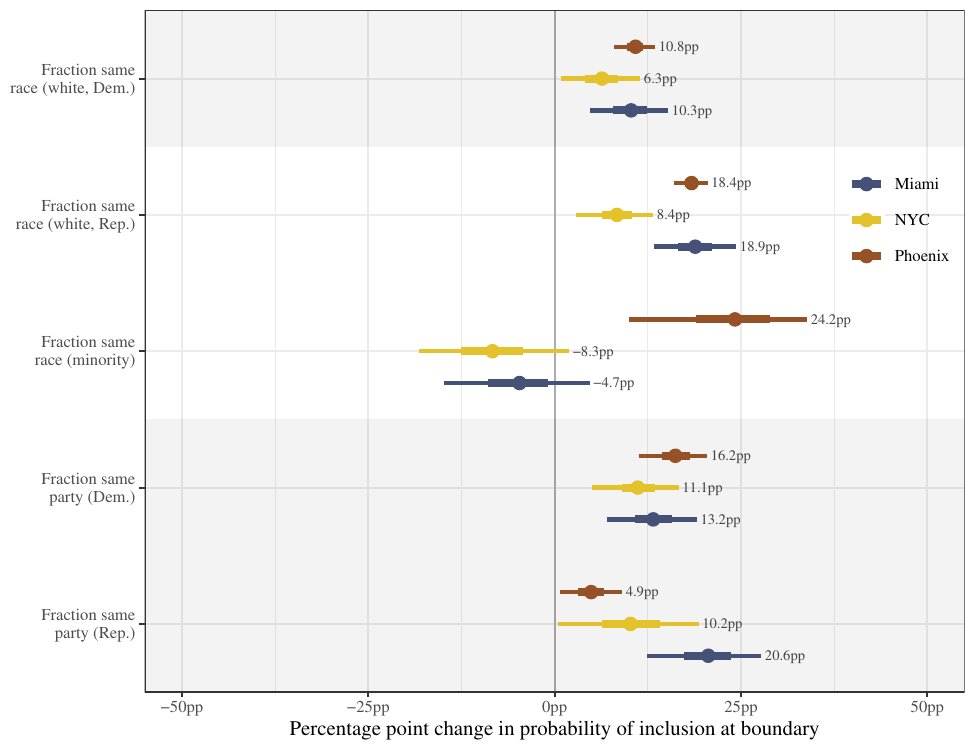} 

}

\caption{Subjective neighborhood survey: Selected full model coefficient posteriors, scaled to show the percentage point change in probability of a block's inclusion for a baseline probability of 50\%. Plotted are 90\% and 50\% credible intervals, with posterior medians displayed to the right of each interval. The full coefficient estimates for this model are available in CSV format as part of the APSR Dataverse files.}\label{fig:party-coef}
\end{figure}

\begin{figure}[htb]

{\centering \includegraphics[width=0.75\linewidth]{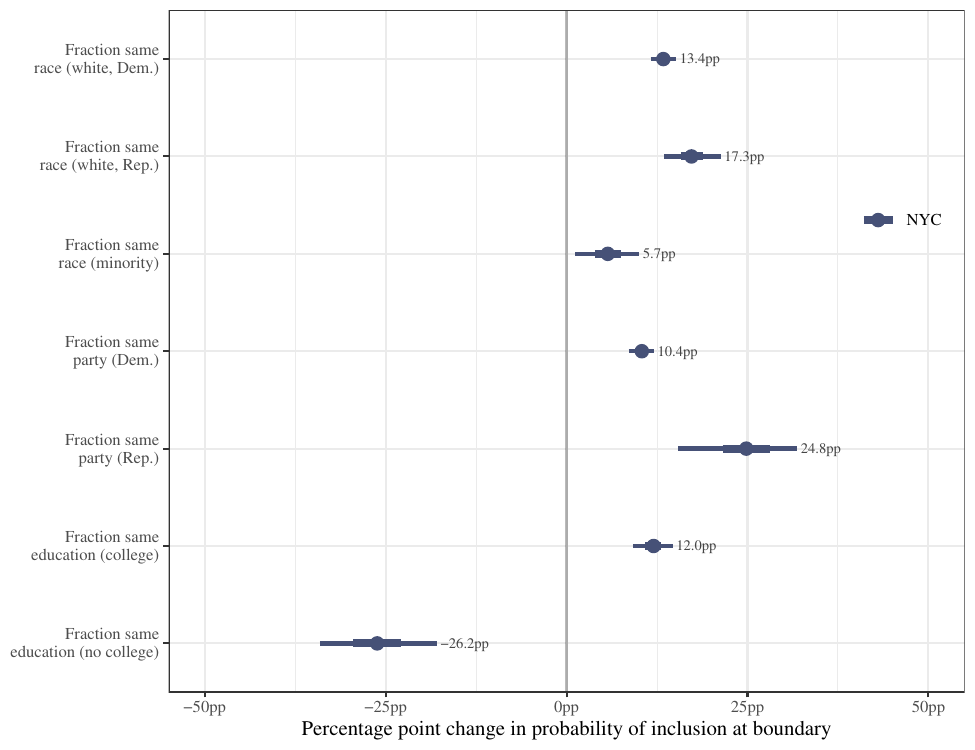} 

}

\caption{City council survey: Selected full model coefficient posteriors, scaled to show the percentage point change in probability of a block's inclusion for a baseline probability of 50\%. Plotted are 90\% and 50\% credible intervals, with posterior medians displayed to the right of each interval. The full coefficient estimates for this model are available in CSV format as part of the APSR Dataverse files.}\label{fig:party-coef-nycc}
\end{figure}

We fit the full model specification again, but include interactions between party and fraction same race in each census block. We do this for the both surveys, the subjective neighborhoods survey and the city council survey. Figure \ref{fig:party-coef} and \ref{fig:party-coef-nycc} report the same race coefficients for Democrats and Republicans. We find that, among whites, that while respondents of both parties prefer census blocks with more same race residents, this preference is stronger for Republicans than for Democrats.


\begin{figure}[htb]

{\centering \includegraphics[width=0.75\linewidth]{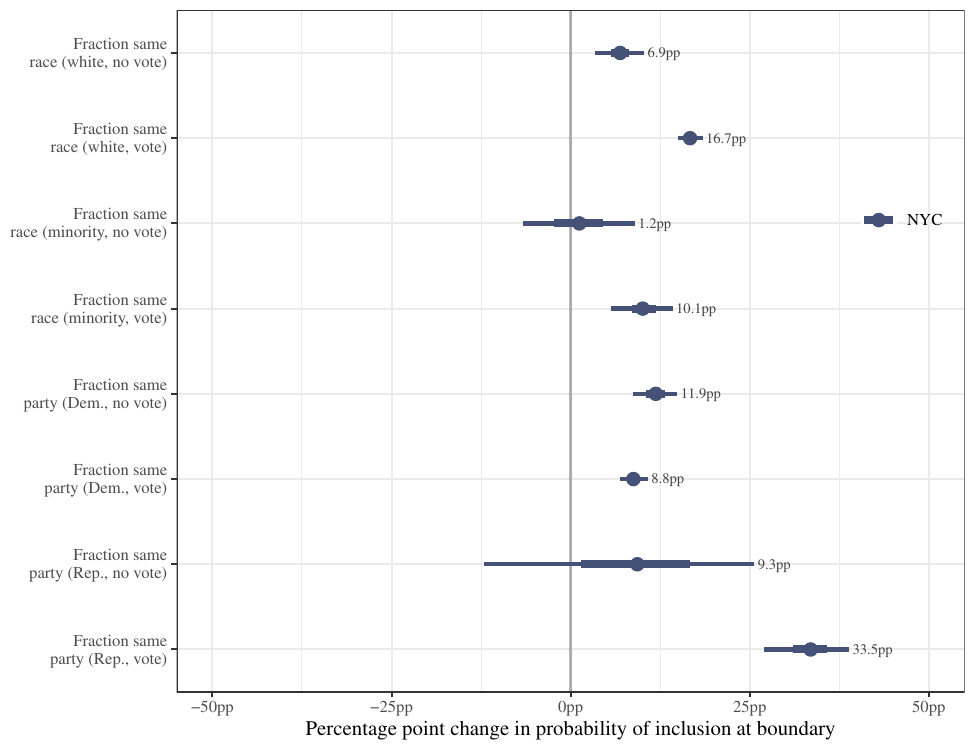} 

}

\caption{City council survey: Selected full model coefficient posteriors, scaled to show the percentage point change in probability of a block's inclusion for a baseline probability of 50\%. Plotted are 90\% and 50\% credible intervals, with posterior medians displayed to the right of each interval. The full coefficient estimates for this model are available in CSV format as part of the APSR Dataverse files.}\label{fig:vote-coef}
\end{figure}

We also fit the full model to the city council survey including an interaction for whether or not the respondent reported that they voted in the 2021 mayoral election. We do not fit the equivalent model to the subjective neighborhood survey because self-reported 2020 turnout was too high (\textgreater95\%) in the sample to estimate this interaction in every city. Figure \ref{fig:vote-coef} reports the results, showing that voters who voted in the 2021 mayoral election also gave greater preference to racial similarity when drawing their communities of interest. We see a similar disparity for voting and preferences for partisan homophily among Republicans but not for Democrats, where Democrats who do not vote are slightly more influenced by party demographics when defining their communities of interest.

\FloatBarrier

\hypertarget{model-fit-on-email-versus-meta-survey}{%
\subsection{Model fit on Email versus Meta survey}\label{model-fit-on-email-versus-meta-survey}}

\begin{figure}[htb]

{\centering \includegraphics[width=0.95\linewidth]{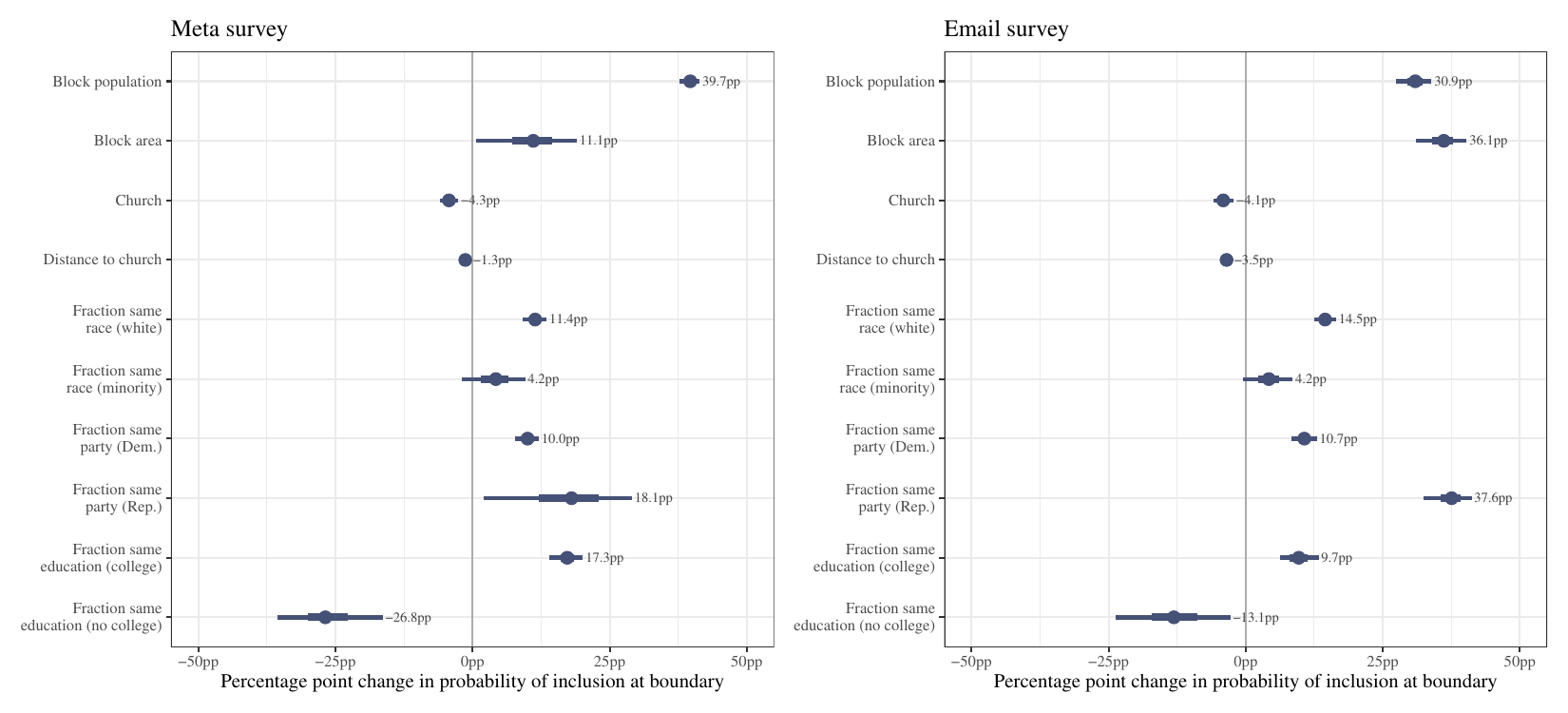} 

}

\caption{City council survey: Selected full model coefficient posteriors, scaled to show the percentage point change in probability of a block's inclusion for a baseline probability of 50\%. Plotted are 90\% and 50\% credible intervals, with posterior medians displayed to the right of each interval. The full coefficient estimates for this model are available in CSV format as part of the APSR Dataverse files.}\label{fig:survey-mode-coef}
\end{figure}

\FloatBarrier

\hypertarget{survey-representativeness}{%
\section{Survey representativeness}\label{survey-representativeness}}

\begin{figure}[htb]

{\centering \includegraphics[width=0.75\linewidth]{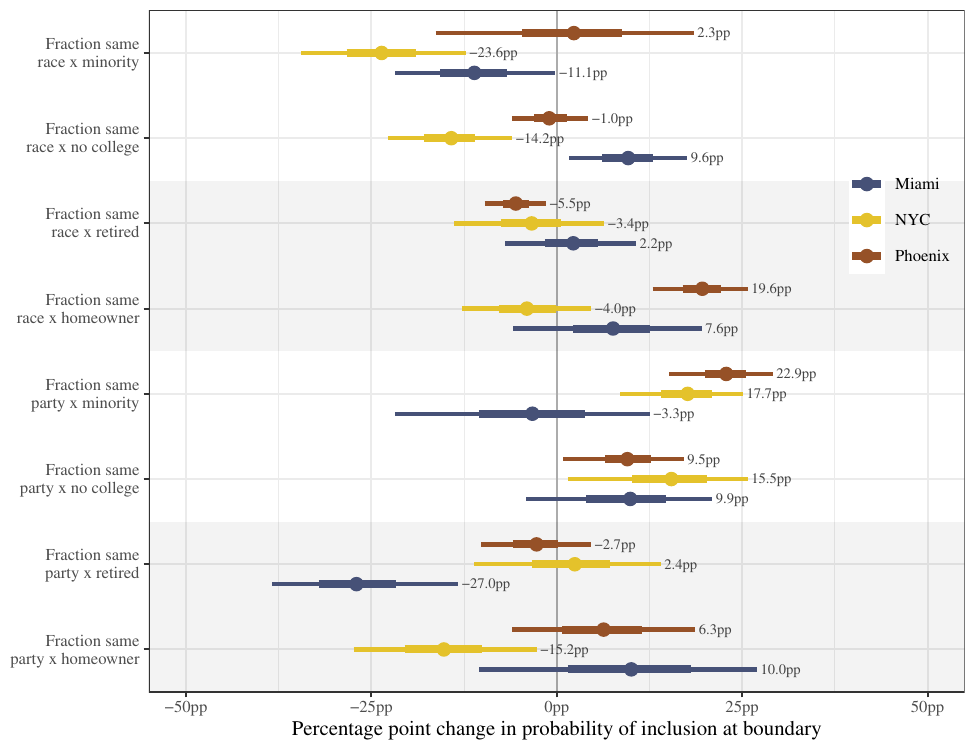} 

}

\caption[Full model estimates for interactions with race, homeownership, retirement, and college.]{Full model estimates for interactions with race, homeownership, retirement, and college.
The full coefficient estimates for this model are available in CSV format
as part of the APSR Dataverse files.
}\label{fig:full-coef-int}
\end{figure}

To assess the sensitivity of our main effects to any unrepresentativeness of the survey, we fit the full model specification, this time adding interactions between race and party homophily variables and respondent race, homeowner status, retirement status, and college education. Figure \ref{fig:full-coef-int} summarizes these interaction coefficients. While the effects can vary by these variables, in almost every case we still observed an overall positive effect of partisan and racial homophily on inclusion (i.e.~even negative interaction coefficients are not large enough to switch the sign of the overall effect). This is remarkable given the relatively small sample sizes for some demographics, and the high amount of individual heterogeneity observed for other aspects of the drawn neighborhoods.

To assess the sensitivity of our main effects to any unrepresentativeness of the survey, we fit the full model specification, this time adding interactions between race and party homophily variables and respondent race, homeowner status, retirement status, and college education. Figure \ref{fig:full-coef-int} summarizes these interaction coefficients. While the effects can vary by these variables, in almost every case we still observed an overall positive effect of partisan and racial homophily on inclusion (i.e.~even negative interaction coefficients are not large enough to switch the sign of the overall effect). This is remarkable given the relatively small sample sizes for some demographics, and the high amount of individual heterogeneity observed for other aspects of the drawn neighborhoods.

\FloatBarrier

\hypertarget{time-spent-drawing-maps}{%
\section{Time spent drawing maps}\label{time-spent-drawing-maps}}

\hypertarget{first-survey-2}{%
\subsection{First survey}\label{first-survey-2}}

\begin{figure}
\includegraphics[width=\textwidth]{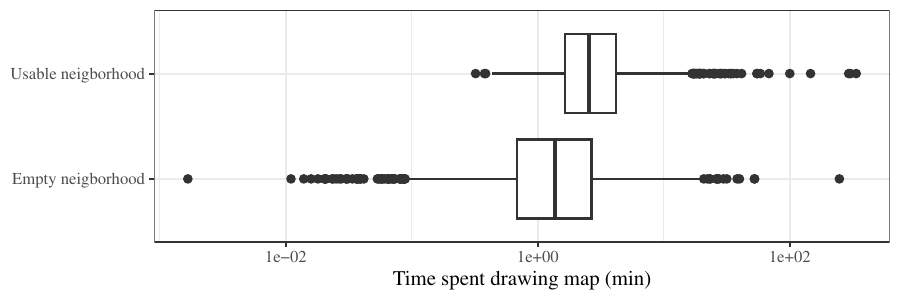}
\caption[Boxplot of time spent drawing map (in minutes)]{
From the subjective neighborhoods survey. Distributions plotted separately for valid and empty neighborhoods.}
\label{fig:map-time}
\end{figure}

Figure \ref{fig:map-time} shows that respondents which drew valid neighborhoods spent more time on average with the drawing tool, as would be expected. The results show that respondents who drew usable neighborhoods on average spent 4.17 minutes on the map (median 2.54 minutes). The subsequent table reports associational measures between various covariates and the time spent drawing the map, which is measured in seconds, binned into three categories, and reported in the table columns. Time spent on the map is not well correlated with individual characteristics or neighborhood features.

\begin{longtable}[]{@{}
  >{\raggedright\arraybackslash}p{(\columnwidth - 8\tabcolsep) * \real{0.3261}}
  >{\raggedright\arraybackslash}p{(\columnwidth - 8\tabcolsep) * \real{0.1848}}
  >{\raggedright\arraybackslash}p{(\columnwidth - 8\tabcolsep) * \real{0.1848}}
  >{\raggedright\arraybackslash}p{(\columnwidth - 8\tabcolsep) * \real{0.1848}}
  >{\raggedright\arraybackslash}p{(\columnwidth - 8\tabcolsep) * \real{0.1196}}@{}}
\caption{Descriptive statistics by total time spent drawing neighborhood (minutes)}\tabularnewline
\toprule\noalign{}
\begin{minipage}[b]{\linewidth}\raggedright
\end{minipage} & \begin{minipage}[b]{\linewidth}\raggedright
(0,1{]} (N=1328)
\end{minipage} & \begin{minipage}[b]{\linewidth}\raggedright
(1,2{]} (N=1556)
\end{minipage} & \begin{minipage}[b]{\linewidth}\raggedright
(2,Inf{]} (N=2732)
\end{minipage} & \begin{minipage}[b]{\linewidth}\raggedright
p value
\end{minipage} \\
\midrule\noalign{}
\endfirsthead
\toprule\noalign{}
\begin{minipage}[b]{\linewidth}\raggedright
\end{minipage} & \begin{minipage}[b]{\linewidth}\raggedright
(0,1{]} (N=1328)
\end{minipage} & \begin{minipage}[b]{\linewidth}\raggedright
(1,2{]} (N=1556)
\end{minipage} & \begin{minipage}[b]{\linewidth}\raggedright
(2,Inf{]} (N=2732)
\end{minipage} & \begin{minipage}[b]{\linewidth}\raggedright
p value
\end{minipage} \\
\midrule\noalign{}
\endhead
\bottomrule\noalign{}
\endlastfoot
\textbf{usable} & & & & \textless{} 0.001\textsuperscript{1} \\
~~~FALSE & 1133 (85.3\%) & 833 (53.5\%) & 1068 (39.1\%) & \\
~~~TRUE & 195 (14.7\%) & 723 (46.5\%) & 1664 (60.9\%) & \\
\textbf{group} & & & & 0.004\textsuperscript{1} \\
~~~C & 293 (22.1\%) & 308 (19.8\%) & 518 (19.0\%) & \\
~~~P & 237 (17.8\%) & 307 (19.7\%) & 611 (22.4\%) & \\
~~~PH & 287 (21.6\%) & 277 (17.8\%) & 490 (17.9\%) & \\
~~~R & 240 (18.1\%) & 316 (20.3\%) & 558 (20.4\%) & \\
~~~RH & 271 (20.4\%) & 348 (22.4\%) & 555 (20.3\%) & \\
\textbf{city} & & & & 0.052\textsuperscript{1} \\
~~~miami & 322 (24.2\%) & 342 (22.0\%) & 549 (20.1\%) & \\
~~~new-york & 241 (18.1\%) & 266 (17.1\%) & 466 (17.1\%) & \\
~~~phoenix & 765 (57.6\%) & 948 (60.9\%) & 1717 (62.8\%) & \\
\textbf{party} & & & & \textless{} 0.001\textsuperscript{1} \\
~~~dem\_strong & 298 (22.4\%) & 423 (27.2\%) & 763 (27.9\%) & \\
~~~dem\_lean & 194 (14.6\%) & 308 (19.8\%) & 561 (20.5\%) & \\
~~~independent & 153 (11.5\%) & 162 (10.4\%) & 209 (7.7\%) & \\
~~~rep\_lean & 288 (21.7\%) & 305 (19.6\%) & 547 (20.0\%) & \\
~~~rep\_strong & 395 (29.7\%) & 358 (23.0\%) & 652 (23.9\%) & \\
\textbf{age} & & & & \textless{} 0.001\textsuperscript{2} \\
~~~Mean (SD) & 59.950 (13.861) & 62.715 (13.057) & 64.484 (13.028) & \\
~~~Range & 18.000 - 105.000 & 21.000 - 105.000 & 18.000 - 105.000 & \\
\textbf{gender} & & & & 0.178\textsuperscript{1} \\
~~~N-Miss & 28 & 16 & 14 & \\
~~~female & 611 (47.0\%) & 674 (43.8\%) & 1255 (46.2\%) & \\
~~~male & 689 (53.0\%) & 866 (56.2\%) & 1463 (53.8\%) & \\
\textbf{education} & & & & \textless{} 0.001\textsuperscript{1} \\
~~~N-Miss & 34 & 23 & 34 & \\
~~~no\_hs & 11 (0.9\%) & 1 (0.1\%) & 2 (0.1\%) & \\
~~~some\_hs & 5 (0.4\%) & 16 (1.0\%) & 16 (0.6\%) & \\
~~~hs & 85 (6.6\%) & 109 (7.1\%) & 183 (6.8\%) & \\
~~~some\_coll & 258 (19.9\%) & 240 (15.7\%) & 478 (17.7\%) & \\
~~~grad\_2yr & 130 (10.0\%) & 164 (10.7\%) & 269 (10.0\%) & \\
~~~grad\_4yr & 427 (33.0\%) & 512 (33.4\%) & 892 (33.1\%) & \\
~~~postgrad & 378 (29.2\%) & 491 (32.0\%) & 858 (31.8\%) & \\
\textbf{retired} & & & & \textless{} 0.001\textsuperscript{1} \\
~~~No & 834 (62.8\%) & 824 (53.0\%) & 1320 (48.3\%) & \\
~~~Yes & 494 (37.2\%) & 732 (47.0\%) & 1412 (51.7\%) & \\
\textbf{race} & & & & \textless{} 0.001\textsuperscript{1} \\
~~~N-Miss & 83 & 42 & 64 & \\
~~~aapi & 34 (2.7\%) & 34 (2.2\%) & 57 (2.1\%) & \\
~~~black & 63 (5.1\%) & 49 (3.2\%) & 95 (3.6\%) & \\
~~~hisp & 190 (15.3\%) & 177 (11.7\%) & 276 (10.3\%) & \\
~~~indig & 12 (1.0\%) & 14 (0.9\%) & 9 (0.3\%) & \\
~~~multi & 32 (2.6\%) & 26 (1.7\%) & 52 (1.9\%) & \\
~~~white & 914 (73.4\%) & 1214 (80.2\%) & 2179 (81.7\%) & \\
\textbf{homeowner} & & & & 0.052\textsuperscript{1} \\
~~~No & 204 (15.4\%) & 214 (13.8\%) & 335 (12.3\%) & \\
~~~Yes & 1124 (84.6\%) & 1342 (86.2\%) & 2397 (87.7\%) & \\
\end{longtable}

\FloatBarrier

\hypertarget{city-council-survey-2}{%
\subsection{City council survey}\label{city-council-survey-2}}

\begin{figure}
\includegraphics[width=\textwidth]{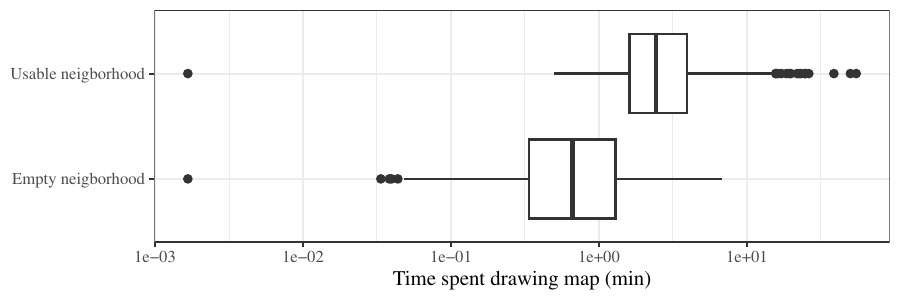}
\caption[Boxplot of time spent drawing map (in minutes)]{
From city council survey. Distributions plotted separately for valid and empty neighborhoods.}
\label{fig:map-time-nycc}
\end{figure}

Figure \ref{fig:map-time-nycc} shows that respondents which drew valid communities of interest spent more time on average with the drawing tool.
The following table reports correlations between various covariates and the time spent drawing the map, which is measured in seconds, binned into three categories, and reported in the table columns.

\begin{longtable}[]{@{}
  >{\raggedright\arraybackslash}p{(\columnwidth - 8\tabcolsep) * \real{0.4020}}
  >{\raggedright\arraybackslash}p{(\columnwidth - 8\tabcolsep) * \real{0.1667}}
  >{\raggedright\arraybackslash}p{(\columnwidth - 8\tabcolsep) * \real{0.1569}}
  >{\raggedright\arraybackslash}p{(\columnwidth - 8\tabcolsep) * \real{0.1667}}
  >{\raggedright\arraybackslash}p{(\columnwidth - 8\tabcolsep) * \real{0.1078}}@{}}
\caption{Descriptive statistics by total time spent drawing neighborhood (minutes)}\tabularnewline
\toprule\noalign{}
\begin{minipage}[b]{\linewidth}\raggedright
\end{minipage} & \begin{minipage}[b]{\linewidth}\raggedright
(0,1{]} (N=326)
\end{minipage} & \begin{minipage}[b]{\linewidth}\raggedright
(1,2{]} (N=393)
\end{minipage} & \begin{minipage}[b]{\linewidth}\raggedright
(2,Inf{]} (N=659)
\end{minipage} & \begin{minipage}[b]{\linewidth}\raggedright
p value
\end{minipage} \\
\midrule\noalign{}
\endfirsthead
\toprule\noalign{}
\begin{minipage}[b]{\linewidth}\raggedright
\end{minipage} & \begin{minipage}[b]{\linewidth}\raggedright
(0,1{]} (N=326)
\end{minipage} & \begin{minipage}[b]{\linewidth}\raggedright
(1,2{]} (N=393)
\end{minipage} & \begin{minipage}[b]{\linewidth}\raggedright
(2,Inf{]} (N=659)
\end{minipage} & \begin{minipage}[b]{\linewidth}\raggedright
p value
\end{minipage} \\
\midrule\noalign{}
\endhead
\bottomrule\noalign{}
\endlastfoot
\textbf{usable} & & & & \textless{} 0.001\textsuperscript{1} \\
~~~FALSE & 270 (82.8\%) & 90 (22.9\%) & 50 (7.6\%) & \\
~~~TRUE & 56 (17.2\%) & 303 (77.1\%) & 609 (92.4\%) & \\
\textbf{group} & & & & 0.759\textsuperscript{1} \\
~~~N-Miss & 84 & 50 & 37 & \\
~~~C & 52 (21.5\%) & 65 (19.0\%) & 122 (19.6\%) & \\
~~~P & 49 (20.2\%) & 73 (21.3\%) & 124 (19.9\%) & \\
~~~PH & 48 (19.8\%) & 71 (20.7\%) & 117 (18.8\%) & \\
~~~R & 39 (16.1\%) & 69 (20.1\%) & 137 (22.0\%) & \\
~~~RH & 54 (22.3\%) & 65 (19.0\%) & 122 (19.6\%) & \\
\textbf{city} & & & & \textless{} 0.001\textsuperscript{2} \\
~~~new-york & 326 (100.0\%) & 393 (100.0\%) & 659 (100.0\%) & \\
\textbf{party} & & & & \textless{} 0.001\textsuperscript{1} \\
~~~N-Miss & 66 & 37 & 22 & \\
~~~dem\_lean & 164 (63.1\%) & 275 (77.2\%) & 495 (77.7\%) & \\
~~~independent & 41 (15.8\%) & 30 (8.4\%) & 51 (8.0\%) & \\
~~~rep\_lean & 55 (21.2\%) & 51 (14.3\%) & 91 (14.3\%) & \\
\textbf{age} & & & & 0.155\textsuperscript{3} \\
~~~N-Miss & 77 & 44 & 29 & \\
~~~Mean (SD) & 50.040 (17.841) & 47.585 (17.655) & 50.605 (18.152) & \\
~~~Range & 18.000 - 105.000 & 18.000 - 89.000 & 18.000 - 105.000 & \\
\textbf{gender} & & & & 0.155\textsuperscript{1} \\
~~~N-Miss & 84 & 56 & 47 & \\
~~~female & 115 (47.5\%) & 127 (37.7\%) & 246 (40.2\%) & \\
~~~male & 127 (52.5\%) & 210 (62.3\%) & 366 (59.8\%) & \\
\textbf{education} & & & & 0.002\textsuperscript{1} \\
~~~N-Miss & 85 & 47 & 35 & \\
~~~no\_hs & 2 (0.8\%) & 1 (0.3\%) & 0 (0.0\%) & \\
~~~some\_hs & 8 (3.3\%) & 4 (1.2\%) & 7 (1.1\%) & \\
~~~hs & 22 (9.1\%) & 26 (7.5\%) & 28 (4.5\%) & \\
~~~some\_coll & 36 (14.9\%) & 29 (8.4\%) & 80 (12.8\%) & \\
~~~grad\_2yr & 21 (8.7\%) & 17 (4.9\%) & 28 (4.5\%) & \\
~~~grad\_4yr & 70 (29.0\%) & 128 (37.0\%) & 216 (34.6\%) & \\
~~~postgrad & 82 (34.0\%) & 141 (40.8\%) & 265 (42.5\%) & \\
\textbf{retired} & & & & 0.119\textsuperscript{1} \\
~~~N-Miss & 79 & 45 & 30 & \\
~~~No & 184 (74.5\%) & 288 (82.8\%) & 479 (76.2\%) & \\
~~~Yes & 63 (25.5\%) & 60 (17.2\%) & 150 (23.8\%) & \\
\textbf{race} & & & & \textless{} 0.001\textsuperscript{1} \\
~~~N-Miss & 109 & 65 & 62 & \\
~~~aapi & 13 (6.0\%) & 12 (3.7\%) & 20 (3.4\%) & \\
~~~black & 37 (17.1\%) & 23 (7.0\%) & 41 (6.9\%) & \\
~~~hisp & 45 (20.7\%) & 35 (10.7\%) & 69 (11.6\%) & \\
~~~indig & 2 (0.9\%) & 2 (0.6\%) & 4 (0.7\%) & \\
~~~multi & 5 (2.3\%) & 13 (4.0\%) & 23 (3.9\%) & \\
~~~white & 115 (53.0\%) & 243 (74.1\%) & 440 (73.7\%) & \\
\textbf{homeowner} & & & & 0.506\textsuperscript{1} \\
~~~N-Miss & 84 & 48 & 34 & \\
~~~Homeowner & 97 (40.1\%) & 133 (38.6\%) & 247 (39.5\%) & \\
~~~Other (please specify) & 6 (2.5\%) & 14 (4.1\%) & 37 (5.9\%) & \\
~~~Renter & 139 (57.4\%) & 198 (57.4\%) & 341 (54.6\%) & \\
\end{longtable}

\FloatBarrier

\hypertarget{support-for-new-housing-construction-analysis}{%
\section{Support for new housing construction analysis}\label{support-for-new-housing-construction-analysis}}

\begin{figure}[htb]

{\centering \includegraphics[width=0.75\linewidth]{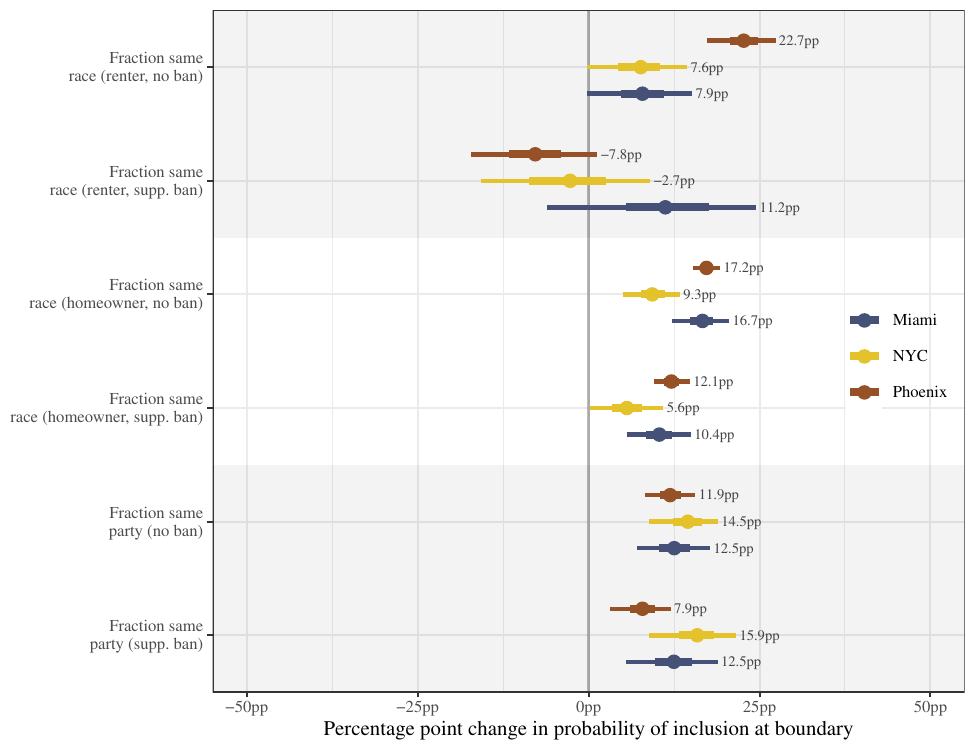} 

}

\caption{Selected full model coefficient posteriors, scaled to show the percentage point change in probability of a block's inclusion for a baseline probability of 50\%. Plotted are 90\% and 50\% credible intervals, with posterior medians displayed to the right of each interval.}\label{fig:housing-coef}
\end{figure}

We also collect respondent attitudes on the construction of new housing in their neighborhoods and test how opposition to new housing intensifies is related preferences. We do so by creating an indicator variable for whether respondents support a ban on the construction of new housing in their neighborhood and interacting it with the fraction same race and fraction same party in each census block. We further interact the homeowner variable with these terms to see how this varies across homeowners and renters. We then report the influence of race and party demographics for respondents who do and do not support a housing ban, separately for homeowners and renters. Figure \ref{fig:housing-coef} reports these coefficients, showing that the influence of homophily by race or party does not substantially vary by whether the respondent supports new housing in their neighborhood.

\FloatBarrier

\hypertarget{neighborhood-trust-analysis}{%
\section{Neighborhood trust analysis}\label{neighborhood-trust-analysis}}

\begin{figure}[htb]

{\centering \includegraphics[width=0.75\linewidth]{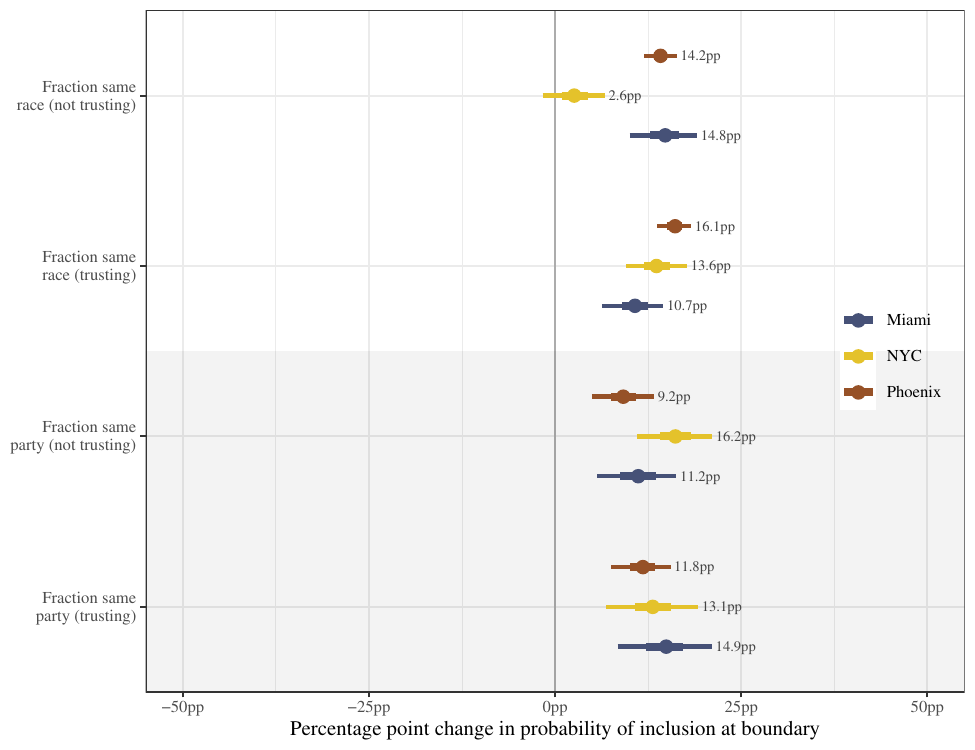} 

}

\caption{Selected full model coefficient posteriors, scaled to show the percentage point change in probability of a block's inclusion for a baseline probability of 50\%. Plotted are 90\% and 50\% credible intervals, with posterior medians displayed to the right of each interval.}\label{fig:trust-coef}
\end{figure}

Next, we conduct a similar exercise for the measures of whether respondents express trust in their neighbors. Figure \ref{fig:trust-coef} reports the effects of fraction same race and party separately for respondents who are above the median level of expressed neighbor trust (median calculated from sample of respondents in each city). Again, we find that the effects on race and party demographics are generally consistent across these comparisons.

\newpage

{\centering\LARGE\bfseries Additional Supplementary Information}

\setstretch{1.1}

\hypertarget{app:experiment}{%
\section{Experimental Application}\label{app:experiment}}

To further explore how race and partisanship influence the drawing of subjective neighborhoods, we design and implement an experiment where we randomly vary whether survey respondents are provided with information as to the racial or partisan demographics of the potential areas to include in their neighborhood. The primary goal of this experiment is methodological: to demonstrate how the proposed methodology can be adapted to a survey experiment with an information treatment, commonly used in political science. The second goal with the experiment is to see whether providing concrete information as to which census blocks are more racially or politically diverse might change how overtly exclusive survey respondents are when defining their neighborhood.
When voters get information on where out-group members live, they may deliberately draw them out of their neighborhoods.
It is also possible that voters are not purposely exclusive but the influence of racial and partisan demographics is subconsciously incorporated into how voters' neighborhood definitions.
This survey experiment, therefore, tests these two possibilities by examining the extent to which voters are deliberately exclusive when defining their neighborhoods.

\hypertarget{experimental-design}{%
\subsection{Experimental design}\label{experimental-design}}

The treatment for the experiment was administered by overlaying a different type of information over the map while the respondent was drawing their neighborhood.
There are five experimental conditions, which were randomly assigned to survey respondents:

\begin{enumerate}[leftmargin=1in]
    \item[\bf Control (C)] No information
    \item[\bf Party Placebo (PH)] Partisan information, but not identified as such
    \item[\bf Party (P)] Partisan information, identified as such
    \item[\bf Race Placebo (RH)] Racial information, but not identified as such
    \item[\bf Race (R)] Racial information, identified as such
\end{enumerate}

\begin{figure}[t]
    \centering
    \begin{subfigure}[b]{0.32\textwidth}
        \centering
        \includegraphics[width=\textwidth]{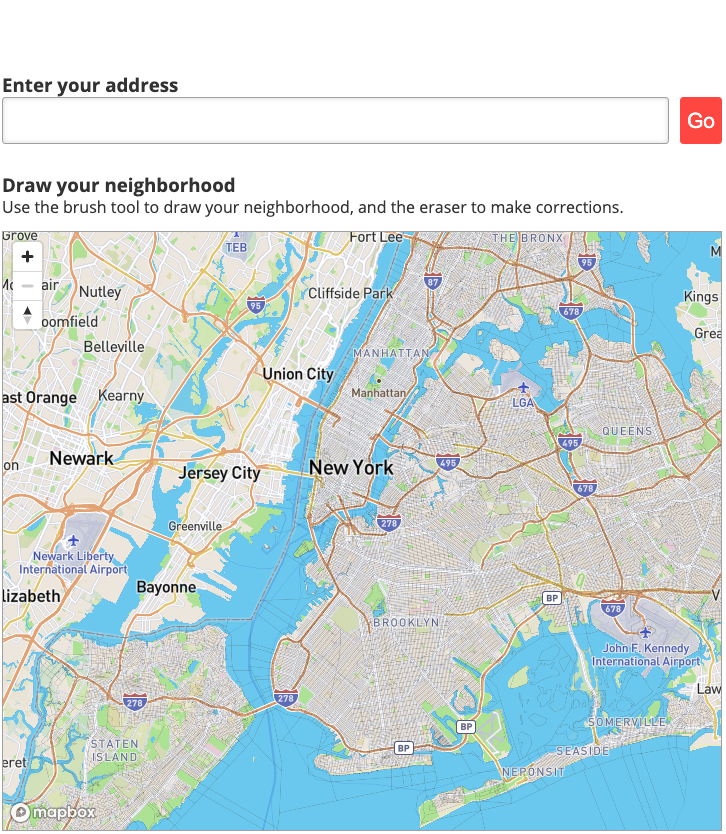}
        \caption{\bf Control}
    \end{subfigure}
    \begin{subfigure}[b]{0.32\textwidth}
        \centering
        \includegraphics[width=\textwidth]{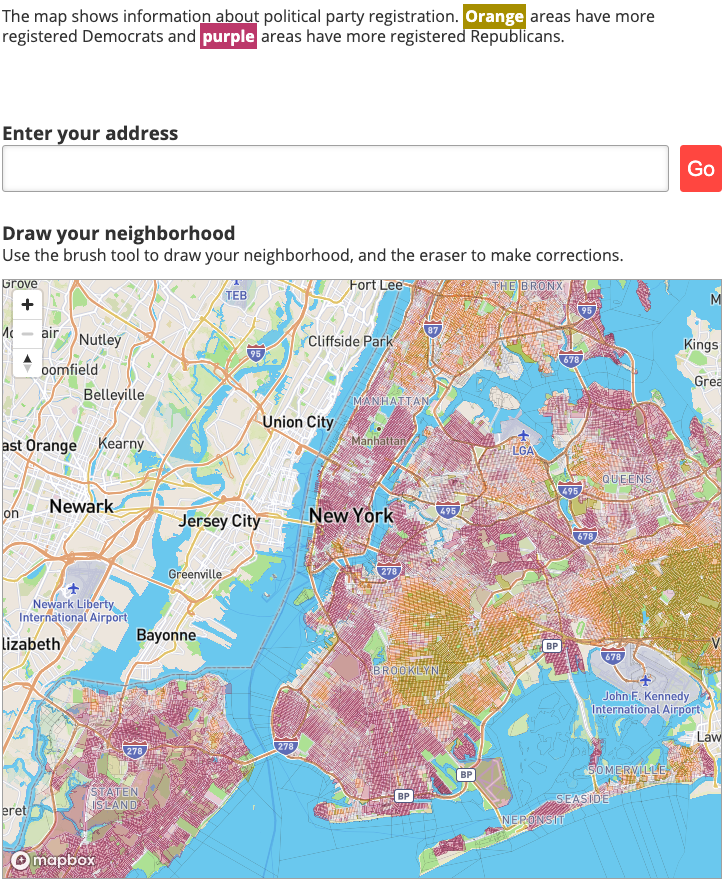}
        \caption{\bf Party}
    \end{subfigure}
    \begin{subfigure}[b]{0.32\textwidth}
        \centering
        \includegraphics[width=\textwidth]{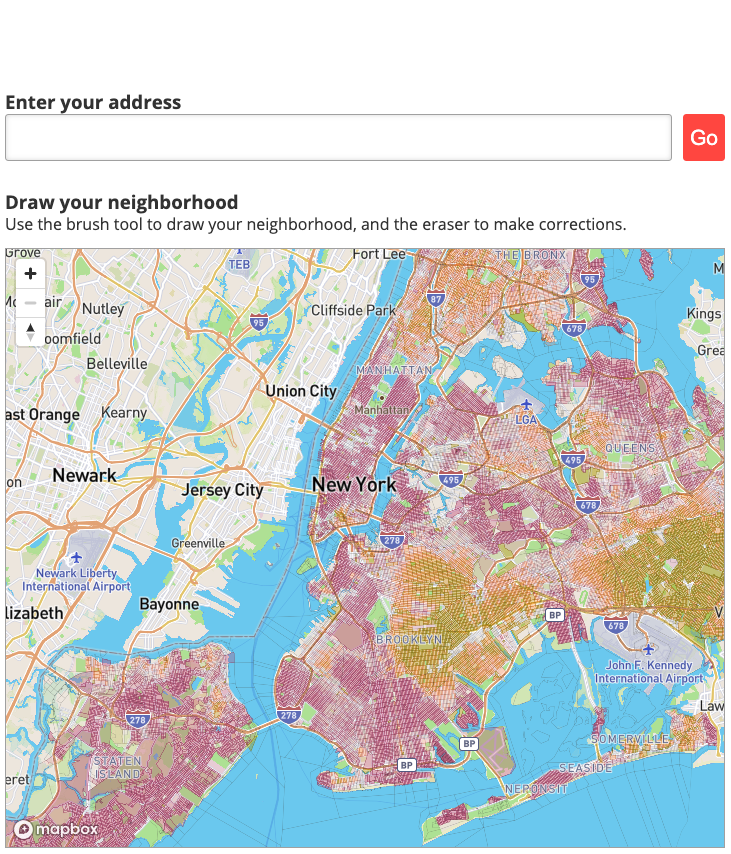}
        \caption{\bf Party Placebo}
    \end{subfigure}
    
    \vspace{0.5cm}
    \begin{subfigure}[b]{0.32\textwidth}
        \centering
        \includegraphics[width=\textwidth]{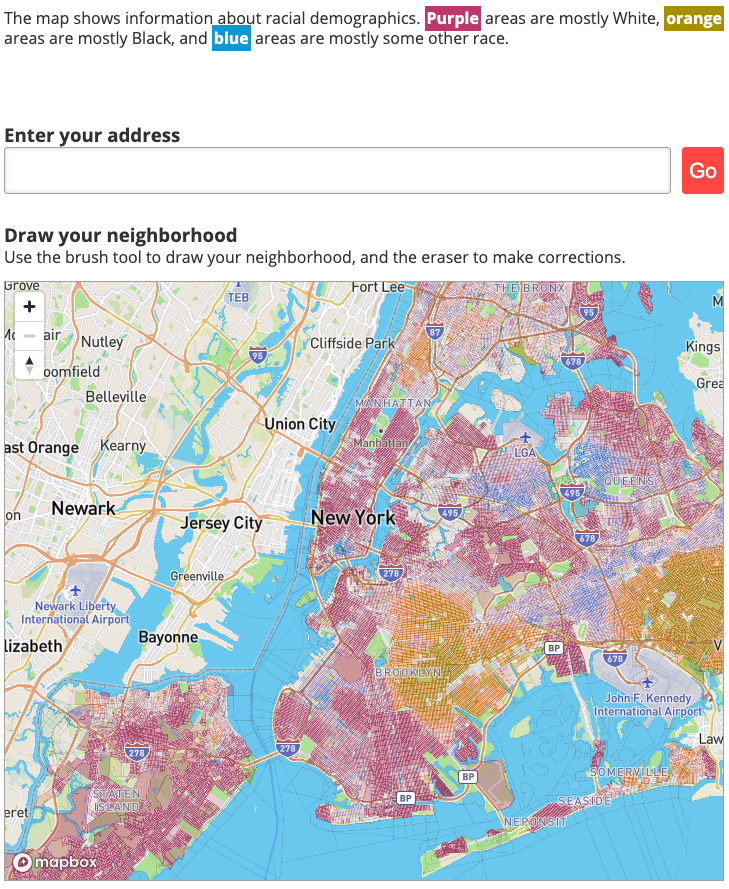}
        \caption{\bf Race}
    \end{subfigure}
    \begin{subfigure}[b]{0.32\textwidth}
        \centering
        \includegraphics[width=\textwidth]{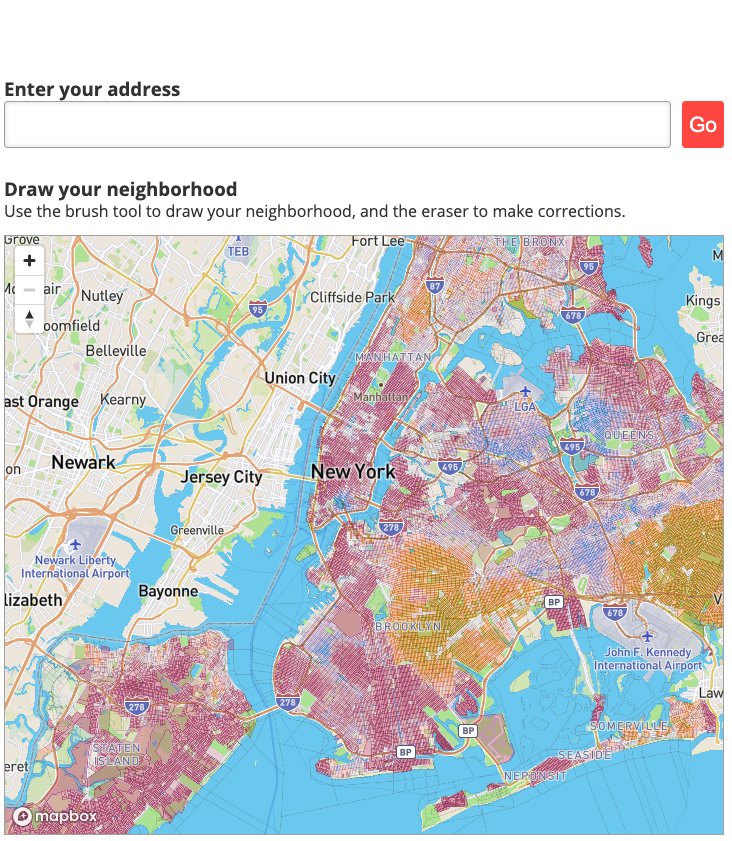}
        \caption{\bf Race Placebo}
    \end{subfigure}
    \caption{Map color schemes and identifying information for the five experimental conditions.}
    \label{fig:conditions}
\end{figure}

The control condition C contains no overlaid shading. For the Party Placebo and Party conditions, blocks were shaded on a gradient scale based on the proportion of major party registrants who are Democrats (\(\hbox{Democrats}/(\hbox{Democrats} + \hbox{Republicans})\)). Respondents in the Party Placebo condition were just shown the colored map, with no explanation for the coloring. Respondents in the Party condition were shown the colored map and an explanation that the purple areas are more Republican, while the orange areas are more Democratic.
For the Race and Race Placebo conditions, the blocks were colored according to the majority racial group (White, Black, or other) and shaded by the proportion of the census block population that is that group. As with the partisan conditions, the difference between the RH and R conditions is whether respondents were told that the colors represent the racial composition of different map areas.
Figure \ref{fig:conditions} shows the overlaid color schemes for each of the five experimental conditions, and the accompanying text that was shown to respondents in the Party and Race groups.

The reason for splitting the partisan and racial conditions in two, creating the intermediate color-but-no-information
condition, is that a map coloring in and of itself may change how respondents draw their neighborhood: it would be
natural for one's neighborhood boundaries to closely match the boundaries created by the artificial map coloring, similar to how people rely on anchors and heuristics in the absence of more concrete information when making decisions or estimating or estimating numerical quantities (Tversky and Kahneman 1974). Creating two different conditions allows us to separate this effect from our main effect.

After completing the mapping module, respondents next were asked whether they would support or oppose a ban on the construction of new housing in their neighborhood. This outcome is meant to measure a policy preference that tracks to general questions of exclusivity and NIMBYism in one's residential environment. Lastly, the respondents were asked to answer three questions comprising a trust battery: measuring the level of trust they express towards their neighbors---which we explicitly define for them as the people living in the neighborhood they just drew. The questions were adapted from well-validated survey items designed to capture general trust (Reeskens and Hooghe 2008). These questions test whether defining one's neighborhood along explicit racial or partisan dimensions cause respondents to express greater trust in their neighborhood (relative to outside their neighborhood) and express greater desire to prevent more housing (and thus new residents) into their defined neighborhood.
We hypothesize that the treatment conditions that provide racial or partisan information will reduce willingness and that defining one's neighborhood along partisan or racial dimensions (i.e.~being assigned to the racial or partisan information condition) will cause people to express greater trust in their neighbors in their drawn neighborhood.

\hypertarget{findings}{%
\subsection{Findings}\label{findings}}

To measure the effect of racial and partisan information on exclusivity, we refit the model on the full sample (\(n = 2,508\) across the three cities) and interact treatment group indicators with the co-partisan, co-ethnic, and co-class interaction variables.\footnote{Due to numerical stability issues in fitting this model to the full Phoenix sample, we used a different inference routine to fit the GLMM (the \texttt{bam} function in the \texttt{mgcv} R package). While this routine fits the same model likelihood, the implied priors are different, and due to the details of the inference function we are unable to perform corrective importance sampling. Given the large number of block-level observations in the sample, we do not believe that final inferences were meaningfully affected. Indeed, using this alternative fitting routine on the Miami and New York samples yielded no noticeable differences other than a narrower credible interval on the intercept parameter.}
From this, we calculate the following posterior quantities of interest:

\begin{itemize}
    \item The difference in the respondent-block co-partisan interaction coefficient between the Party and Party Placebo groups
    \item The difference in the respondent-block co-ethnic interaction coefficient between the Race and Race Placebo groups
\end{itemize}

\begin{figure}[p!]
\includegraphics[width=\textwidth]{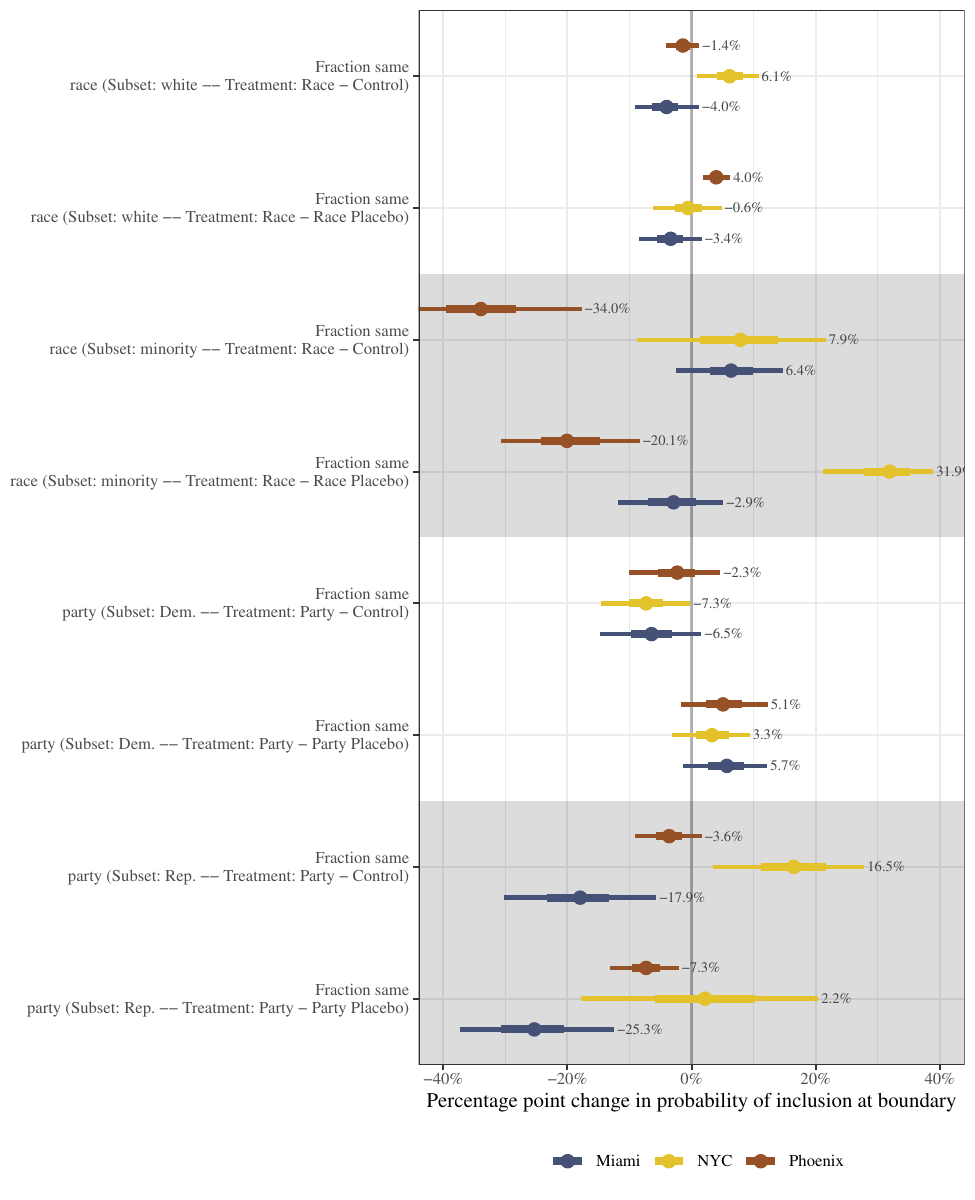}
\caption{Change in racial and partisan homophily coefficients between treatment groups, scaled to show the percentage pointchange in probability of a block's inclusion for a baseline probability of 50\%. Plotted are 95\% and 50\% credible intervals, with posterior medians displayed to the right of each interval.}
\label{fig:experiment-coef}
\end{figure}

These quantities of interest are shown in Figure \ref{fig:experiment-coef}, along with the difference in the respondent-block co-ethnic and co-partisan interaction coefficient between the Race, Party, and Control groups.
Appendix \ref{app:experiment} contains the full results table.
In general, we find little overall difference in the influence of co-ethnicity or co-partisanship.
Respondents in Miami and New York who were assigned to view the map shaded to show racial demographics, do not exhibit a stronger preference for co-ethnicity than respondents assigned to the placebo condition where they are shown the same shaded map but not told what the shading signifies.
In Phoenix, however, we do find a statistically significant effect, with the racial information increasing the coefficient on racial homophily by 4.0 percentage points.
For minority voters, there are conflicting results across cities, with little treatment effect in Miami, a positive (more exclusivity) effect in New York, but a negative effect in Phoenix.
We see little effect of partisan information for Democrats in each of the cities, but conflicting results for Republicans.
Republicans in New York become much more exclusionary along partisan lines in response to partisan information, while Republicans in Miami and Phoenix become much less exclusionary.

\begin{figure}
\includegraphics[width=\textwidth]{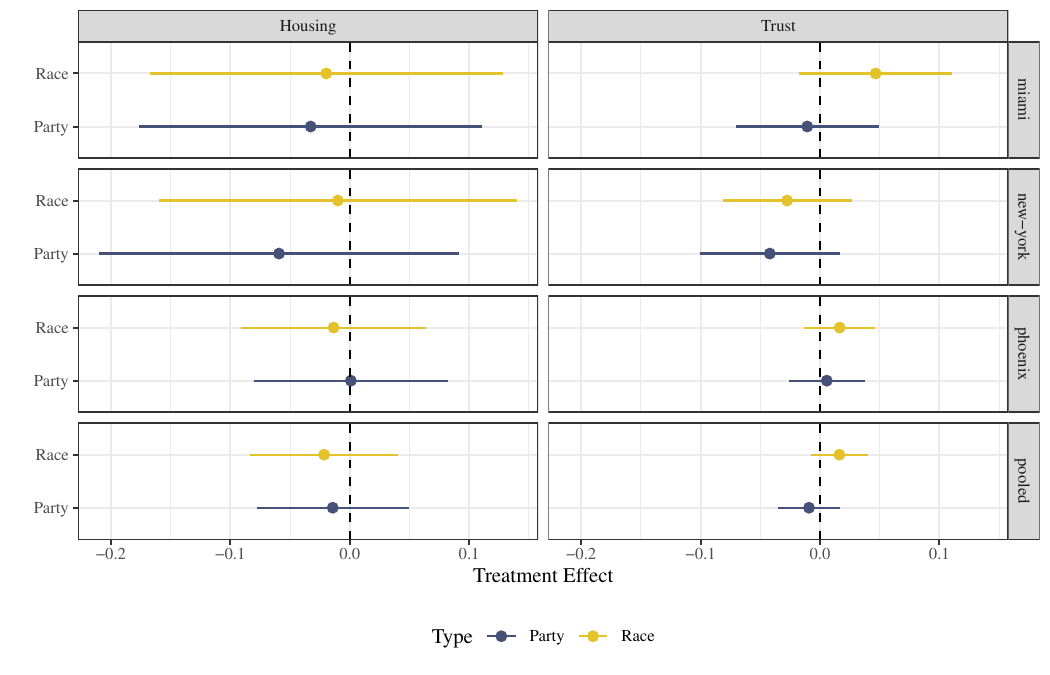}
\caption[Treatment effects on housing and trust survey outcomes]{
Treatment effects on housing and trust survey outcomes. }
\label{fig:experiment-other}
\end{figure}

These null or conflicting results do not support the conclusion that voters are explicitly motivated to draw more homogeneous neighborhoods.
When given the information to more starkly define their neighborhood along racial or partisan definitions, voters do not generally do so.
But across treatment groups, the model demonstrates that---net of other variables in the model---racial and partisan homophily are important determinants of how neighborhoods are defined in voters' minds.
This is consistent with the influence of local demographics being already subconsciously incorporated into how voters form attachments to their local area.
In Figure \ref{fig:experiment-other}, we present the treatment effects for these outcomes, which are unchanged across treatment groups.

\begin{figure}
\includegraphics[width=\textwidth]{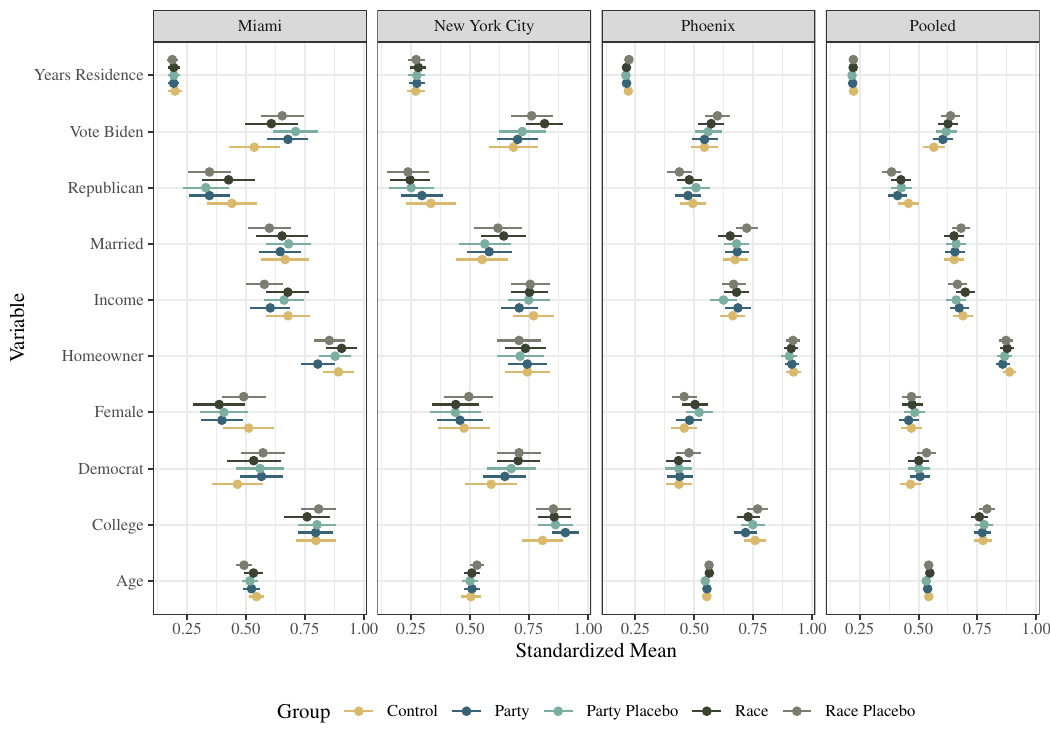}
\caption[Average levels of pre-treatment variables by treatment group]{
Average levels of pre-treatment variables by treatment group. Variables are standardized to be between 0 and 1.}
\label{fig:balance}
\end{figure}

This tables contains posterior summaries for all model coefficients on the original model scale.
These models were fit separately to each city:

\begin{itemize}
\tightlist
\item
  Miami: 473 survey respondents (91,435 individual block-level observations)
\item
  New York: 450 survey respondents (82,842 individual block-level observations)
\item
  Phoenix: 1,585 survey respondents (272,511 individual block-level observations)
\end{itemize}

\begingroup\fontsize{9}{11}\selectfont

\begin{longtable}[t]{llrrrrr}
\caption{Full experimental sample model estimates.}\\
\toprule
Coefficient & City & Mean & Std. Dev. & Q5 & Median & Q95\\
\midrule
\cellcolor{gray!6}{(Intercept)} & \cellcolor{gray!6}{Miami} & \cellcolor{gray!6}{-8.06} & \cellcolor{gray!6}{6.71} & \cellcolor{gray!6}{-18.88} & \cellcolor{gray!6}{-8.20} & \cellcolor{gray!6}{2.84}\\
\cellcolor{gray!6}{(Intercept)} & \cellcolor{gray!6}{NYC} & \cellcolor{gray!6}{-5.59} & \cellcolor{gray!6}{6.15} & \cellcolor{gray!6}{-15.88} & \cellcolor{gray!6}{-5.33} & \cellcolor{gray!6}{4.09}\\
\cellcolor{gray!6}{(Intercept)} & \cellcolor{gray!6}{Phoenix} & \cellcolor{gray!6}{-11.87} & \cellcolor{gray!6}{0.40} & \cellcolor{gray!6}{-12.53} & \cellcolor{gray!6}{-11.87} & \cellcolor{gray!6}{-11.21}\\
church & Miami & 0.01 & 0.03 & -0.04 & 0.01 & 0.06\\
church & NYC & 0.08 & 0.02 & 0.05 & 0.08 & 0.12\\
\addlinespace
church & Phoenix & 0.20 & 0.03 & 0.16 & 0.20 & 0.24\\
\cellcolor{gray!6}{distance} & \cellcolor{gray!6}{Miami} & \cellcolor{gray!6}{0.06} & \cellcolor{gray!6}{0.01} & \cellcolor{gray!6}{0.04} & \cellcolor{gray!6}{0.06} & \cellcolor{gray!6}{0.07}\\
\cellcolor{gray!6}{distance} & \cellcolor{gray!6}{Miami} & \cellcolor{gray!6}{0.19} & \cellcolor{gray!6}{0.02} & \cellcolor{gray!6}{0.17} & \cellcolor{gray!6}{0.19} & \cellcolor{gray!6}{0.22}\\
\cellcolor{gray!6}{distance} & \cellcolor{gray!6}{NYC} & \cellcolor{gray!6}{0.07} & \cellcolor{gray!6}{0.01} & \cellcolor{gray!6}{0.05} & \cellcolor{gray!6}{0.07} & \cellcolor{gray!6}{0.08}\\
distance & NYC & 0.07 & 0.01 & 0.05 & 0.07 & 0.09\\
\addlinespace
distance & Phoenix & 0.12 & 0.01 & 0.11 & 0.12 & 0.13\\
distance & Phoenix & 0.06 & 0.01 & 0.05 & 0.06 & 0.08\\
\cellcolor{gray!6}{park} & \cellcolor{gray!6}{Miami} & \cellcolor{gray!6}{-0.12} & \cellcolor{gray!6}{0.04} & \cellcolor{gray!6}{-0.18} & \cellcolor{gray!6}{-0.12} & \cellcolor{gray!6}{-0.05}\\
\cellcolor{gray!6}{park} & \cellcolor{gray!6}{NYC} & \cellcolor{gray!6}{0.10} & \cellcolor{gray!6}{0.02} & \cellcolor{gray!6}{0.07} & \cellcolor{gray!6}{0.10} & \cellcolor{gray!6}{0.13}\\
\cellcolor{gray!6}{park} & \cellcolor{gray!6}{Phoenix} & \cellcolor{gray!6}{0.13} & \cellcolor{gray!6}{0.02} & \cellcolor{gray!6}{0.09} & \cellcolor{gray!6}{0.13} & \cellcolor{gray!6}{0.16}\\
\addlinespace
school & Miami & 0.33 & 0.07 & 0.20 & 0.33 & 0.44\\
school & NYC & 0.03 & 0.05 & -0.06 & 0.03 & 0.11\\
school & Phoenix & 0.11 & 0.04 & 0.04 & 0.11 & 0.17\\
\cellcolor{gray!6}{children} & \cellcolor{gray!6}{Miami} & \cellcolor{gray!6}{1.37} & \cellcolor{gray!6}{2.43} & \cellcolor{gray!6}{-2.58} & \cellcolor{gray!6}{1.33} & \cellcolor{gray!6}{5.33}\\
\cellcolor{gray!6}{children} & \cellcolor{gray!6}{NYC} & \cellcolor{gray!6}{-0.13} & \cellcolor{gray!6}{1.89} & \cellcolor{gray!6}{-3.19} & \cellcolor{gray!6}{-0.16} & \cellcolor{gray!6}{2.94}\\
\addlinespace
\cellcolor{gray!6}{children} & \cellcolor{gray!6}{Phoenix} & \cellcolor{gray!6}{-0.93} & \cellcolor{gray!6}{0.15} & \cellcolor{gray!6}{-1.19} & \cellcolor{gray!6}{-0.93} & \cellcolor{gray!6}{-0.68}\\
same block group & Miami & 0.01 & 0.03 & -0.04 & 0.01 & 0.06\\
same block group & NYC & 0.06 & 0.04 & 0.00 & 0.06 & 0.12\\
same block group & Phoenix & -0.01 & 0.02 & -0.04 & -0.01 & 0.02\\
\cellcolor{gray!6}{same tract} & \cellcolor{gray!6}{Miami} & \cellcolor{gray!6}{-0.23} & \cellcolor{gray!6}{0.05} & \cellcolor{gray!6}{-0.31} & \cellcolor{gray!6}{-0.23} & \cellcolor{gray!6}{-0.15}\\
\addlinespace
\cellcolor{gray!6}{same tract} & \cellcolor{gray!6}{NYC} & \cellcolor{gray!6}{-0.04} & \cellcolor{gray!6}{0.06} & \cellcolor{gray!6}{-0.13} & \cellcolor{gray!6}{-0.04} & \cellcolor{gray!6}{0.04}\\
\cellcolor{gray!6}{same tract} & \cellcolor{gray!6}{Phoenix} & \cellcolor{gray!6}{-0.43} & \cellcolor{gray!6}{0.07} & \cellcolor{gray!6}{-0.54} & \cellcolor{gray!6}{-0.43} & \cellcolor{gray!6}{-0.32}\\
same road region & Miami & -0.30 & 0.02 & -0.33 & -0.30 & -0.26\\
same road region & NYC & -0.08 & 0.02 & -0.11 & -0.08 & -0.05\\
same road region & Phoenix & -0.34 & 0.02 & -0.37 & -0.34 & -0.31\\
\addlinespace
\cellcolor{gray!6}{population} & \cellcolor{gray!6}{Miami} & \cellcolor{gray!6}{-1.63} & \cellcolor{gray!6}{0.13} & \cellcolor{gray!6}{-1.83} & \cellcolor{gray!6}{-1.63} & \cellcolor{gray!6}{-1.42}\\
\cellcolor{gray!6}{population} & \cellcolor{gray!6}{NYC} & \cellcolor{gray!6}{-0.86} & \cellcolor{gray!6}{0.08} & \cellcolor{gray!6}{-1.00} & \cellcolor{gray!6}{-0.86} & \cellcolor{gray!6}{-0.73}\\
\cellcolor{gray!6}{population} & \cellcolor{gray!6}{Phoenix} & \cellcolor{gray!6}{-1.74} & \cellcolor{gray!6}{0.10} & \cellcolor{gray!6}{-1.90} & \cellcolor{gray!6}{-1.74} & \cellcolor{gray!6}{-1.58}\\
area & Miami & -0.11 & 0.04 & -0.17 & -0.11 & -0.05\\
area & NYC & -0.01 & 0.05 & -0.09 & -0.01 & 0.07\\
\addlinespace
area & Phoenix & -0.17 & 0.03 & -0.21 & -0.17 & -0.13\\
\cellcolor{gray!6}{age} & \cellcolor{gray!6}{Miami} & \cellcolor{gray!6}{0.00} & \cellcolor{gray!6}{0.12} & \cellcolor{gray!6}{-0.19} & \cellcolor{gray!6}{0.00} & \cellcolor{gray!6}{0.20}\\
\cellcolor{gray!6}{age} & \cellcolor{gray!6}{NYC} & \cellcolor{gray!6}{0.00} & \cellcolor{gray!6}{0.10} & \cellcolor{gray!6}{-0.15} & \cellcolor{gray!6}{0.00} & \cellcolor{gray!6}{0.16}\\
\cellcolor{gray!6}{age} & \cellcolor{gray!6}{Phoenix} & \cellcolor{gray!6}{0.01} & \cellcolor{gray!6}{0.01} & \cellcolor{gray!6}{0.01} & \cellcolor{gray!6}{0.01} & \cellcolor{gray!6}{0.02}\\
education = No College & Miami & -2.82 & 5.82 & -12.39 & -2.75 & 6.45\\
\addlinespace
education = No College & NYC & -0.44 & 4.94 & -8.61 & -0.40 & 7.38\\
education = No College & Phoenix & -0.80 & 1.17 & -2.73 & -0.80 & 1.13\\
\cellcolor{gray!6}{retired} & \cellcolor{gray!6}{Miami} & \cellcolor{gray!6}{0.14} & \cellcolor{gray!6}{2.86} & \cellcolor{gray!6}{-4.59} & \cellcolor{gray!6}{0.07} & \cellcolor{gray!6}{4.84}\\
\cellcolor{gray!6}{retired} & \cellcolor{gray!6}{NYC} & \cellcolor{gray!6}{0.04} & \cellcolor{gray!6}{2.37} & \cellcolor{gray!6}{-3.83} & \cellcolor{gray!6}{-0.10} & \cellcolor{gray!6}{4.09}\\
\cellcolor{gray!6}{retired} & \cellcolor{gray!6}{Phoenix} & \cellcolor{gray!6}{-0.05} & \cellcolor{gray!6}{0.12} & \cellcolor{gray!6}{-0.26} & \cellcolor{gray!6}{-0.05} & \cellcolor{gray!6}{0.15}\\
\addlinespace
tenure & Miami & -0.04 & 0.89 & -1.50 & -0.06 & 1.34\\
tenure & NYC & 0.00 & 0.66 & -1.08 & -0.02 & 1.02\\
tenure & Phoenix & 0.02 & 0.02 & -0.02 & 0.02 & 0.06\\
\cellcolor{gray!6}{party = IND} & \cellcolor{gray!6}{Miami} & \cellcolor{gray!6}{0.17} & \cellcolor{gray!6}{8.43} & \cellcolor{gray!6}{-13.58} & \cellcolor{gray!6}{-0.15} & \cellcolor{gray!6}{14.14}\\
\cellcolor{gray!6}{party = IND} & \cellcolor{gray!6}{NYC} & \cellcolor{gray!6}{0.06} & \cellcolor{gray!6}{7.32} & \cellcolor{gray!6}{-11.44} & \cellcolor{gray!6}{0.13} & \cellcolor{gray!6}{12.69}\\
\addlinespace
\cellcolor{gray!6}{party = IND} & \cellcolor{gray!6}{Phoenix} & \cellcolor{gray!6}{-0.41} & \cellcolor{gray!6}{0.42} & \cellcolor{gray!6}{-1.10} & \cellcolor{gray!6}{-0.41} & \cellcolor{gray!6}{0.27}\\
party = REP & Miami & -0.06 & 5.01 & -7.86 & -0.16 & 8.23\\
party = REP & NYC & -0.43 & 4.37 & -7.54 & -0.48 & 7.20\\
party = REP & Phoenix & -0.01 & 0.22 & -0.36 & -0.01 & 0.34\\
\cellcolor{gray!6}{minority} & \cellcolor{gray!6}{Miami} & \cellcolor{gray!6}{0.01} & \cellcolor{gray!6}{5.66} & \cellcolor{gray!6}{-9.08} & \cellcolor{gray!6}{-0.07} & \cellcolor{gray!6}{9.03}\\
\addlinespace
\cellcolor{gray!6}{minority} & \cellcolor{gray!6}{NYC} & \cellcolor{gray!6}{0.13} & \cellcolor{gray!6}{4.13} & \cellcolor{gray!6}{-6.64} & \cellcolor{gray!6}{0.07} & \cellcolor{gray!6}{6.93}\\
\cellcolor{gray!6}{minority} & \cellcolor{gray!6}{Phoenix} & \cellcolor{gray!6}{-0.23} & \cellcolor{gray!6}{0.31} & \cellcolor{gray!6}{-0.74} & \cellcolor{gray!6}{-0.23} & \cellcolor{gray!6}{0.28}\\
homeowner & Miami & 1.30 & 4.60 & -6.22 & 1.17 & 8.78\\
homeowner & NYC & -0.36 & 3.50 & -6.07 & -0.39 & 5.59\\
homeowner & Phoenix & -0.34 & 0.33 & -0.88 & -0.34 & 0.20\\
\addlinespace
\cellcolor{gray!6}{school * children} & \cellcolor{gray!6}{Miami} & \cellcolor{gray!6}{-0.50} & \cellcolor{gray!6}{0.11} & \cellcolor{gray!6}{-0.68} & \cellcolor{gray!6}{-0.49} & \cellcolor{gray!6}{-0.31}\\
\cellcolor{gray!6}{school * children} & \cellcolor{gray!6}{NYC} & \cellcolor{gray!6}{0.00} & \cellcolor{gray!6}{0.08} & \cellcolor{gray!6}{-0.13} & \cellcolor{gray!6}{0.00} & \cellcolor{gray!6}{0.13}\\
\cellcolor{gray!6}{school * children} & \cellcolor{gray!6}{Phoenix} & \cellcolor{gray!6}{0.22} & \cellcolor{gray!6}{0.07} & \cellcolor{gray!6}{0.11} & \cellcolor{gray!6}{0.22} & \cellcolor{gray!6}{0.33}\\
children * distance & Miami & -0.15 & 0.02 & -0.19 & -0.15 & -0.11\\
children * distance & NYC & 0.01 & 0.02 & -0.02 & 0.01 & 0.04\\
\addlinespace
children * distance & Phoenix & 0.15 & 0.02 & 0.13 & 0.15 & 0.18\\
\cellcolor{gray!6}{same tract * same road region} & \cellcolor{gray!6}{Miami} & \cellcolor{gray!6}{0.11} & \cellcolor{gray!6}{0.05} & \cellcolor{gray!6}{0.02} & \cellcolor{gray!6}{0.11} & \cellcolor{gray!6}{0.19}\\
\cellcolor{gray!6}{same tract * same road region} & \cellcolor{gray!6}{NYC} & \cellcolor{gray!6}{-0.11} & \cellcolor{gray!6}{0.06} & \cellcolor{gray!6}{-0.20} & \cellcolor{gray!6}{-0.11} & \cellcolor{gray!6}{-0.02}\\
\cellcolor{gray!6}{same tract * same road region} & \cellcolor{gray!6}{Phoenix} & \cellcolor{gray!6}{0.01} & \cellcolor{gray!6}{0.07} & \cellcolor{gray!6}{-0.10} & \cellcolor{gray!6}{0.01} & \cellcolor{gray!6}{0.12}\\
Fraction same race * C group & Miami & -0.38 & 0.06 & -0.48 & -0.38 & -0.28\\
\addlinespace
Fraction same race * C group & NYC & -0.25 & 0.04 & -0.32 & -0.25 & -0.18\\
Fraction same race * C group & Phoenix & -0.46 & 0.03 & -0.51 & -0.46 & -0.40\\
\cellcolor{gray!6}{Fraction same race * P group} & \cellcolor{gray!6}{Miami} & \cellcolor{gray!6}{-0.45} & \cellcolor{gray!6}{0.06} & \cellcolor{gray!6}{-0.54} & \cellcolor{gray!6}{-0.46} & \cellcolor{gray!6}{-0.36}\\
\cellcolor{gray!6}{Fraction same race * P group} & \cellcolor{gray!6}{NYC} & \cellcolor{gray!6}{-0.48} & \cellcolor{gray!6}{0.05} & \cellcolor{gray!6}{-0.56} & \cellcolor{gray!6}{-0.48} & \cellcolor{gray!6}{-0.40}\\
\cellcolor{gray!6}{Fraction same race * P group} & \cellcolor{gray!6}{Phoenix} & \cellcolor{gray!6}{-0.39} & \cellcolor{gray!6}{0.03} & \cellcolor{gray!6}{-0.44} & \cellcolor{gray!6}{-0.39} & \cellcolor{gray!6}{-0.34}\\
\addlinespace
Fraction same race * PH group & Miami & -0.17 & 0.06 & -0.27 & -0.17 & -0.07\\
Fraction same race * PH group & NYC & -0.35 & 0.06 & -0.44 & -0.35 & -0.26\\
Fraction same race * PH group & Phoenix & -0.45 & 0.04 & -0.51 & -0.45 & -0.39\\
\cellcolor{gray!6}{Fraction same race * R group} & \cellcolor{gray!6}{Miami} & \cellcolor{gray!6}{-0.28} & \cellcolor{gray!6}{0.06} & \cellcolor{gray!6}{-0.37} & \cellcolor{gray!6}{-0.27} & \cellcolor{gray!6}{-0.18}\\
\cellcolor{gray!6}{Fraction same race * R group} & \cellcolor{gray!6}{NYC} & \cellcolor{gray!6}{-0.38} & \cellcolor{gray!6}{0.05} & \cellcolor{gray!6}{-0.47} & \cellcolor{gray!6}{-0.38} & \cellcolor{gray!6}{-0.30}\\
\addlinespace
\cellcolor{gray!6}{Fraction same race * R group} & \cellcolor{gray!6}{Phoenix} & \cellcolor{gray!6}{-0.42} & \cellcolor{gray!6}{0.03} & \cellcolor{gray!6}{-0.47} & \cellcolor{gray!6}{-0.42} & \cellcolor{gray!6}{-0.36}\\
Fraction same race * RH group & Miami & -0.36 & 0.06 & -0.45 & -0.37 & -0.27\\
Fraction same race * RH group & NYC & -0.39 & 0.05 & -0.47 & -0.39 & -0.31\\
Fraction same race * RH group & Phoenix & -0.30 & 0.03 & -0.35 & -0.30 & -0.25\\
\cellcolor{gray!6}{C group * Fraction same party} & \cellcolor{gray!6}{Miami} & \cellcolor{gray!6}{-0.22} & \cellcolor{gray!6}{0.09} & \cellcolor{gray!6}{-0.37} & \cellcolor{gray!6}{-0.22} & \cellcolor{gray!6}{-0.07}\\
\addlinespace
\cellcolor{gray!6}{C group * Fraction same party} & \cellcolor{gray!6}{NYC} & \cellcolor{gray!6}{-0.40} & \cellcolor{gray!6}{0.07} & \cellcolor{gray!6}{-0.51} & \cellcolor{gray!6}{-0.41} & \cellcolor{gray!6}{-0.30}\\
\cellcolor{gray!6}{C group * Fraction same party} & \cellcolor{gray!6}{Phoenix} & \cellcolor{gray!6}{-0.36} & \cellcolor{gray!6}{0.09} & \cellcolor{gray!6}{-0.51} & \cellcolor{gray!6}{-0.36} & \cellcolor{gray!6}{-0.22}\\
P group * Fraction same party & Miami & -0.06 & 0.09 & -0.20 & -0.06 & 0.08\\
P group * Fraction same party & NYC & -0.26 & 0.06 & -0.35 & -0.26 & -0.16\\
P group * Fraction same party & Phoenix & -0.30 & 0.09 & -0.45 & -0.30 & -0.14\\
\addlinespace
\cellcolor{gray!6}{PH group * Fraction same party} & \cellcolor{gray!6}{Miami} & \cellcolor{gray!6}{0.09} & \cellcolor{gray!6}{0.08} & \cellcolor{gray!6}{-0.04} & \cellcolor{gray!6}{0.09} & \cellcolor{gray!6}{0.22}\\
\cellcolor{gray!6}{PH group * Fraction same party} & \cellcolor{gray!6}{NYC} & \cellcolor{gray!6}{-0.19} & \cellcolor{gray!6}{0.06} & \cellcolor{gray!6}{-0.29} & \cellcolor{gray!6}{-0.19} & \cellcolor{gray!6}{-0.08}\\
\cellcolor{gray!6}{PH group * Fraction same party} & \cellcolor{gray!6}{Phoenix} & \cellcolor{gray!6}{-0.14} & \cellcolor{gray!6}{0.09} & \cellcolor{gray!6}{-0.28} & \cellcolor{gray!6}{-0.14} & \cellcolor{gray!6}{0.00}\\
R group * Fraction same party & Miami & -0.19 & 0.07 & -0.31 & -0.19 & -0.06\\
R group * Fraction same party & NYC & -0.31 & 0.06 & -0.41 & -0.31 & -0.22\\
\addlinespace
R group * Fraction same party & Phoenix & -0.53 & 0.08 & -0.66 & -0.53 & -0.40\\
\cellcolor{gray!6}{RH group * Fraction same party} & \cellcolor{gray!6}{Miami} & \cellcolor{gray!6}{-0.32} & \cellcolor{gray!6}{0.07} & \cellcolor{gray!6}{-0.43} & \cellcolor{gray!6}{-0.32} & \cellcolor{gray!6}{-0.20}\\
\cellcolor{gray!6}{RH group * Fraction same party} & \cellcolor{gray!6}{NYC} & \cellcolor{gray!6}{-0.18} & \cellcolor{gray!6}{0.06} & \cellcolor{gray!6}{-0.28} & \cellcolor{gray!6}{-0.18} & \cellcolor{gray!6}{-0.08}\\
\cellcolor{gray!6}{RH group * Fraction same party} & \cellcolor{gray!6}{Phoenix} & \cellcolor{gray!6}{-0.18} & \cellcolor{gray!6}{0.07} & \cellcolor{gray!6}{-0.30} & \cellcolor{gray!6}{-0.18} & \cellcolor{gray!6}{-0.07}\\
C group * Fraction same ownership & Miami & -0.43 & 0.18 & -0.72 & -0.43 & -0.13\\
\addlinespace
C group * Fraction same ownership & NYC & -0.40 & 0.13 & -0.60 & -0.40 & -0.19\\
C group * Fraction same ownership & Phoenix & -0.41 & 0.15 & -0.65 & -0.41 & -0.17\\
\cellcolor{gray!6}{P group * Fraction same ownership} & \cellcolor{gray!6}{Miami} & \cellcolor{gray!6}{0.20} & \cellcolor{gray!6}{0.13} & \cellcolor{gray!6}{0.00} & \cellcolor{gray!6}{0.20} & \cellcolor{gray!6}{0.41}\\
\cellcolor{gray!6}{P group * Fraction same ownership} & \cellcolor{gray!6}{NYC} & \cellcolor{gray!6}{-0.41} & \cellcolor{gray!6}{0.13} & \cellcolor{gray!6}{-0.63} & \cellcolor{gray!6}{-0.41} & \cellcolor{gray!6}{-0.20}\\
\cellcolor{gray!6}{P group * Fraction same ownership} & \cellcolor{gray!6}{Phoenix} & \cellcolor{gray!6}{0.54} & \cellcolor{gray!6}{0.16} & \cellcolor{gray!6}{0.28} & \cellcolor{gray!6}{0.54} & \cellcolor{gray!6}{0.81}\\
\addlinespace
PH group * Fraction same ownership & Miami & -0.63 & 0.26 & -1.05 & -0.63 & -0.21\\
PH group * Fraction same ownership & NYC & -0.41 & 0.11 & -0.59 & -0.40 & -0.22\\
PH group * Fraction same ownership & Phoenix & 0.18 & 0.13 & -0.03 & 0.18 & 0.39\\
\cellcolor{gray!6}{R group * Fraction same ownership} & \cellcolor{gray!6}{Miami} & \cellcolor{gray!6}{1.39} & \cellcolor{gray!6}{0.79} & \cellcolor{gray!6}{0.10} & \cellcolor{gray!6}{1.37} & \cellcolor{gray!6}{2.69}\\
\cellcolor{gray!6}{R group * Fraction same ownership} & \cellcolor{gray!6}{NYC} & \cellcolor{gray!6}{-0.07} & \cellcolor{gray!6}{0.12} & \cellcolor{gray!6}{-0.27} & \cellcolor{gray!6}{-0.07} & \cellcolor{gray!6}{0.13}\\
\addlinespace
\cellcolor{gray!6}{R group * Fraction same ownership} & \cellcolor{gray!6}{Phoenix} & \cellcolor{gray!6}{0.41} & \cellcolor{gray!6}{0.15} & \cellcolor{gray!6}{0.16} & \cellcolor{gray!6}{0.41} & \cellcolor{gray!6}{0.66}\\
RH group * Fraction same ownership & Miami & -0.28 & 0.23 & -0.65 & -0.29 & 0.09\\
RH group * Fraction same ownership & NYC & -0.07 & 0.10 & -0.23 & -0.07 & 0.09\\
RH group * Fraction same ownership & Phoenix & -0.56 & 0.17 & -0.84 & -0.56 & -0.28\\
\cellcolor{gray!6}{C group * Fraction same education} & \cellcolor{gray!6}{Miami} & \cellcolor{gray!6}{0.01} & \cellcolor{gray!6}{0.17} & \cellcolor{gray!6}{-0.27} & \cellcolor{gray!6}{0.01} & \cellcolor{gray!6}{0.28}\\
\addlinespace
\cellcolor{gray!6}{C group * Fraction same education} & \cellcolor{gray!6}{NYC} & \cellcolor{gray!6}{-0.53} & \cellcolor{gray!6}{0.10} & \cellcolor{gray!6}{-0.69} & \cellcolor{gray!6}{-0.53} & \cellcolor{gray!6}{-0.36}\\
\cellcolor{gray!6}{C group * Fraction same education} & \cellcolor{gray!6}{Phoenix} & \cellcolor{gray!6}{-1.37} & \cellcolor{gray!6}{0.09} & \cellcolor{gray!6}{-1.53} & \cellcolor{gray!6}{-1.37} & \cellcolor{gray!6}{-1.22}\\
P group * Fraction same education & Miami & -0.26 & 0.14 & -0.48 & -0.26 & -0.04\\
P group * Fraction same education & NYC & 0.12 & 0.09 & -0.02 & 0.12 & 0.26\\
P group * Fraction same education & Phoenix & -0.94 & 0.10 & -1.11 & -0.94 & -0.77\\
\addlinespace
\cellcolor{gray!6}{PH group * Fraction same education} & \cellcolor{gray!6}{Miami} & \cellcolor{gray!6}{0.19} & \cellcolor{gray!6}{0.17} & \cellcolor{gray!6}{-0.08} & \cellcolor{gray!6}{0.18} & \cellcolor{gray!6}{0.46}\\
\cellcolor{gray!6}{PH group * Fraction same education} & \cellcolor{gray!6}{NYC} & \cellcolor{gray!6}{-0.57} & \cellcolor{gray!6}{0.10} & \cellcolor{gray!6}{-0.73} & \cellcolor{gray!6}{-0.57} & \cellcolor{gray!6}{-0.42}\\
\cellcolor{gray!6}{PH group * Fraction same education} & \cellcolor{gray!6}{Phoenix} & \cellcolor{gray!6}{-0.17} & \cellcolor{gray!6}{0.11} & \cellcolor{gray!6}{-0.34} & \cellcolor{gray!6}{-0.17} & \cellcolor{gray!6}{0.01}\\
R group * Fraction same education & Miami & -0.22 & 0.16 & -0.48 & -0.22 & 0.04\\
R group * Fraction same education & NYC & 0.14 & 0.07 & 0.01 & 0.14 & 0.26\\
\addlinespace
R group * Fraction same education & Phoenix & -1.07 & 0.10 & -1.24 & -1.07 & -0.90\\
\cellcolor{gray!6}{RH group * Fraction same education} & \cellcolor{gray!6}{Miami} & \cellcolor{gray!6}{-0.80} & \cellcolor{gray!6}{0.13} & \cellcolor{gray!6}{-1.02} & \cellcolor{gray!6}{-0.80} & \cellcolor{gray!6}{-0.58}\\
\cellcolor{gray!6}{RH group * Fraction same education} & \cellcolor{gray!6}{NYC} & \cellcolor{gray!6}{-0.08} & \cellcolor{gray!6}{0.08} & \cellcolor{gray!6}{-0.21} & \cellcolor{gray!6}{-0.08} & \cellcolor{gray!6}{0.04}\\
\cellcolor{gray!6}{RH group * Fraction same education} & \cellcolor{gray!6}{Phoenix} & \cellcolor{gray!6}{-0.79} & \cellcolor{gray!6}{0.08} & \cellcolor{gray!6}{-0.93} & \cellcolor{gray!6}{-0.79} & \cellcolor{gray!6}{-0.65}\\
C group * income & Miami & -0.19 & 0.06 & -0.29 & -0.19 & -0.10\\
\addlinespace
C group * income & NYC & -0.13 & 0.04 & -0.19 & -0.13 & -0.06\\
C group * income & Phoenix & 0.19 & 0.03 & 0.14 & 0.19 & 0.24\\
\cellcolor{gray!6}{P group * income} & \cellcolor{gray!6}{Miami} & \cellcolor{gray!6}{0.09} & \cellcolor{gray!6}{0.06} & \cellcolor{gray!6}{0.00} & \cellcolor{gray!6}{0.09} & \cellcolor{gray!6}{0.18}\\
\cellcolor{gray!6}{P group * income} & \cellcolor{gray!6}{NYC} & \cellcolor{gray!6}{-0.16} & \cellcolor{gray!6}{0.04} & \cellcolor{gray!6}{-0.23} & \cellcolor{gray!6}{-0.16} & \cellcolor{gray!6}{-0.09}\\
\cellcolor{gray!6}{P group * income} & \cellcolor{gray!6}{Phoenix} & \cellcolor{gray!6}{0.15} & \cellcolor{gray!6}{0.03} & \cellcolor{gray!6}{0.10} & \cellcolor{gray!6}{0.15} & \cellcolor{gray!6}{0.21}\\
\addlinespace
PH group * income & Miami & -0.06 & 0.06 & -0.15 & -0.06 & 0.04\\
PH group * income & NYC & -0.04 & 0.04 & -0.10 & -0.04 & 0.03\\
PH group * income & Phoenix & 0.13 & 0.03 & 0.08 & 0.13 & 0.19\\
\cellcolor{gray!6}{R group * income} & \cellcolor{gray!6}{Miami} & \cellcolor{gray!6}{-0.19} & \cellcolor{gray!6}{0.06} & \cellcolor{gray!6}{-0.29} & \cellcolor{gray!6}{-0.19} & \cellcolor{gray!6}{-0.09}\\
\cellcolor{gray!6}{R group * income} & \cellcolor{gray!6}{NYC} & \cellcolor{gray!6}{-0.13} & \cellcolor{gray!6}{0.04} & \cellcolor{gray!6}{-0.19} & \cellcolor{gray!6}{-0.13} & \cellcolor{gray!6}{-0.07}\\
\addlinespace
\cellcolor{gray!6}{R group * income} & \cellcolor{gray!6}{Phoenix} & \cellcolor{gray!6}{0.14} & \cellcolor{gray!6}{0.03} & \cellcolor{gray!6}{0.09} & \cellcolor{gray!6}{0.14} & \cellcolor{gray!6}{0.20}\\
RH group * income & Miami & 0.05 & 0.05 & -0.03 & 0.05 & 0.13\\
RH group * income & NYC & -0.19 & 0.04 & -0.26 & -0.19 & -0.13\\
RH group * income & Phoenix & 0.15 & 0.03 & 0.10 & 0.15 & 0.19\\
\cellcolor{gray!6}{education = No College * P group} & \cellcolor{gray!6}{Miami} & \cellcolor{gray!6}{2.29} & \cellcolor{gray!6}{7.93} & \cellcolor{gray!6}{-10.29} & \cellcolor{gray!6}{2.08} & \cellcolor{gray!6}{15.20}\\
\addlinespace
\cellcolor{gray!6}{education = No College * P group} & \cellcolor{gray!6}{NYC} & \cellcolor{gray!6}{-7.76} & \cellcolor{gray!6}{8.06} & \cellcolor{gray!6}{-20.66} & \cellcolor{gray!6}{-7.84} & \cellcolor{gray!6}{5.85}\\
\cellcolor{gray!6}{education = No College * P group} & \cellcolor{gray!6}{Phoenix} & \cellcolor{gray!6}{-2.44} & \cellcolor{gray!6}{1.57} & \cellcolor{gray!6}{-5.02} & \cellcolor{gray!6}{-2.44} & \cellcolor{gray!6}{0.14}\\
education = No College * PH group & Miami & 5.15 & 8.35 & -8.14 & 4.72 & 18.93\\
education = No College * PH group & NYC & -0.20 & 8.12 & -13.32 & -0.44 & 14.47\\
education = No College * PH group & Phoenix & 9.03 & 1.72 & 6.20 & 9.03 & 11.85\\
\addlinespace
\cellcolor{gray!6}{education = No College * R group} & \cellcolor{gray!6}{Miami} & \cellcolor{gray!6}{-5.21} & \cellcolor{gray!6}{8.18} & \cellcolor{gray!6}{-18.74} & \cellcolor{gray!6}{-5.07} & \cellcolor{gray!6}{7.71}\\
\cellcolor{gray!6}{education = No College * R group} & \cellcolor{gray!6}{NYC} & \cellcolor{gray!6}{-6.27} & \cellcolor{gray!6}{7.51} & \cellcolor{gray!6}{-18.56} & \cellcolor{gray!6}{-6.26} & \cellcolor{gray!6}{6.02}\\
\cellcolor{gray!6}{education = No College * R group} & \cellcolor{gray!6}{Phoenix} & \cellcolor{gray!6}{3.47} & \cellcolor{gray!6}{1.52} & \cellcolor{gray!6}{0.98} & \cellcolor{gray!6}{3.47} & \cellcolor{gray!6}{5.96}\\
education = No College * RH group & Miami & 11.38 & 8.09 & -1.80 & 11.72 & 24.71\\
education = No College * RH group & NYC & 1.03 & 7.01 & -10.34 & 1.21 & 12.34\\
\addlinespace
education = No College * RH group & Phoenix & -3.98 & 1.44 & -6.34 & -3.98 & -1.62\\
\cellcolor{gray!6}{party = IND * P group} & \cellcolor{gray!6}{Miami} & \cellcolor{gray!6}{-0.43} & \cellcolor{gray!6}{11.37} & \cellcolor{gray!6}{-18.68} & \cellcolor{gray!6}{-0.32} & \cellcolor{gray!6}{18.36}\\
\cellcolor{gray!6}{party = IND * P group} & \cellcolor{gray!6}{NYC} & \cellcolor{gray!6}{0.39} & \cellcolor{gray!6}{10.71} & \cellcolor{gray!6}{-17.55} & \cellcolor{gray!6}{0.35} & \cellcolor{gray!6}{17.23}\\
\cellcolor{gray!6}{party = IND * P group} & \cellcolor{gray!6}{Phoenix} & \cellcolor{gray!6}{0.77} & \cellcolor{gray!6}{0.56} & \cellcolor{gray!6}{-0.15} & \cellcolor{gray!6}{0.77} & \cellcolor{gray!6}{1.69}\\
party = REP * P group & Miami & 0.06 & 6.74 & -11.30 & 0.44 & 10.64\\
\addlinespace
party = REP * P group & NYC & 0.52 & 5.87 & -9.26 & 0.63 & 9.99\\
party = REP * P group & Phoenix & 0.14 & 0.30 & -0.36 & 0.14 & 0.63\\
\cellcolor{gray!6}{party = IND * PH group} & \cellcolor{gray!6}{Miami} & \cellcolor{gray!6}{1.55} & \cellcolor{gray!6}{11.60} & \cellcolor{gray!6}{-16.96} & \cellcolor{gray!6}{1.73} & \cellcolor{gray!6}{19.99}\\
\cellcolor{gray!6}{party = IND * PH group} & \cellcolor{gray!6}{NYC} & \cellcolor{gray!6}{0.32} & \cellcolor{gray!6}{11.14} & \cellcolor{gray!6}{-18.32} & \cellcolor{gray!6}{0.80} & \cellcolor{gray!6}{17.62}\\
\cellcolor{gray!6}{party = IND * PH group} & \cellcolor{gray!6}{Phoenix} & \cellcolor{gray!6}{1.28} & \cellcolor{gray!6}{0.64} & \cellcolor{gray!6}{0.23} & \cellcolor{gray!6}{1.28} & \cellcolor{gray!6}{2.32}\\
\addlinespace
party = REP * PH group & Miami & 0.25 & 7.07 & -10.93 & 0.14 & 12.14\\
party = REP * PH group & NYC & 0.53 & 6.42 & -10.38 & 0.72 & 10.64\\
party = REP * PH group & Phoenix & -0.22 & 0.31 & -0.73 & -0.22 & 0.28\\
\cellcolor{gray!6}{party = IND * R group} & \cellcolor{gray!6}{Miami} & \cellcolor{gray!6}{0.34} & \cellcolor{gray!6}{15.84} & \cellcolor{gray!6}{-25.60} & \cellcolor{gray!6}{0.34} & \cellcolor{gray!6}{26.06}\\
\cellcolor{gray!6}{party = IND * R group} & \cellcolor{gray!6}{NYC} & \cellcolor{gray!6}{-0.46} & \cellcolor{gray!6}{11.12} & \cellcolor{gray!6}{-18.74} & \cellcolor{gray!6}{-0.28} & \cellcolor{gray!6}{17.84}\\
\addlinespace
\cellcolor{gray!6}{party = IND * R group} & \cellcolor{gray!6}{Phoenix} & \cellcolor{gray!6}{1.16} & \cellcolor{gray!6}{0.56} & \cellcolor{gray!6}{0.24} & \cellcolor{gray!6}{1.16} & \cellcolor{gray!6}{2.08}\\
party = REP * R group & Miami & 0.37 & 7.72 & -12.80 & 0.39 & 12.83\\
party = REP * R group & NYC & 0.22 & 5.77 & -9.16 & 0.22 & 9.38\\
party = REP * R group & Phoenix & 0.09 & 0.30 & -0.40 & 0.09 & 0.58\\
\cellcolor{gray!6}{party = IND * RH group} & \cellcolor{gray!6}{Miami} & \cellcolor{gray!6}{1.37} & \cellcolor{gray!6}{11.74} & \cellcolor{gray!6}{-18.91} & \cellcolor{gray!6}{1.83} & \cellcolor{gray!6}{19.39}\\
\addlinespace
\cellcolor{gray!6}{party = IND * RH group} & \cellcolor{gray!6}{NYC} & \cellcolor{gray!6}{-0.40} & \cellcolor{gray!6}{11.26} & \cellcolor{gray!6}{-19.48} & \cellcolor{gray!6}{0.01} & \cellcolor{gray!6}{16.58}\\
\cellcolor{gray!6}{party = IND * RH group} & \cellcolor{gray!6}{Phoenix} & \cellcolor{gray!6}{-0.01} & \cellcolor{gray!6}{0.55} & \cellcolor{gray!6}{-0.91} & \cellcolor{gray!6}{-0.01} & \cellcolor{gray!6}{0.89}\\
party = REP * RH group & Miami & 0.48 & 6.85 & -10.09 & 0.31 & 11.85\\
party = REP * RH group & NYC & 0.24 & 6.44 & -10.47 & 0.29 & 10.75\\
party = REP * RH group & Phoenix & 0.25 & 0.30 & -0.24 & 0.25 & 0.74\\
\addlinespace
\cellcolor{gray!6}{minority * P group} & \cellcolor{gray!6}{Miami} & \cellcolor{gray!6}{-0.37} & \cellcolor{gray!6}{6.85} & \cellcolor{gray!6}{-11.30} & \cellcolor{gray!6}{-0.36} & \cellcolor{gray!6}{10.66}\\
\cellcolor{gray!6}{minority * P group} & \cellcolor{gray!6}{NYC} & \cellcolor{gray!6}{0.21} & \cellcolor{gray!6}{5.38} & \cellcolor{gray!6}{-8.83} & \cellcolor{gray!6}{0.18} & \cellcolor{gray!6}{8.88}\\
\cellcolor{gray!6}{minority * P group} & \cellcolor{gray!6}{Phoenix} & \cellcolor{gray!6}{0.25} & \cellcolor{gray!6}{0.44} & \cellcolor{gray!6}{-0.47} & \cellcolor{gray!6}{0.25} & \cellcolor{gray!6}{0.97}\\
minority * PH group & Miami & -0.24 & 7.20 & -12.10 & -0.53 & 11.81\\
minority * PH group & NYC & -0.72 & 5.64 & -10.23 & -0.75 & 8.22\\
\addlinespace
minority * PH group & Phoenix & -0.11 & 0.44 & -0.82 & -0.11 & 0.61\\
\cellcolor{gray!6}{minority * R group} & \cellcolor{gray!6}{Miami} & \cellcolor{gray!6}{0.27} & \cellcolor{gray!6}{7.98} & \cellcolor{gray!6}{-12.95} & \cellcolor{gray!6}{0.22} & \cellcolor{gray!6}{12.88}\\
\cellcolor{gray!6}{minority * R group} & \cellcolor{gray!6}{NYC} & \cellcolor{gray!6}{-0.06} & \cellcolor{gray!6}{5.78} & \cellcolor{gray!6}{-9.32} & \cellcolor{gray!6}{-0.04} & \cellcolor{gray!6}{9.64}\\
\cellcolor{gray!6}{minority * R group} & \cellcolor{gray!6}{Phoenix} & \cellcolor{gray!6}{0.05} & \cellcolor{gray!6}{0.43} & \cellcolor{gray!6}{-0.66} & \cellcolor{gray!6}{0.05} & \cellcolor{gray!6}{0.75}\\
minority * RH group & Miami & -0.19 & 6.82 & -11.80 & -0.06 & 10.74\\
\addlinespace
minority * RH group & NYC & -0.47 & 5.77 & -9.91 & -0.48 & 9.53\\
minority * RH group & Phoenix & 0.33 & 0.41 & -0.34 & 0.33 & 1.00\\
\cellcolor{gray!6}{homeowner * P group} & \cellcolor{gray!6}{Miami} & \cellcolor{gray!6}{-1.92} & \cellcolor{gray!6}{4.74} & \cellcolor{gray!6}{-9.85} & \cellcolor{gray!6}{-1.95} & \cellcolor{gray!6}{5.77}\\
\cellcolor{gray!6}{homeowner * P group} & \cellcolor{gray!6}{NYC} & \cellcolor{gray!6}{0.25} & \cellcolor{gray!6}{3.89} & \cellcolor{gray!6}{-5.93} & \cellcolor{gray!6}{0.26} & \cellcolor{gray!6}{6.68}\\
\cellcolor{gray!6}{homeowner * P group} & \cellcolor{gray!6}{Phoenix} & \cellcolor{gray!6}{0.83} & \cellcolor{gray!6}{0.43} & \cellcolor{gray!6}{0.13} & \cellcolor{gray!6}{0.83} & \cellcolor{gray!6}{1.53}\\
\addlinespace
homeowner * PH group & Miami & -1.08 & 5.12 & -9.44 & -1.08 & 7.32\\
homeowner * PH group & NYC & -0.60 & 4.13 & -7.30 & -0.63 & 6.31\\
homeowner * PH group & Phoenix & 0.51 & 0.43 & -0.21 & 0.51 & 1.22\\
\cellcolor{gray!6}{homeowner * R group} & \cellcolor{gray!6}{Miami} & \cellcolor{gray!6}{-0.26} & \cellcolor{gray!6}{5.27} & \cellcolor{gray!6}{-8.67} & \cellcolor{gray!6}{-0.54} & \cellcolor{gray!6}{8.53}\\
\cellcolor{gray!6}{homeowner * R group} & \cellcolor{gray!6}{NYC} & \cellcolor{gray!6}{-0.08} & \cellcolor{gray!6}{4.08} & \cellcolor{gray!6}{-6.90} & \cellcolor{gray!6}{-0.09} & \cellcolor{gray!6}{7.02}\\
\addlinespace
\cellcolor{gray!6}{homeowner * R group} & \cellcolor{gray!6}{Phoenix} & \cellcolor{gray!6}{0.54} & \cellcolor{gray!6}{0.42} & \cellcolor{gray!6}{-0.15} & \cellcolor{gray!6}{0.54} & \cellcolor{gray!6}{1.23}\\
homeowner * RH group & Miami & -2.03 & 4.77 & -9.99 & -2.07 & 5.44\\
homeowner * RH group & NYC & 0.88 & 4.22 & -5.92 & 0.65 & 7.86\\
homeowner * RH group & Phoenix & 0.12 & 0.41 & -0.56 & 0.12 & 0.81\\
\cellcolor{gray!6}{minority * Fraction same race * C group} & \cellcolor{gray!6}{Miami} & \cellcolor{gray!6}{0.43} & \cellcolor{gray!6}{0.10} & \cellcolor{gray!6}{0.26} & \cellcolor{gray!6}{0.44} & \cellcolor{gray!6}{0.60}\\
\addlinespace
\cellcolor{gray!6}{minority * Fraction same race * C group} & \cellcolor{gray!6}{NYC} & \cellcolor{gray!6}{0.43} & \cellcolor{gray!6}{0.12} & \cellcolor{gray!6}{0.23} & \cellcolor{gray!6}{0.43} & \cellcolor{gray!6}{0.63}\\
\cellcolor{gray!6}{minority * Fraction same race * C group} & \cellcolor{gray!6}{Phoenix} & \cellcolor{gray!6}{-0.28} & \cellcolor{gray!6}{0.27} & \cellcolor{gray!6}{-0.73} & \cellcolor{gray!6}{-0.28} & \cellcolor{gray!6}{0.17}\\
minority * Fraction same race * P group & Miami & 0.38 & 0.11 & 0.21 & 0.38 & 0.56\\
minority * Fraction same race * P group & NYC & 0.32 & 0.14 & 0.10 & 0.31 & 0.56\\
minority * Fraction same race * P group & Phoenix & 0.42 & 0.24 & 0.04 & 0.42 & 0.81\\
\addlinespace
\cellcolor{gray!6}{minority * Fraction same race * PH group} & \cellcolor{gray!6}{Miami} & \cellcolor{gray!6}{-0.10} & \cellcolor{gray!6}{0.09} & \cellcolor{gray!6}{-0.25} & \cellcolor{gray!6}{-0.10} & \cellcolor{gray!6}{0.06}\\
\cellcolor{gray!6}{minority * Fraction same race * PH group} & \cellcolor{gray!6}{NYC} & \cellcolor{gray!6}{0.24} & \cellcolor{gray!6}{0.11} & \cellcolor{gray!6}{0.06} & \cellcolor{gray!6}{0.24} & \cellcolor{gray!6}{0.42}\\
\cellcolor{gray!6}{minority * Fraction same race * PH group} & \cellcolor{gray!6}{Phoenix} & \cellcolor{gray!6}{0.33} & \cellcolor{gray!6}{0.13} & \cellcolor{gray!6}{0.12} & \cellcolor{gray!6}{0.33} & \cellcolor{gray!6}{0.55}\\
minority * Fraction same race * R group & Miami & 0.16 & 0.12 & -0.05 & 0.16 & 0.36\\
minority * Fraction same race * R group & NYC & 0.39 & 0.17 & 0.11 & 0.38 & 0.67\\
\addlinespace
minority * Fraction same race * R group & Phoenix & 0.69 & 0.13 & 0.47 & 0.69 & 0.91\\
\cellcolor{gray!6}{minority * Fraction same race * RH group} & \cellcolor{gray!6}{Miami} & \cellcolor{gray!6}{0.17} & \cellcolor{gray!6}{0.09} & \cellcolor{gray!6}{0.02} & \cellcolor{gray!6}{0.16} & \cellcolor{gray!6}{0.32}\\
\cellcolor{gray!6}{minority * Fraction same race * RH group} & \cellcolor{gray!6}{NYC} & \cellcolor{gray!6}{1.30} & \cellcolor{gray!6}{0.17} & \cellcolor{gray!6}{1.02} & \cellcolor{gray!6}{1.29} & \cellcolor{gray!6}{1.58}\\
\cellcolor{gray!6}{minority * Fraction same race * RH group} & \cellcolor{gray!6}{Phoenix} & \cellcolor{gray!6}{0.02} & \cellcolor{gray!6}{0.14} & \cellcolor{gray!6}{-0.21} & \cellcolor{gray!6}{0.02} & \cellcolor{gray!6}{0.25}\\
party = IND * C group * Fraction same party & Miami & -0.21 & 0.24 & -0.58 & -0.21 & 0.17\\
\addlinespace
party = IND * C group * Fraction same party & NYC & 0.00 & 0.25 & -0.41 & 0.00 & 0.41\\
party = IND * C group * Fraction same party & Phoenix & 0.37 & 0.18 & 0.07 & 0.37 & 0.68\\
\cellcolor{gray!6}{party = REP * C group * Fraction same party} & \cellcolor{gray!6}{Miami} & \cellcolor{gray!6}{-0.38} & \cellcolor{gray!6}{0.16} & \cellcolor{gray!6}{-0.63} & \cellcolor{gray!6}{-0.39} & \cellcolor{gray!6}{-0.11}\\
\cellcolor{gray!6}{party = REP * C group * Fraction same party} & \cellcolor{gray!6}{NYC} & \cellcolor{gray!6}{0.20} & \cellcolor{gray!6}{0.13} & \cellcolor{gray!6}{-0.02} & \cellcolor{gray!6}{0.20} & \cellcolor{gray!6}{0.42}\\
\cellcolor{gray!6}{party = REP * C group * Fraction same party} & \cellcolor{gray!6}{Phoenix} & \cellcolor{gray!6}{0.00} & \cellcolor{gray!6}{0.10} & \cellcolor{gray!6}{-0.17} & \cellcolor{gray!6}{0.00} & \cellcolor{gray!6}{0.17}\\
\addlinespace
party = IND * P group * Fraction same party & Miami & -0.61 & 0.27 & -1.04 & -0.61 & -0.16\\
party = IND * P group * Fraction same party & NYC & -0.34 & 0.54 & -1.24 & -0.36 & 0.54\\
party = IND * P group * Fraction same party & Phoenix & 0.00 & 0.18 & -0.29 & 0.00 & 0.29\\
\cellcolor{gray!6}{party = REP * P group * Fraction same party} & \cellcolor{gray!6}{Miami} & \cellcolor{gray!6}{-0.09} & \cellcolor{gray!6}{0.15} & \cellcolor{gray!6}{-0.34} & \cellcolor{gray!6}{-0.09} & \cellcolor{gray!6}{0.15}\\
\cellcolor{gray!6}{party = REP * P group * Fraction same party} & \cellcolor{gray!6}{NYC} & \cellcolor{gray!6}{-0.34} & \cellcolor{gray!6}{0.17} & \cellcolor{gray!6}{-0.61} & \cellcolor{gray!6}{-0.33} & \cellcolor{gray!6}{-0.08}\\
\addlinespace
\cellcolor{gray!6}{party = REP * P group * Fraction same party} & \cellcolor{gray!6}{Phoenix} & \cellcolor{gray!6}{0.03} & \cellcolor{gray!6}{0.11} & \cellcolor{gray!6}{-0.14} & \cellcolor{gray!6}{0.03} & \cellcolor{gray!6}{0.21}\\
party = IND * PH group * Fraction same party & Miami & -0.19 & 0.35 & -0.75 & -0.19 & 0.39\\
party = IND * PH group * Fraction same party & NYC & 0.70 & 0.33 & 0.16 & 0.70 & 1.26\\
party = IND * PH group * Fraction same party & Phoenix & -0.16 & 0.30 & -0.65 & -0.16 & 0.33\\
\cellcolor{gray!6}{party = REP * PH group * Fraction same party} & \cellcolor{gray!6}{Miami} & \cellcolor{gray!6}{-0.88} & \cellcolor{gray!6}{0.17} & \cellcolor{gray!6}{-1.17} & \cellcolor{gray!6}{-0.88} & \cellcolor{gray!6}{-0.61}\\
\addlinespace
\cellcolor{gray!6}{party = REP * PH group * Fraction same party} & \cellcolor{gray!6}{NYC} & \cellcolor{gray!6}{-0.36} & \cellcolor{gray!6}{0.23} & \cellcolor{gray!6}{-0.72} & \cellcolor{gray!6}{-0.36} & \cellcolor{gray!6}{0.01}\\
\cellcolor{gray!6}{party = REP * PH group * Fraction same party} & \cellcolor{gray!6}{Phoenix} & \cellcolor{gray!6}{-0.33} & \cellcolor{gray!6}{0.10} & \cellcolor{gray!6}{-0.50} & \cellcolor{gray!6}{-0.33} & \cellcolor{gray!6}{-0.16}\\
party = IND * R group * Fraction same party & Miami & 0.21 & 0.80 & -1.08 & 0.24 & 1.55\\
party = IND * R group * Fraction same party & NYC & -0.28 & 0.39 & -0.95 & -0.27 & 0.33\\
party = IND * R group * Fraction same party & Phoenix & 0.40 & 0.25 & -0.02 & 0.40 & 0.82\\
\addlinespace
\cellcolor{gray!6}{party = REP * R group * Fraction same party} & \cellcolor{gray!6}{Miami} & \cellcolor{gray!6}{-0.13} & \cellcolor{gray!6}{0.16} & \cellcolor{gray!6}{-0.39} & \cellcolor{gray!6}{-0.13} & \cellcolor{gray!6}{0.12}\\
\cellcolor{gray!6}{party = REP * R group * Fraction same party} & \cellcolor{gray!6}{NYC} & \cellcolor{gray!6}{-0.08} & \cellcolor{gray!6}{0.15} & \cellcolor{gray!6}{-0.35} & \cellcolor{gray!6}{-0.07} & \cellcolor{gray!6}{0.17}\\
\cellcolor{gray!6}{party = REP * R group * Fraction same party} & \cellcolor{gray!6}{Phoenix} & \cellcolor{gray!6}{0.22} & \cellcolor{gray!6}{0.10} & \cellcolor{gray!6}{0.07} & \cellcolor{gray!6}{0.22} & \cellcolor{gray!6}{0.38}\\
party = IND * RH group * Fraction same party & Miami & -0.01 & 0.29 & -0.49 & -0.01 & 0.46\\
party = IND * RH group * Fraction same party & NYC & -0.21 & 0.21 & -0.55 & -0.22 & 0.14\\
\addlinespace
party = IND * RH group * Fraction same party & Phoenix & -0.73 & 0.16 & -0.99 & -0.73 & -0.47\\
\cellcolor{gray!6}{party = REP * RH group * Fraction same party} & \cellcolor{gray!6}{Miami} & \cellcolor{gray!6}{-0.49} & \cellcolor{gray!6}{0.15} & \cellcolor{gray!6}{-0.75} & \cellcolor{gray!6}{-0.49} & \cellcolor{gray!6}{-0.23}\\
\cellcolor{gray!6}{party = REP * RH group * Fraction same party} & \cellcolor{gray!6}{NYC} & \cellcolor{gray!6}{-0.13} & \cellcolor{gray!6}{0.21} & \cellcolor{gray!6}{-0.47} & \cellcolor{gray!6}{-0.13} & \cellcolor{gray!6}{0.20}\\
\cellcolor{gray!6}{party = REP * RH group * Fraction same party} & \cellcolor{gray!6}{Phoenix} & \cellcolor{gray!6}{-0.32} & \cellcolor{gray!6}{0.09} & \cellcolor{gray!6}{-0.46} & \cellcolor{gray!6}{-0.32} & \cellcolor{gray!6}{-0.18}\\
homeowner * C group * Fraction same ownership & Miami & 0.81 & 0.21 & 0.46 & 0.81 & 1.16\\
\addlinespace
homeowner * C group * Fraction same ownership & NYC & 0.67 & 0.16 & 0.40 & 0.67 & 0.92\\
homeowner * C group * Fraction same ownership & Phoenix & 0.38 & 0.16 & 0.12 & 0.38 & 0.64\\
\cellcolor{gray!6}{homeowner * P group * Fraction same ownership} & \cellcolor{gray!6}{Miami} & \cellcolor{gray!6}{-0.38} & \cellcolor{gray!6}{0.17} & \cellcolor{gray!6}{-0.65} & \cellcolor{gray!6}{-0.37} & \cellcolor{gray!6}{-0.12}\\
\cellcolor{gray!6}{homeowner * P group * Fraction same ownership} & \cellcolor{gray!6}{NYC} & \cellcolor{gray!6}{0.46} & \cellcolor{gray!6}{0.16} & \cellcolor{gray!6}{0.20} & \cellcolor{gray!6}{0.46} & \cellcolor{gray!6}{0.72}\\
\cellcolor{gray!6}{homeowner * P group * Fraction same ownership} & \cellcolor{gray!6}{Phoenix} & \cellcolor{gray!6}{-0.96} & \cellcolor{gray!6}{0.17} & \cellcolor{gray!6}{-1.24} & \cellcolor{gray!6}{-0.96} & \cellcolor{gray!6}{-0.67}\\
\addlinespace
homeowner * PH group * Fraction same ownership & Miami & 0.38 & 0.28 & -0.06 & 0.38 & 0.84\\
homeowner * PH group * Fraction same ownership & NYC & 0.32 & 0.15 & 0.07 & 0.32 & 0.56\\
homeowner * PH group * Fraction same ownership & Phoenix & -0.13 & 0.15 & -0.37 & -0.13 & 0.11\\
\cellcolor{gray!6}{homeowner * R group * Fraction same ownership} & \cellcolor{gray!6}{Miami} & \cellcolor{gray!6}{-1.00} & \cellcolor{gray!6}{0.79} & \cellcolor{gray!6}{-2.30} & \cellcolor{gray!6}{-0.98} & \cellcolor{gray!6}{0.27}\\
\cellcolor{gray!6}{homeowner * R group * Fraction same ownership} & \cellcolor{gray!6}{NYC} & \cellcolor{gray!6}{0.13} & \cellcolor{gray!6}{0.15} & \cellcolor{gray!6}{-0.13} & \cellcolor{gray!6}{0.14} & \cellcolor{gray!6}{0.37}\\
\addlinespace
\cellcolor{gray!6}{homeowner * R group * Fraction same ownership} & \cellcolor{gray!6}{Phoenix} & \cellcolor{gray!6}{-0.47} & \cellcolor{gray!6}{0.17} & \cellcolor{gray!6}{-0.74} & \cellcolor{gray!6}{-0.47} & \cellcolor{gray!6}{-0.20}\\
homeowner * RH group * Fraction same ownership & Miami & 0.28 & 0.26 & -0.13 & 0.29 & 0.69\\
homeowner * RH group * Fraction same ownership & NYC & 0.10 & 0.13 & -0.12 & 0.10 & 0.32\\
homeowner * RH group * Fraction same ownership & Phoenix & 0.54 & 0.18 & 0.25 & 0.54 & 0.83\\
\cellcolor{gray!6}{education = No College * C group * Fraction same educ} & \cellcolor{gray!6}{Miami} & \cellcolor{gray!6}{0.81} & \cellcolor{gray!6}{0.31} & \cellcolor{gray!6}{0.30} & \cellcolor{gray!6}{0.81} & \cellcolor{gray!6}{1.31}\\
\addlinespace
\cellcolor{gray!6}{education = No College * C group * Fraction same educ} & \cellcolor{gray!6}{NYC} & \cellcolor{gray!6}{1.48} & \cellcolor{gray!6}{0.21} & \cellcolor{gray!6}{1.13} & \cellcolor{gray!6}{1.49} & \cellcolor{gray!6}{1.82}\\
\cellcolor{gray!6}{education = No College * C group * Fraction same educ} & \cellcolor{gray!6}{Phoenix} & \cellcolor{gray!6}{1.03} & \cellcolor{gray!6}{0.27} & \cellcolor{gray!6}{0.59} & \cellcolor{gray!6}{1.03} & \cellcolor{gray!6}{1.47}\\
education = No College * P group * Fraction same educ & Miami & 0.24 & 0.35 & -0.34 & 0.24 & 0.81\\
education = No College * P group * Fraction same educ & NYC & 1.23 & 0.35 & 0.66 & 1.22 & 1.78\\
education = No College * P group * Fraction same educ & Phoenix & 1.74 & 0.25 & 1.32 & 1.74 & 2.16\\
\addlinespace
\cellcolor{gray!6}{education = No College * PH group * Fraction same educ} & \cellcolor{gray!6}{Miami} & \cellcolor{gray!6}{0.03} & \cellcolor{gray!6}{0.34} & \cellcolor{gray!6}{-0.54} & \cellcolor{gray!6}{0.01} & \cellcolor{gray!6}{0.60}\\
\cellcolor{gray!6}{education = No College * PH group * Fraction same educ} & \cellcolor{gray!6}{NYC} & \cellcolor{gray!6}{2.77} & \cellcolor{gray!6}{0.42} & \cellcolor{gray!6}{2.07} & \cellcolor{gray!6}{2.79} & \cellcolor{gray!6}{3.42}\\
\cellcolor{gray!6}{education = No College * PH group * Fraction same educ} & \cellcolor{gray!6}{Phoenix} & \cellcolor{gray!6}{1.18} & \cellcolor{gray!6}{0.29} & \cellcolor{gray!6}{0.70} & \cellcolor{gray!6}{1.18} & \cellcolor{gray!6}{1.65}\\
education = No College * R group * Fraction same educ & Miami & 2.02 & 0.35 & 1.44 & 2.03 & 2.56\\
education = No College * R group * Fraction same educ & NYC & -0.53 & 0.29 & -1.01 & -0.54 & -0.07\\
\addlinespace
education = No College * R group * Fraction same educ & Phoenix & 0.36 & 0.23 & -0.02 & 0.36 & 0.73\\
\cellcolor{gray!6}{education = No College * RH group * Fraction same educ} & \cellcolor{gray!6}{Miami} & \cellcolor{gray!6}{-0.49} & \cellcolor{gray!6}{0.36} & \cellcolor{gray!6}{-1.09} & \cellcolor{gray!6}{-0.49} & \cellcolor{gray!6}{0.10}\\
\cellcolor{gray!6}{education = No College * RH group * Fraction same educ} & \cellcolor{gray!6}{NYC} & \cellcolor{gray!6}{0.51} & \cellcolor{gray!6}{0.24} & \cellcolor{gray!6}{0.13} & \cellcolor{gray!6}{0.50} & \cellcolor{gray!6}{0.92}\\
\cellcolor{gray!6}{education = No College * RH group * Fraction same educ} & \cellcolor{gray!6}{Phoenix} & \cellcolor{gray!6}{1.53} & \cellcolor{gray!6}{0.19} & \cellcolor{gray!6}{1.22} & \cellcolor{gray!6}{1.53} & \cellcolor{gray!6}{1.83}\\
education = No College * C group * income & Miami & 0.26 & 0.10 & 0.08 & 0.26 & 0.43\\
\addlinespace
education = No College * C group * income & NYC & -0.02 & 0.09 & -0.16 & -0.02 & 0.12\\
education = No College * C group * income & Phoenix & 0.07 & 0.10 & -0.08 & 0.07 & 0.23\\
\cellcolor{gray!6}{education = No College * P group * income} & \cellcolor{gray!6}{Miami} & \cellcolor{gray!6}{0.00} & \cellcolor{gray!6}{0.12} & \cellcolor{gray!6}{-0.22} & \cellcolor{gray!6}{0.00} & \cellcolor{gray!6}{0.19}\\
\cellcolor{gray!6}{education = No College * P group * income} & \cellcolor{gray!6}{NYC} & \cellcolor{gray!6}{0.64} & \cellcolor{gray!6}{0.14} & \cellcolor{gray!6}{0.41} & \cellcolor{gray!6}{0.64} & \cellcolor{gray!6}{0.87}\\
\cellcolor{gray!6}{education = No College * P group * income} & \cellcolor{gray!6}{Phoenix} & \cellcolor{gray!6}{0.20} & \cellcolor{gray!6}{0.09} & \cellcolor{gray!6}{0.06} & \cellcolor{gray!6}{0.20} & \cellcolor{gray!6}{0.34}\\
\addlinespace
education = No College * PH group * income & Miami & -0.22 & 0.12 & -0.41 & -0.21 & -0.01\\
education = No College * PH group * income & NYC & -0.05 & 0.17 & -0.32 & -0.05 & 0.23\\
education = No College * PH group * income & Phoenix & -0.76 & 0.10 & -0.93 & -0.76 & -0.59\\
\cellcolor{gray!6}{education = No College * R group * income} & \cellcolor{gray!6}{Miami} & \cellcolor{gray!6}{0.66} & \cellcolor{gray!6}{0.13} & \cellcolor{gray!6}{0.45} & \cellcolor{gray!6}{0.66} & \cellcolor{gray!6}{0.87}\\
\cellcolor{gray!6}{education = No College * R group * income} & \cellcolor{gray!6}{NYC} & \cellcolor{gray!6}{0.63} & \cellcolor{gray!6}{0.12} & \cellcolor{gray!6}{0.42} & \cellcolor{gray!6}{0.62} & \cellcolor{gray!6}{0.84}\\
\addlinespace
\cellcolor{gray!6}{education = No College * R group * income} & \cellcolor{gray!6}{Phoenix} & \cellcolor{gray!6}{-0.24} & \cellcolor{gray!6}{0.08} & \cellcolor{gray!6}{-0.37} & \cellcolor{gray!6}{-0.24} & \cellcolor{gray!6}{-0.11}\\
education = No College * RH group * income & Miami & -0.73 & 0.13 & -0.94 & -0.74 & -0.52\\
education = No College * RH group * income & NYC & -0.08 & 0.10 & -0.26 & -0.08 & 0.09\\
education = No College * RH group * income & Phoenix & 0.36 & 0.07 & 0.24 & 0.36 & 0.47\\
\cellcolor{gray!6}{alpha} & \cellcolor{gray!6}{Miami} & \cellcolor{gray!6}{1.12} & \cellcolor{gray!6}{0.02} & \cellcolor{gray!6}{1.09} & \cellcolor{gray!6}{1.12} & \cellcolor{gray!6}{1.15}\\
\addlinespace
\cellcolor{gray!6}{alpha} & \cellcolor{gray!6}{NYC} & \cellcolor{gray!6}{1.39} & \cellcolor{gray!6}{0.02} & \cellcolor{gray!6}{1.36} & \cellcolor{gray!6}{1.39} & \cellcolor{gray!6}{1.42}\\
\cellcolor{gray!6}{alpha} & \cellcolor{gray!6}{Phoenix} & \cellcolor{gray!6}{1.34} & \cellcolor{gray!6}{0.01} & \cellcolor{gray!6}{1.33} & \cellcolor{gray!6}{1.34} & \cellcolor{gray!6}{1.36}\\
\bottomrule
\end{longtable}
\endgroup{}

\break

\hypertarget{using-drawn-maps-to-measure-contextual-variables}{%
\section{Using drawn maps to measure contextual variables}\label{using-drawn-maps-to-measure-contextual-variables}}

Here, we demonstrate how researchers can use subjective neighborhoods as an improved measure of contextual variables. As described in the manuscript, researchers may include our survey application in their survey, collect drawn maps, and then calculate any geographic variables they want to include in their analysis using the map a given respondent drew. For example, if a researcher cares about how local exposure to people of different races or political orientations influences political attitudes or behavior, they can calculate racial and partisan composition within each subjective neighborhood and include those variables in a regression predicting their outcomes of interest. To illustrate this, we model levels of neighborhood trust -- survey respondents were asked three questions about how much they trust their neighbors, each with a 10 point scale, and we took the average of these questions as a neighor trust scale -- and self-reported turnout in the 2020 presidential election as a function of the proportion white, proportion college educated, and proportion Democrat in each respondent's drawn neighborhood. We run OLS models on the control group in the first survey.

Table \ref{tab:trust-reg} reports the results. We find that trust in one's neighbors is generally increasing as the proportion of college educated residents in one's subjective neighborhood increases. We also find that Republicans report lower trust in their neighbors as the proportion of Democrats in their subjective neighborhood increases. We do not observe statistically significant predictors of self-reported turnout in this sample.

\begin{table} \centering 
  \caption{Neighborhood trust by subjective neighborhood race and partisan composition} 
  \label{tab:trust-reg} 
  \scriptsize
\begin{tabular}{@{\extracolsep{0pt}}lc} 
\\\hline 
\hline \\ 
 & \multicolumn{1}{c}{\textit{Dependent variable:}} \\ 
\cline{2-2} 
\\ & Trust \\ 
\\ & (1) \\ 
\hline \\ 
 Prop. White & 0.249 \\ 
  & (0.290)  \\ 
  & \\ 
 White & 0.127  \\ 
  & (0.191)  \\ 
  &  \\ 
 Prop. Dem & 0.108  \\ 
  & (0.286)  \\ 
  &  \\ 
 Independent & 0.476  \\ 
  & (0.318)  \\ 
  &  \\ 
 Republican & 0.367$^{**}$  \\ 
  & (0.154) \\ 
  &  \\ 
Prop. College & 0.465$^{**}$  \\ 
  & (0.191)  \\ 
  &  \\ 
 College & $-$0.073  \\ 
  & (0.065)  \\ 
  &  \\ 
 Male & $-$0.034  \\ 
  & (0.054)  \\ 
  &  \\ 
 Age & 0.010$^{***}$  \\ 
  & (0.002) \\ 
  &  \\ 
 Married & 0.023  \\ 
  & (0.060)  \\ 
  & \\ 
 Children in Home & 0.076 \\ 
  & (0.063)  \\ 
  &  \\ 
 Homeowner & $-$0.042 \\ 
  & (0.090) \\ 
  &  \\ 
 Prop. White * White  & $-$0.089  \\ 
  & (0.301) \\ 
  &  \\ 
Prop. Democrat * Independent & $-$1.226  \\ 
  & (0.784)  \\ 
  &  \\ 
Prop. Democrat * Republican  & $-$1.032$^{**}$  \\ 
  & (0.414)  \\ 
  &  \\ 
 Constant & 1.241$^{***}$  \\ 
  & (0.259)  \\ 
  &  \\ 
\hline \\ 
Observations & 437  \\ 
R$^{2}$ & 0.121  \\ 
Adjusted R$^{2}$ & 0.090 \\ 
Residual Std. Error & 0.545 (df = 421)  \\ 
F Statistic & 3.869$^{***}$ (df = 15; 421)  \\ 
\hline 
\hline \\ 
\multicolumn{2}{r}{\textit{Note:}   $^{*}$p$<$0.1; $^{**}$p$<$0.05; $^{***}$p$<$0.01} \\ 
\end{tabular} 
\end{table}

\end{document}